\documentclass[twocolumn,amsmath,amssymb,aps,rmp,reprint,longbibliography]{revtex4-1}

\usepackage{graphicx}
\usepackage{bm}
\begin{document}

\title{
Statics and Dynamics of Skyrmions Interacting with Pinning: A Review  
} 
\author{C. Reichhardt}
\author{C. J. O. Reichhardt}
\affiliation{
Theoretical Division and Center for Nonlinear Studies,
Los Alamos National Laboratory, Los Alamos, NM 87545, USA}
\author{M. V. Milo\v{s}evi\' c}
\affiliation{
NANOlab Center of Excellence, Department of Physics, University of Antwerp, Belgium
} 

\date{\today{}}

\begin{abstract}
  Magnetic skyrmions are topologically stable nanoscale particle-like
  objects that were discovered in 2009. 
  Since that time, intense research interest in the field has led to
  the identification of
  numerous compounds that
  support skyrmions over a range of conditions spanning cryogenic
  to room temperatures.
  Skyrmions can be set into motion under
  various types of driving, and
  the combination of their size, stability, and dynamics
  makes
  them ideal candidates for numerous applications.
  At the same time,
skyrmions represent a new class of system
in which the energy scales of the
skyrmion-skyrmion interactions, 
sample disorder,
temperature, and drive can compete.
A growing body of work indicates that the
static and dynamic states
of skyrmions can be influenced strongly
by pinning or disorder in the sample; thus,
an understanding of such effects
is essential for the eventual use of skyrmions in applications.
In this article we review the current state of knowledge
regarding
individual skyrmions and skyrmion assemblies
interacting with quenched disorder
or pinning.
We outline the microscopic
mechanisms for skyrmion pinning, including the repulsive and
attractive interactions that can arise from
impurities, grain boundaries, or nanostructures.
This is followed by descriptions of
depinning phenomena, sliding states over disorder, the 
effect of pinning on the skyrmion Hall angle,
the competition between thermal and pinning effects,
the control of skyrmion motion using ordered potential landscapes
such as one- or two-dimensional periodic asymmetric substrates,
the creation of skyrmion diodes, and skyrmion ratchet effects.   
We highlight the distinctions
arising from internal modes and the strong gyroscopic or Magnus forces
that cause
the dynamical states of skyrmions
to differ from those of
other systems with pinning,
such as vortices in type-II superconductors, charge density waves, or colloidal particles.
Throughout this work
we also discuss future directions and open questions
related to the pinning and dynamics in skyrmion 
systems.
\end{abstract}

\maketitle

\tableofcontents{}

\section{Introduction}

\begin{figure}
\includegraphics[width=\columnwidth]{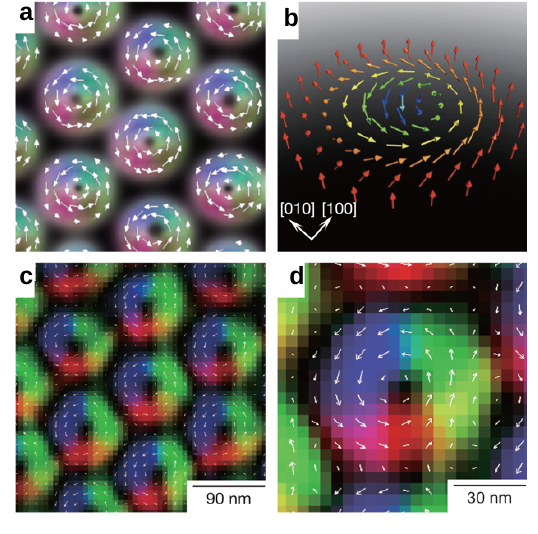}
\caption{
  Skyrmion crystal image obtained using
  Lorentz microscopy on thin film Fe$_{0.5}$Co$_{0.5}$Si
near $T =25$K from Ref.~\cite{Yu10}. 
(a) The spin structures predicted by simulation.
(b) Schematic of the spin configuration in a single skyrmion.
(c) Lorentz image of the skyrmion lattice.
(d) Magnified view of panel (c).
Here the skyrmions are on the order of 90 nm in diameter.
Reprinted by permission from: Springer Nature, 
X. Z. Yu {\it et al.},
``Real-space observation of a two-dimensional skyrmion crystal'',
Nature (London) {\bf 465}, 901 (2010),
\copyright 2010.
}
\label{fig:1}
\end{figure}

\begin{figure}
\includegraphics[width=\columnwidth]{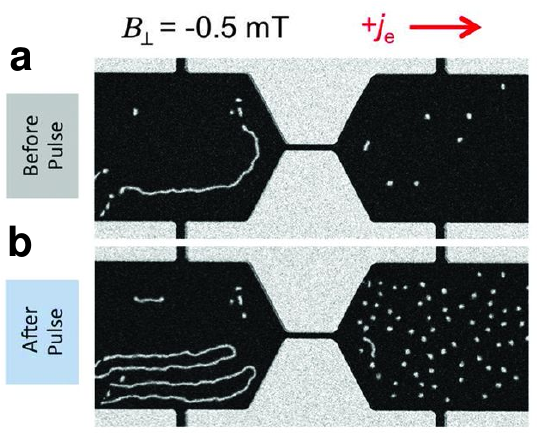}
\caption{
  Image of skyrmion creation at room temperature by passing current
  through a constriction \cite{Jiang15}.
  Here the skyrmions are approximately a micron in diameter.
  From W. Jiang {\it et al.}, Science {\bf 349}, 283 (2015). Reprinted with
  permission from AAAS.
}
\label{fig:2}
\end{figure}

The idea of a particle like magnetic texture called a skyrmion
was initially proposed theoretically \cite{Bogdanov89,Rossler06}, and
was confirmed experimentally
in 2009 when
neutron scattering experiments revealed a six-fold scattering pattern 
in the chiral magnet MnSi, indicating the presence of a collection of
lines forming a two-dimensional (2D)
hexagonal skyrmion lattice \cite{Muhlbauer09}.
Shortly afterward, direct images of 
the skyrmion lattice
in thin film samples
were obtained using Lorentz microscopy \cite{Yu10}.
Since this initial discovery, 
skyrmions
with sizes ranging from micron scale down to 10 nm 
have been identified in a growing number of
2D, three-dimensional (3D), and layered materials
\cite{Yu11,Heinze11,Seki12,Nagaosa13,Milde13,Romming13,Jiang15,Wiesendanger16,Wang18}.

As an applied magnetic field is
increased, skyrmions emerge from the helical state in the form of a lattice,
remain stable over a range of temperatures and fields,
and then disappear at sufficiently high fields when
the system transitions into a ferromagnetic state
\cite{Muhlbauer09,Nagaosa13}.   
The predicted 
spin structure of a
skyrmion lattice and of an individual skyrmion, shown schematically
in 
Fig.~\ref{fig:1}(a, b), agrees well with 
the initial Lorentz microscopy images
in Fig.~\ref{fig:1}(c, d)
of skyrmions
that are approximately 90 nm in diameter
\cite{Yu10}.
These first observations of skyrmions
were performed
at temperatures near $T = 30$ K,
but
since that time
numerous systems
have been identified which support skyrmions
at and above room temperature
\cite{Jiang15,Wiesendanger16,Tokunaga15,MoreauLuchaire16,Woo16,Boulle16,Soumyanarayanan17}.
Figure~\ref{fig:2} shows
images of skyrmion bubbles
of diameter close to a micron
created in a room temperature system
\cite{Jiang15}. 
It is also possible to observe
transitions from hexagonal to square skyrmion
lattices \cite{Yi09,Karube16,Nakajima17},
as well as new types of particle like textures such as a square meron lattice 
that transitions into a triangular skyrmion lattice \cite{Yu18}. 

\begin{figure}
\includegraphics[width=\columnwidth]{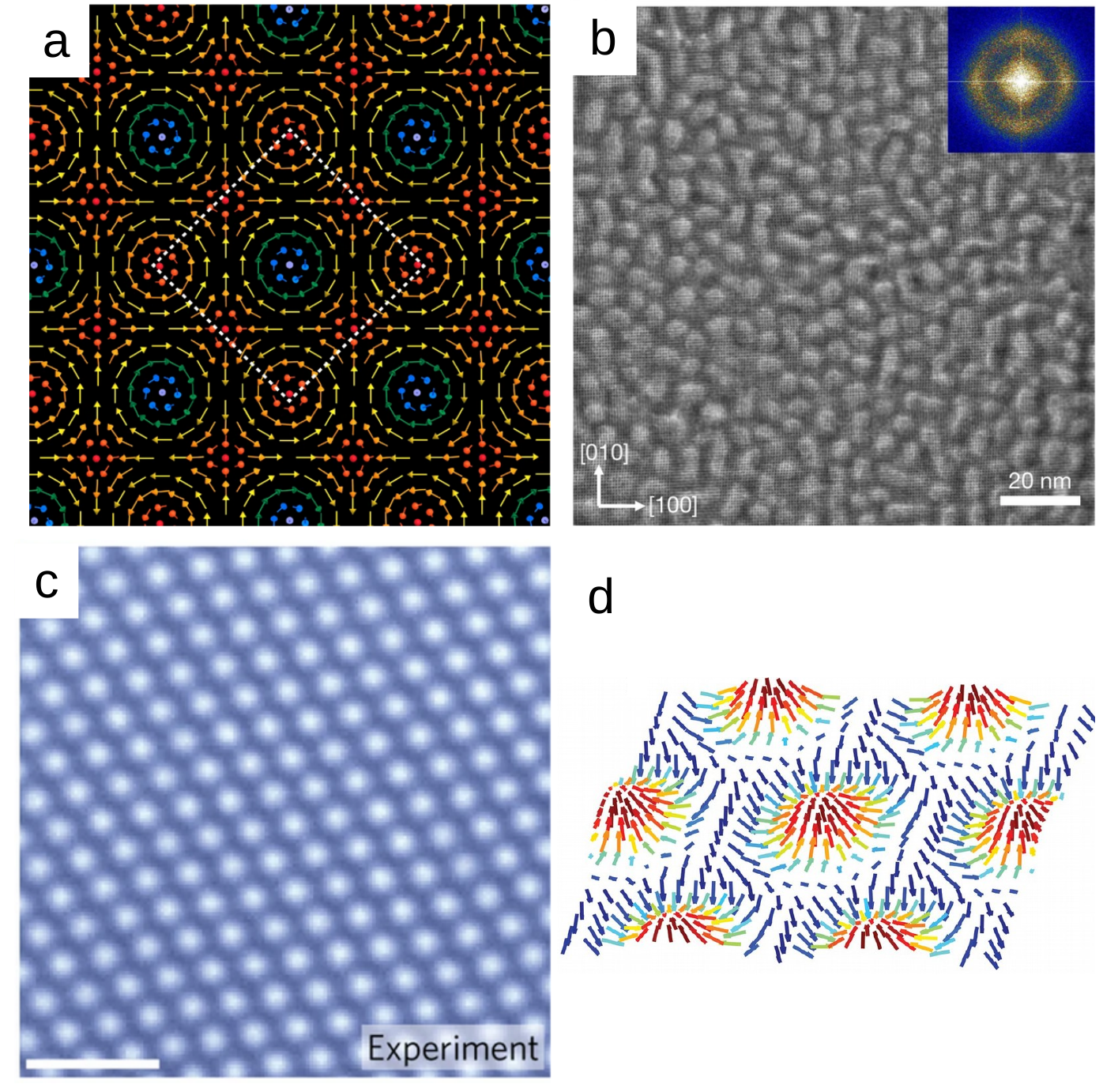}
\caption{
Different types of skyrmionic textures in real space.
(a) Schematic
magnetization texture of a square meron lattice \cite{Yu18}.
Reprinted by permission from: Springer Nature,
X. Z. Yu {\it et al.},
``Transformation between meron and skyrmion topological spin textures in a
chiral magnet'',
Nature (London) {\bf 564}, 95 (2018),
\copyright 2018.
(b) Image of a polar skyrmion structure \cite{Das19}.
Reprinted by permission from: Springer Nature,
S. Das {\it et al.},
``Observation of room-temperature polar skyrmions'',
Nature (London) {\bf 568}, 368 (2019),
\copyright 2019.
(c)
Image of a half skyrmion lattice in a liquid crystal system \cite{Nych17}.
Reprinted by permission from: Springer Nature,
A. Nych {\it et al.},
``Spontaneous formation and dynamics of half-skyrmions in a chiral
liquid-crystal film'',
Nature Phys. {\bf 13}, 1215 (2017),
\copyright 2017.
(d) Vector representation of the electric field for a
Neel-type optical skyrmion \cite{Tsesses18}.
  From S. Tsesses {\it et al.}, Science {\bf 361}, 993 (2018). Reprinted with
  permission from AAAS.
}
\label{fig:figNew1}
\end{figure}

Skyrmions can be two dimensional in thin films,
\cite{Yu10,Muhlbauer09},
have a layered
or pancake-like
structure in layered materials,
form 3D lines in bulk materials \cite{Milde13,Zhang18,Park14,Birch20},
and even
assemble into 3D lattices of particle-like hedgehogs
in certain bulk systems \cite{Lin18,Fujishiro19}.
Different species of skyrmions can exist \cite{Leonov15},
including bi-skyrmions \cite{Yu14,Wang16,Takagi18}, multiply charged
skyrmions \cite{Rybakov19}, chiral bobbers \cite{Rybakov15,Zheng18},
antiskyrmions \cite{Nayak17,Hoffmann17},
antiferromagnetic skyrmions \cite{Barker16,Akosa18}, magnetic bi-layer
skyrmions \cite{Zhang16c},
elliptical skyrmions \cite{Jena20},
meron lattices \cite{Yu18,Wang20,Gao20},
bi-merons \cite{Jani21},
hopfions \cite{Wang19a,Liu20},
hedgehog textures \cite{Fujishiro19,Zou20} and polar skyrmions \cite{Das19}.  
Skyrmions and similar quasiparticle textures
can arise in many other non-magnetic systems including
graphene \cite{Zhou20,Bomerich20} 
liquid crystals \cite{Duzgun18,Nych17,Foster19},
and optical \cite{Tsesses18} and plasmonic systems \cite{Davis20}. 
We highlight a variety of
the possible textures in Fig.~\ref{fig:figNew1},
including
a real space image of a square meron 
lattice \cite{Yu18} in
Fig.~\ref{fig:figNew1}(a),
an image of polar skyrmions \cite{Das19} in Fig.~\ref{fig:figNew1}(b),
a half skyrmion lattice in a chiral liquid crystal system \cite{Nych17}
in Fig.~\ref{fig:figNew1}(c),
and
the electric fields for an optical skyrmion \cite{Tsesses18}
in Fig.~\ref{fig:figNew1}(d).

Skyrmions can be set into motion
with an applied drive,
such as a current
which creates a spin torque Hall effect.
The skyrmion motion can be deduced
from changes in the topological Hall effect \cite{Schulz12,Liang15} 
or observed through direct imaging
\cite{Jiang15,Woo16,Yu14,Yu12,Jiang17,Litzius17,Legrand17,Woo18,Tolley18}.
It is also possible to 
move skyrmions with temperature gradients
\cite{Kong13,Mochizuki14,Pollath17,Wang20a},
magnetic fields \cite{Zhang18a,Shen19,Casiraghi19}, 
electric fields \cite{White14,Kruchkov18}, microwaves \cite{Wang15,Ikka18},
spin waves \cite{Shen18}, magnons \cite{Psaroudaki18}, 
or acoustic waves \cite{Nepal18}.
Due to their size scale, mobility, 
and stability at room temperature,
skyrmions have great potential for use in a wide range of applications 
such as race track memory
\cite{Fert13,Tomasello14,Muller17,Fert17,Suess18,EverschorSitte18,Suess19}, 
logic devices \cite{Zhang15,Luo18,Mankalale19,Liu19}
or novel computing  architectures
\cite{Prychynenko18,Pinna18,Grollier20,Song20}. 
Many of the proposed skyrmion-based devices
would require the skyrmions to
move through a nanostructured landscape 
in a highly controlled fashion.

A growing body of work indicates
that in many 
skyrmion systems, pinning and the effects of quenched disorder
are very important in determining both the static and dynamic
skyrmion response
\cite{Nagaosa13,Wiesendanger16,Woo16,Jiang17,Litzius17,Fert17}.
Initial transport studies revealed only weak skyrmion pinning
effects,
with a critical depinning force $j_c$
in MnSi at $T = 28K$ 
of only
$j_{c} \propto 10^6$ A/m$^2$
\cite{Schulz12,Jonietz10},
nearly five orders of magnitude smaller than the depinning force
for magnetic domain walls.
In contrast, recent work
by Woo {\it et al.}
on room temperature skyrmions in thin films
showed that the pinning can be very strong,
with $j_{c} \propto 2.2\times 10^{11}$ A/m$^2$ \cite{Woo18}.
Similar high depinning thresholds observed in other systems
\cite{Hrabec17} indicate that a variety of pinning
effects and skyrmion-pin interaction mechanisms can
be important
in different materials that  support skyrmions,
depending on the skyrmion size, dimensionality,
and the characteristics of the disorder in the sample.

Skyrmion motion can be strongly modified when pinning is present.
For example, there is evidence that the skyrmion Hall effect
is heavily impacted by pinning.
The skyrmion
Hall effect
arises when
the gyroscopic nature of the skyrmion 
dynamics causes the skyrmions to move at an angle
called the skyrmion Hall angle
$\theta_{\rm SkH}$ with respect to the applied drive  
\cite{Nagaosa13,Zang11,Iwasaki13,Everschorsitte14}.
Due to the skyrmion Hall effect,
a skyrmion driven along a narrow strip
by a current parallel to the strip
does not move
in the direction of the current.
Instead, it translates toward the edge
of the strip, where it is annihilated.
This behavior imposes a limitation on the
use of skyrmions in strip-based devices
\cite{Iwasaki13a}.
It may, however, be possible to use pinning to change the motion of
skyrmions through a strip.
Simulations \cite{Legrand17,Muller15,Reichhardt15,Reichhardt15a,Reichhardt16,Kim17,Litzius20} 
and experiments \cite{Jiang17,Litzius17,Woo18,Litzius20}
have shown that the introduction of pinning to a system
not only produces
a finite depinning threshold for skyrmion motion,
but also generates a strong drive dependence of
the skyrmion Hall angle,
which increases from
a very small value at low drives
to the pin-free intrinsic value $\theta_{\rm SkH}^{\rm int}$
as the drive increases.

There are a variety of
other cases in which pinning effects can be beneficial.
Thermal and diffusive motion of skyrmions has been observed
in experiment \cite{Zazvorka18,Nozaki19,Zhao19},
and it will soon be important to take
thermal effects into account
for device creation, particularly for smaller skyrmions. 
For example, a skyrmion serving as an information carrier
in a memory device
may need to be locked in a specific location for long times,
but
the thermal motion present at room temperature
could cause the skyrmion to wander 
away gradually and lose the memory of
its 
initial position.
Long term stable memory
could be achieved through pinning, which could
overcome the thermal effects
over arbitrarily long times.
It would be ideal
to have tunable pinning 
that would be absent
when rapid motion of skyrmions is needed but strong when
long time stability of the skyrmion configuration
is required to create a desired memory state. 
Already, different types of pinning have been identified
that have attractive, repulsive, radially symmetric, or
radially asymmetric behavior.
Devices could be created by
using nanoscale techniques to fabricate
controlled pinning patterns 
in the form of
lines or channels that guide skyrmions,
periodic arrays that stabilize certain skyrmion configurations, 
or asymmetric pinning that produces skyrmion
diodes, rectifiers and logic devices.
For future
applications it is
important to develop a
thorough understanding
of skyrmion pinning and dynamics. 

Beyond applications,
interacting skyrmions driven over pinning represent a
fascinating class of systems
in which collective
and competing effects can produce a rich variety
of nonequilibrium dynamical phases \cite{Fisher98,Reichhardt17}.  
The skyrmion-skyrmion interactions usually favor the formation of a
triangular skyrmion lattice, while the interactions of
skyrmions with random pinning
favor a disordered skyrmion structure,
producing
a competition between crystalline and glassy states even for static
skyrmion configurations.  
Under an applied drive,
the pinning opposes the skyrmion motion,
and the competition between the pinning and driving forces
generates complex dynamics
near the depinning threshold. 
Additional competing effects appear when thermal fluctuations are important.
Temperature can reduce the effectiveness of the pinning, favoring an
ordered state, but 
can also disorder the skyrmion lattice.

In this review, we focus on aspects of pinning and dynamics in skyrmion systems.
We highlight what is known currently about
skyrmion pinning and the variety of mechanisms that can produce it,
including
changes in the Dzyaloshinskii-Moriya interaction (DMI),
atomic impurities, local anisotropy, sample thickness,
damage tracks, missing spins, holes, or blind holes.
We outline the
microscopic models for pinning and skyrmion dynamics
currently in use, and
show that skyrmions can have attractive,
repulsive, or combined attractive and repulsive interactions
with pointlike or linelike disorder.
Throughout this review  we discuss similarities and differences between
skyrmions and
other systems with pinning
such as superconducting vortices, sliding charge density waves,
Wigner crystals, and colloidal particles.
In the absence of driving,
we consider disorder-induced transitions from a
skyrmion crystal to different types
of glassy states.
When a drive is added, we describe the
different types of depinning that occur, ranging
from elastic to plastic,
as well as the effect of disorder on bulk transport measures such
as velocity-force curves, the role of temperature, and creep effects.
The effects of pinning on
fluctuations, the skyrmion Hall angle,
and the skyrmion-skyrmion interactions are also covered.
In addition to sources of random disorder,
we describe the pinning and dynamics of skyrmions on
ordered structures such as 2D periodic,
quasiperiodic, quasi-one-dimensional (1D) periodic, and 1D asymmetric
substrates, which can produce commensurate
and incommensurate states, soliton motion, and diode and ratchet effects.

In each section we discuss
future directions
including studies of skyrmions in bulk materials,
skyrmion behavior in thin films with extended or
point defects,
the effects of nanostructured
arrays with periodic or 1D modulation,
the behavior of layered materials,
the coupling of skyrmions to other topological defects such as
vortices in type-II superconductors,
or even the effect of having different species of skyrmions coexist.
In the case of 3D skyrmions, introduction of a columnar pinning landscape
could create a state analogous to the
Bose glass
found in type-II superconductors,
and could also lead to cutting and entanglement
effects as well as the possibility of creating transformer geometries.
We outline potential new measures for
characterizing the skyrmion structures and dynamics
that are borrowed from work in vortex dynamics,
soft matter, and statistical physics,
such as structural measures, force chains,
jamming concepts, glassy effects, and defect proliferation.

For more general reviews of other aspects of
skyrmions, we refer the reader to Refs.~\cite{Jiang17a,Bogdanov20,Tokura20}
for materials that support skyrmions, 
Ref.~\cite{Li21} for
skyrmion devices and skyrmionlike textures such as swirling quasiparticles,
and to Refs.~\cite{Back20,Gobel21a} for further future directions.

\begin{figure}
\includegraphics[width=\columnwidth]{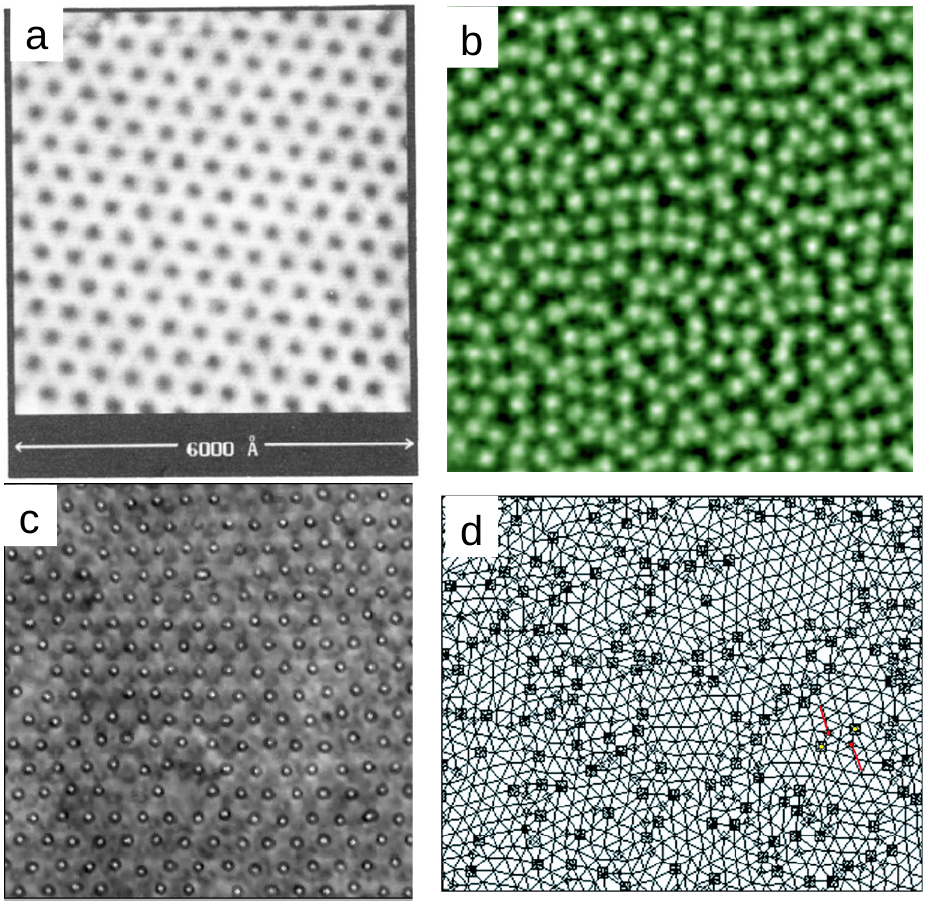}
\caption{
  (a) Image of an ordered superconducting vortex lattice obtained using a scanning
  tunneling microscope \cite{Hess89}.
Reprinted with permission from H. F. Hess {\it et al.}, Phys. Rev. Lett.
{\bf 62}, 214 (1989). Copyright 1989 by the American Physical Society.
(b) Image of a disordered superconducting vortex lattice obtained using
magneto-optical imaging \cite{Goa01}.
Used with permission of IOP Publishing, Ltd, from ``Real-time magneto-optical imaging of vortices in superconducting NbSe$_2$,'' P. E. Goa {\it et al.}, Supercond. Sci. Technol. {\bf 14}, 729, 2001; permission conveyed through Copyright Clearance Center, Inc.
(c) Image of a colloidal lattice obtained with an optical microscope \cite{Weiss98}.
Reprinted from J. A. Weiss {\it et al.}, J. Chem. Phys. {\bf 109}, 8659 (1998), with the permission of AIP publishing.
(d) Delaunay triangulation of colloidal
particle positions in a colloidal glass state based
on an optical microscope image obtained experimentally \cite{Pertsinidis08}.
Reprinted with permission from A. Pertsinidis {\it et al.}, Phys. Rev. Lett.
{\bf 100}, 028303 (2008). Copyright 2008 by the American Physical Society.
}
\label{fig:3}
\end{figure}

\section{Pinning in particle Like systems}
Systems with many
interacting particles
coupled to some form of disorder or pinning 
are known to exhibit very rich static and dynamic phase
behavior as a function of changing particle-particle interactions, 
disorder strength, and temperature.
One of the best studied examples of such systems
is
magnetic vortices
in type-II 
superconductors \cite{Blatter94}.
In the absence of driving, the vortices
can form a triangular lattice,
a weakly pinned Bragg glass
in which the
vortices remain elastic with topological order
but still have glassy properties \cite{Giamarchi95,Klein01},
topologically disordered vortex glass states
\cite{Fisher91,Henderson96,Natterman00,ToftPetersen18,Ganguli15}, entangled 
vortex lines \cite{Nelson88,Giller97}, liquid states \cite{Safar92,Cubitt93,Zeldov95},  
or reentrant liquid states \cite{Banerjee00,Avraham01}.
Vortices
in the presence  of an external 
drive can exhibit elastic 
depinning,
where the system transitions from
a pinned crystal into a moving crystal state
\cite{Reichhardt17,Bhattacharya93,DiScala12},  
or plastic depinning, where the moving state has a liquid structure
\cite{Reichhardt17,Bhattacharya93,Jensen88,Matsuda96,Olson98,Fily10,Shaw12}.  
Plastically moving vortices at higher drives can transition into  a
moving
crystalline \cite{Reichhardt17,Bhattacharya93,Koshelev94,Giamarchi96,Olson98a}
or moving smectic phase \cite{Olson98a,Balents98,Pardo98}.
These different depinning and dynamical phase transitions
produce distinct signatures in the bulk transport measures and
velocity-force curves
as well as changes in the vortex structure and fluctuations
\cite{Fisher98,Reichhardt17,Bhattacharya93,DiScala12,Jensen88,Fily10,Shaw12,Koshelev94,Olson98a}.
Similar depinning and sliding dynamics occur in other systems of particle-like objects
moving through quenched disorder,
such as colloidal particles \cite{Hu95,Reichhardt02,Pertsinidis08,Tierno12},
Wigner crystals \cite{Williams91,Cha94,Cha98,Reichhardt01,Kumar18},
and certain pattern forming systems \cite{Sengupta10,Morin17,Sandor17}. 

To highlight the similarities between skyrmions and other systems
with pinning,
in Fig.~\ref{fig:3}(a) we show an image of a
triangular superconducting vortex 
lattice obtained using scanning tunneling microscopy
\cite{Hess89}. In Fig.~\ref{fig:3}(b),
a disordered vortex structure appears in an image obtained
using a magnetooptical
technique
\cite{Goa01}.
Figure~\ref{fig:3}(c) shows a colloidal triangular lattice
observed with optical microscopy \cite{Weiss98},
while in Fig.~\ref{fig:3}(d), the colloidal lattice is distorted by strong pinning,
there are numerous topological defects,
and the system
forms a pinned glass \cite{Pertsinidis08}.
If the disorder is weak, as in
Fig.~\ref{fig:3}(a) and (c),
the particles
depin without the generation of topological defects
and flow elastically,
while for strong disorder, as in
Fig.~\ref{fig:3}(b) and (d),
the particles depin plastically with large
lattice distortions or 
with a coexistence  
of pinned and moving particles.

\begin{figure}
\includegraphics[width=\columnwidth]{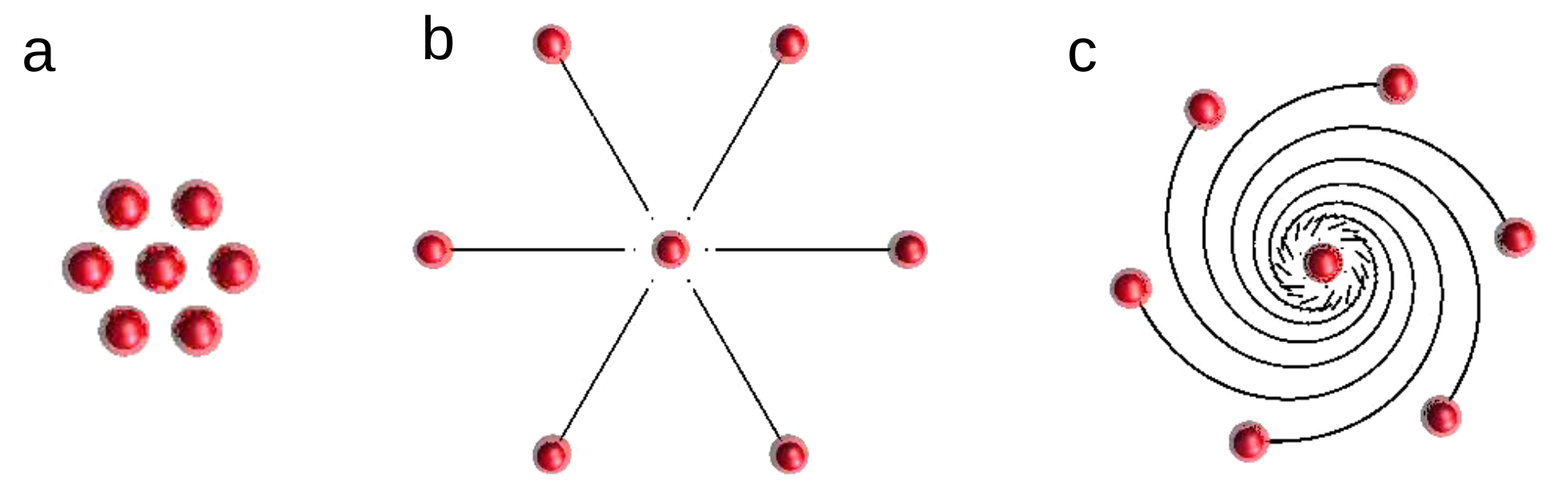}
\caption{ Illustration of the difference between purely overdamped motion and
  motion with a Magnus force
  of strength $\alpha_m$ for particles with finite damping, $\alpha_d>0$.
  (a) Initial dense cluster of particles.
  (b) Trajectories of overdamped particles with $\alpha_m=0$
  moving away from the center.
  (c) Trajectories of particles with a Magnus force $\alpha_m>0$
  moving away from the center,
  showing the emergence of nonconservative rotation.
}
\label{fig:4}
\end{figure}

A crucial difference
between skyrmions and the
vortices or colloidal particles illustrated in Fig.~\ref{fig:3}
is the fact that
skyrmions experience
a  strong non-dissipative gyroscopic or Magnus force which
generates a velocity component
{\it perpendicular} to the net external
forces acting on the skyrmion.
In many of the previously studied systems, the dynamics
are overdamped
and the particle velocity ${\bf v}_d$ is strictly 
aligned with the net external force ${\bf F}_{\rm ext}$,
${\bf v}_{d} = \alpha_{d}{\bf F}_{\rm ext}$, where $\alpha_{d}$ is a damping constant.
In a skyrmion system, the damping term is accompanied by a
Magnus force contribution to the velocity,
${\bf v}_m = \alpha_{m}{\bf \hat{z}} \times {\bf F}_{\rm ext}$, which
generates a velocity
component perpendicular to the applied force.
Here $\alpha_{m}$ is the strength of the Magnus term. 
The ratio $\alpha_{m}/\alpha_{d}$ for skyrmions
can
be as large as ten or even higher \cite{Nagaosa13}. 
One consequence of the Magnus force
is the appearance of
a skyrmion Hall effect
in which
the skyrmion moves at 
an angle $\theta_{\rm SkH}$ with respect to the applied driving force.
The intrinsic value of this angle is given by
$\theta_{\rm SkH}^{\rm int} = \tan^{-1}(\alpha_{m}/\alpha_{d})$.
The Magnus force affects the skyrmion-skyrmion
interactions as well as the motion of skyrmions through pinning sites.
In Fig.~\ref{fig:4}(a)
we show repulsively interacting particles that have been initialized in a dense
cluster and are then allowed to move away from each other.
In the overdamped limit with $\alpha_m=0$, Fig.~\ref{fig:4}(b) shows that
the particles move radially, in the direction of the forces generated by the
particle-particle interactions.
In contrast, the particles in Fig.~\ref{fig:4}(c) have a finite Magnus
force, $\alpha_m>0$,
so that in addition to the radial displacement, there is a strong rotational
component of the motion.  Here the dissipative term $\alpha_d$ is still finite, but
if it were zero, only rotational motion of the particles would occur with no radial
motion.

Many of the previously studied systems with pinning,
including superconducting vortices, classical charges, and colloidal
particles, 
are composed of relatively stiff
particle-like objects
in which the internal degrees of freedom are unimportant,
making a particle-based treatment of their dynamics appropriate.
In contrast, skyrmions can exhibit excitations of internal modes
\cite{Ikka18,Onose12,Garst17,Beg17}
or large distortions \cite{Litzius17,Zeissler17,Gross18}
that activate additional degrees of freedom,
significantly impacting the statics and dynamics. 
Furthermore, moving skyrmions can emit spin waves
that could modify the
effective skyrmion-skyrmion 
interactions \cite{Schutte14,Koshibae18}.
The uniformity often associated with particle-based models may also not
capture the behavior of a skyrmion system well.
It is possible for skyrmions to coexist with a stripe phase or ferromagnetic
domains
\cite{Muller17a,Shibata18,Loudon18,Yu18a},
and in some systems, there is considerable dispersion in the size of the
skyrmions, making the skyrmion assembly effectively polydisperse
\cite{Karube18}
This contrasts strongly with superconducting vortices,
which are all the same size
in a given sample.

\section{Models of Skyrmions and Mechanisms of Skyrmion Pinning}

\begin{figure}
\includegraphics[width=\columnwidth]{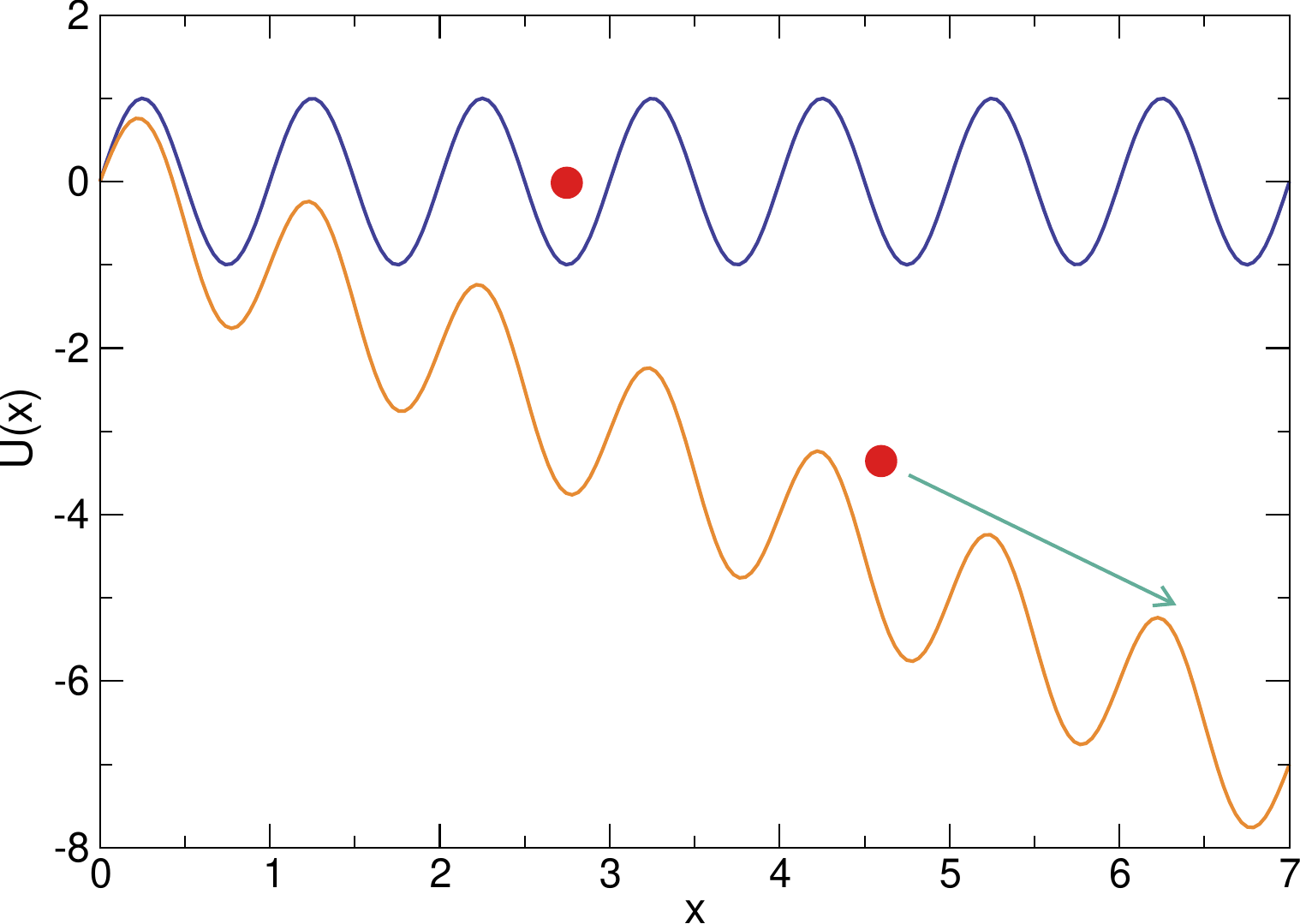}
\caption{The simplest system exhibiting depinning is an
overdamped particle (red circle) in a sinusoidal
potential $U(x)=A\cos(kx)$
that has been tilted by a driving force $F_{D}$.
The particle is pinned when $F_{D} < F_{c}$ (blue curve),
where $F_{c}$ is the critical driving force
that must be applied to enable the particle
to slide.
Steady state particle motion occurs when $F_{D} > F_{c}$ (orange curve).   
}
\label{fig:5}
\end{figure}

\begin{figure}
\includegraphics[width=\columnwidth]{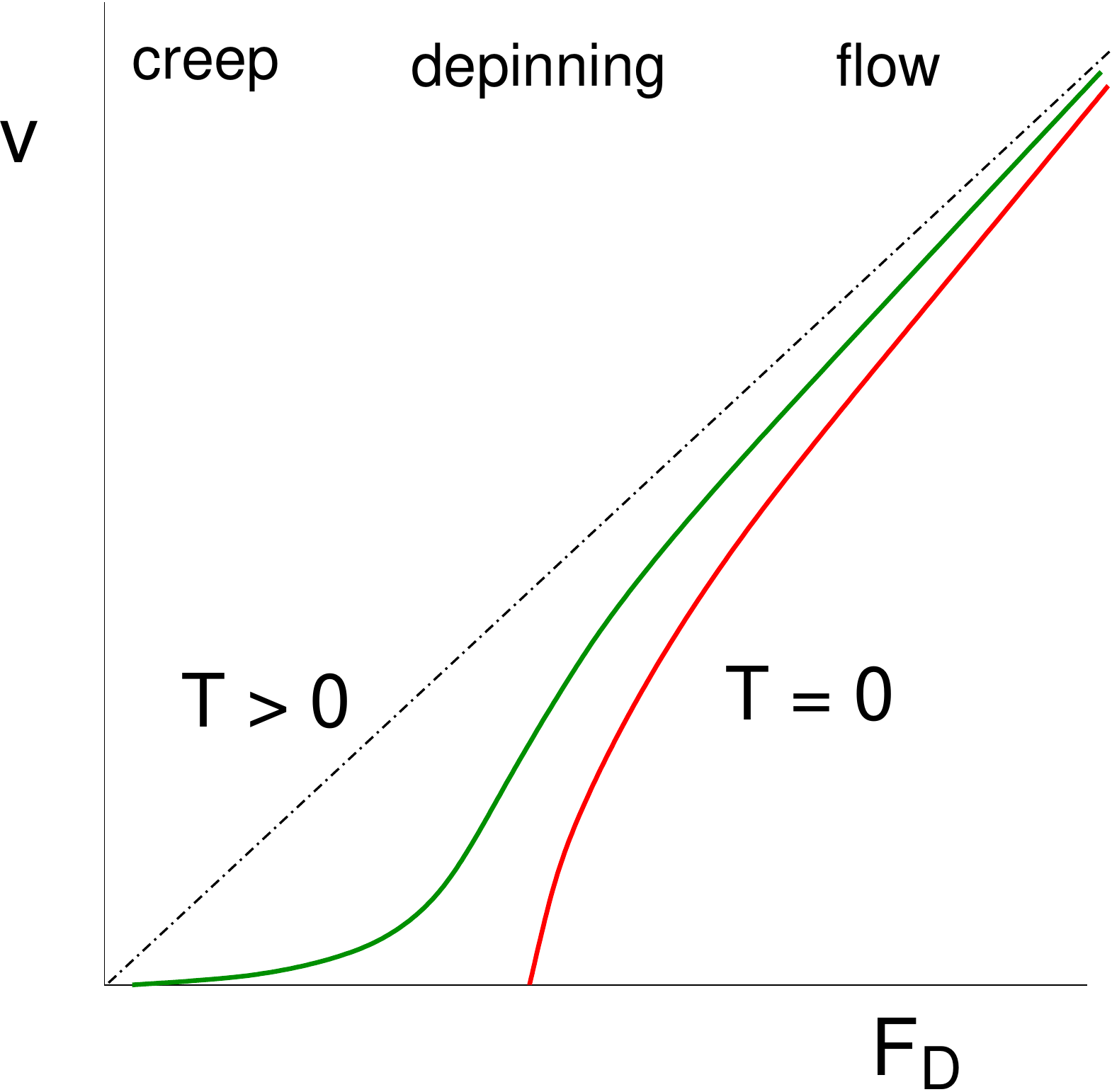}
\caption{ Schematic velocity $v$ vs drive $F_{D}$
curves for a system with a finite depinning threshold $F_{c}$
at zero temperature $T = 0$ (red curve) and finite temperature
$T>0$ (green curve).
Creep behavior occurs for finite temperature
when the velocity remains nonzero 
for $F_D<F_c$.
There is a change in the shape of the $T>0$ velocity-force curve
near $F_c$ due to
a crossover from creep
to flow.
The dashed line indicates the free-flow limit $v \propto F_D$ for a
system with no pinning.
}
\label{fig:6}
\end{figure}

One of the simplest pictures of pinning and sliding dynamics 
is
a model of a single particle 
in a tilted sinusoidal potential with period $L$.
To further simplify the problem,
consider an overdamped particle that
obeys the following equation of motion:
\begin{equation}
\alpha_d \frac{dx}{dt} = -\frac{dU(x)}{dx} + F_{D} .
\label{eqn1}
\end{equation}
Here $\alpha_d$ is the damping constant, $F_D$ is the external dc drive,
and $U(x) = A\cos(kx)$,
where $k = 2\pi/L$.
When $A = 0$, the substrate disappears and the particle moves in the direction
of the drive with a velocity
$v = F_{D}/\alpha_d$.
When $A>0$, there is a finite depinning threshold $F_c$,
indicating that there is no steady state motion
unless $F_D>F_c$.
If we set $A = A_{0}/k$, we obtain a critical force of $F_c=A_0$.
For drives close to but above the critical force,
$F_{D} \gtrsim F_{c}$, the particle
begins to slide with
a velocity $v \propto (F_{D} - F_{c})^\beta$
where $\beta = 1/2$ \cite{Fisher85}.
At higher drives, the velocity 
crosses over
to the clean value limit of
$v \propto F_{D}$, similar to
what is shown in Fig.~\ref{fig:5}.    

Additional effects can be included in the single particle picture,
such as a coupling to a thermal bath modeled using
a fluctuating force term
$\eta(t)$ representing Langevin kicks.
These obey the correlations $\langle \eta(t)\rangle = 0$
and $\langle \eta(t)\eta(t^\prime)\rangle = 2k_B T\delta(t-t^{\prime})$,
where $k_B$ is the Boltzmann constant.
When $F_{D} = 0$, the particle can thermally hop with equal
probability to the left or right
according to an Arrhenius law,
with an instantaneous velocity $|v| \propto \exp(-U/k_{B}T)$
and an average velocity of zero; 
however, under a finite applied drive, the time-averaged
velocity becomes finite since the Arrhenius jumps are now biased.
In this case, the potential $U(x)$ is replaced by
$U(x) \pm  U_{D}(x)$, where for 
a linear drive $U_{D}(x) = U(x) - F_{D}x$.
This increases the average size of jumps in the driving direction
and produces a finite velocity
even for drives that are below the $T = 0$ value of $F_{c}$,
leading to the emergence of a creep regime.
The creep velocity for $F<F_c$ is of the form
\begin{equation}
v \propto C_A\exp\left(\frac{A - F_{D}}{kT}\right)  
\end{equation}
where $C_A$ is the attempt frequency.
Figure~\ref{fig:6}
shows schematic velocity-force curves at $T = 0$ and $T > 0$.
Even at finite temperatures,
there can be a noticeable change in the velocity-force
curves upon crossing
the $T=0$ value of $F_c$ due to a crossover
from the creep motion of intermittently hopping particles
for $F_D<F_c$
to continuous flow for $F_D>F_c$.
When multiple interacting particles are present,
collective creep, plastic creep, and glassy effects can occur,
which typically
introduce a power law prefactor to the exponential velocity term
\cite{Feigelman89,Luo07}. 

It is possible to add other terms to Eq.~(\ref{eqn1}),
such as an inertial
term $Md^2x/dt^2$, where $M$ is the mass of the particle,
as well as asymmetry or disorder in the substrate.
If the dc drive is supplemented with an additional
ac drive
of the form $F^{ac} = A_{ac}\sin(\omega t)$,
the well known Shapiro step phenomenon appears in the form of
steps in the velocity-force curves \cite{Shapiro63}.   
The complexity of the substrate can be increased by introducing
spatial variations in two dimensions, such as square or triangular
pinning lattices, or by adding random disorder.
For an overdamped system, the 1D picture of depinning 
generally captures the effects that are found even for a substrate
with 2D features.
Interestingly, this is not case for skyrmions, since the
presence of the Magnus force causes 2D skyrmions to
exhibit different dynamics 
than their completely 1D counterparts. 

The next level of complexity is to
include multiple interacting
or coupled particles.
For example, a dimer or trimer arrangement of particles connected by
springs could be placed on a periodic 1D substrate.
The best known model
of such a system is the
Frenkel-Kontorova
model consisting of a 1D chain of elastically coupled particles
moving over a 1D periodic substrate \cite{Braun98}.
This model can be extended to describe
a 1D string of particles or a 2D array of particles moving in
2D or 3D and coupled to a random substrate.
For example, a 2D triangular array of skyrmions
could be modeled as a 2D elastic lattice.
In 3D, a single 1D linelike string could be modeled as
an elastically coupled array of elements extending
along the length of the string.  
Additional terms can be incorporated into the equation of motion
to capture effects appropriate for the specific system.
When the particles are coupled by unbreakable elastic springs that
do not
allow for exchange of neighbors, phase slips, or breaking of the lattice,
the system is said to be
in an elastic limit.
One advantage of working in the elastic limit is that
it may be possible to ignore
the exact details of the specific  
particle-particle interactions, since
the system can be modeled effectively as a collection
of harmonic springs.
This approximation can be used when
both the pinning and the temperature are sufficiently weak
that only small distortions of the lattice can occur.
It has been applied for understanding
depinning phenomena in systems such as
directed lines \cite{Ertas96,Kardar98},
superconducting vortices \cite{Dobramysl14},
sliding charge density waves \cite{Fisher85}, 
models of friction \cite{Vanossi13},
and even plate tectonics \cite{Carlson94}. 
For skyrmions,
elastically coupled particle models are appropriate for
2D
skyrmion lattices moving over weak disorder well below the temperature at which
dislocations can be created thermally, as well as for
3D coupled skyrmion lines or even a single 3D skyrmion line in 3D systems.
Additional terms such as the Magnus force can be
inserted into the Frenkel-Kontorova
model
to capture long wavelength features of the 
depinning and sliding states. 

The next step beyond an elastically coupled system is to consider
particle based models with pairwise particle-particle interactions.
Here, the particles respond not only to the positions of their nearest neighbors,
but also to those of more distant neighbors
or even to all other particles.
Models of this type
allow
neighbor exchange,
dislocation generation, and
other plastic or nonaffine events
\cite{Fisher98,Reichhardt17}.
Driven particle
based models
that undergo depinning have been used extensively
in a wide range of studies of
both hard matter systems, such as
superconducting vortices, and soft matter systems,
such as colloidal particles and granular matter \cite{Reichhardt17}.
Particle based models have the advantages of
permitting transitions between elastic and plastic motion as well as
the ability to incorporate realistic pairwise particle-particle interactions.
They are also generally more computationally
efficient than fully continuum models, such as micromagnetic models
of skyrmions in which the full spin degrees of freedom are included.
The particle-particle interaction potentials are typically more complex than
simple nearest neighbor harmonic interactions,
and have a range that
can depend strongly
on the microscopic details of the system.
For example, in
thin film superconductors, the vortex-vortex pairwise interactions are
logarithmic, so all the particles
interact with all other particles in the system as well as with image charges,
while in colloidal systems with strong screening, the particles
interact only out to their first or second nearest neighbors.

\subsection{Particle Based Approaches to Skyrmion Dynamics and Pinning}

One approach for modeling skyrmions is to treat them as
point particles with dynamics that evolve according to an equation of
motion proposed by Thiele \cite{Thiele73} to
describe a driven magnetic particle:
\begin{equation} 
{\bf  \mathcal{G}}\times {\dot {\bf R}} + \alpha \mathcal{D} {\dot {\bf R}} 
+ m{\ddot {\bf R}} = {\bf F}_{D}.
\end{equation}
Here ${\bf F}_D$ is the driving force,
$\alpha$ is the Gilbert damping of an individual spin,
$\alpha \mathcal{D}$
is the friction experienced by the skyrmion,
and $\mathcal{G}$
is the gyrocoupling term which
acts like a magnetic field
applied perpendicular to the plane. This
term is analogous to the Coriolis force.  
The inertial term is proportional to the skyrmion mass $m$; however,
when $m$ is small, this term can be neglected.
Additional
second derivative terms can arise due to the excitation of
internal modes of the skyrmion.
The Thiele equation can be
extended using
terms that represent a
substrate potential, field gradients, thermal forces, 
or gyrodamping \cite{Schutte14a}.   
Due to its flexibility, the Thiele approach 
has been used extensively to model
the dynamics of single rigid skyrmions
\cite{Buttner15}.

\begin{figure}
\includegraphics[width=\columnwidth]{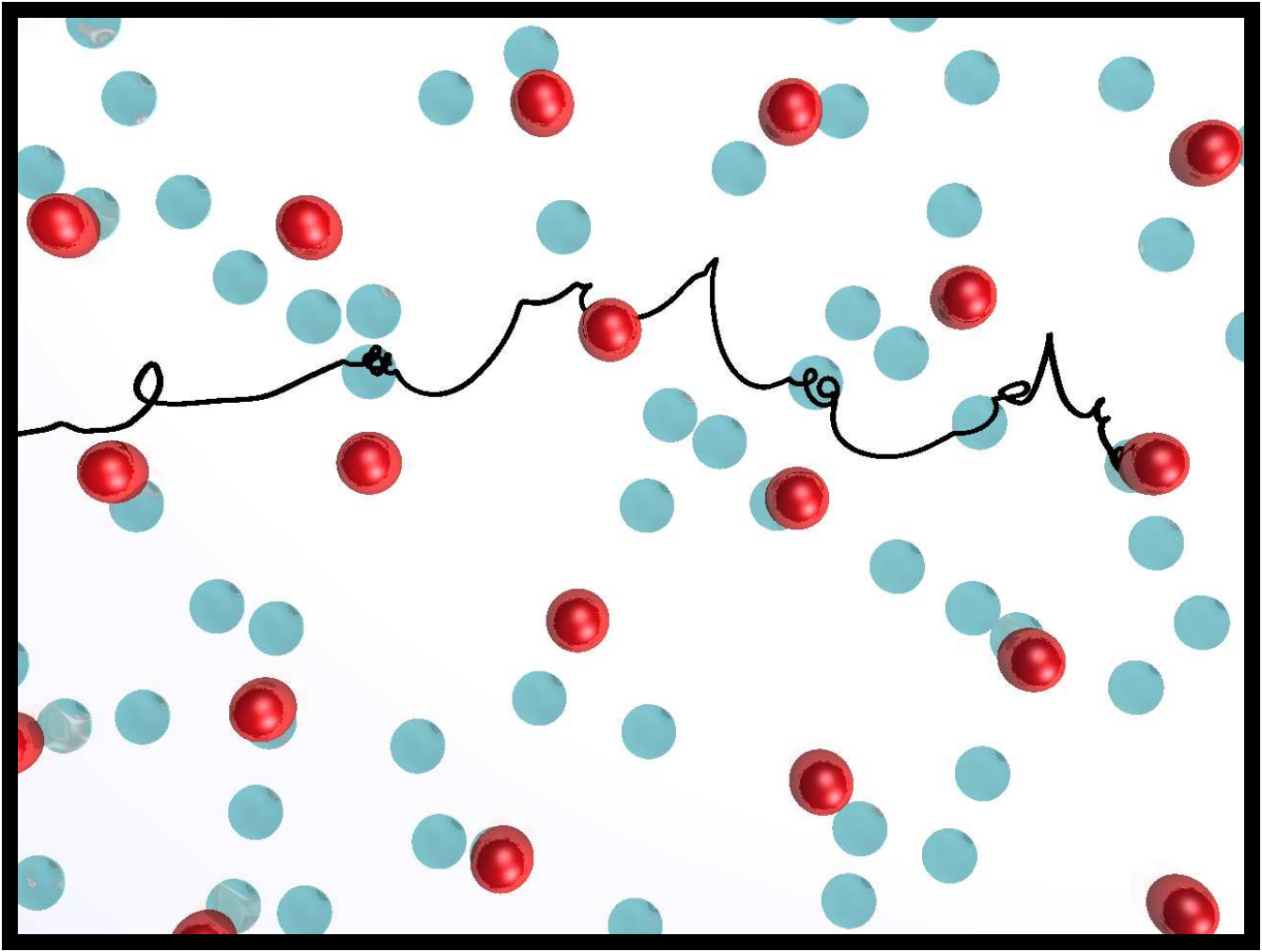}
\caption{ 
Real-space image of skyrmions (dark red dots) in a particle based model
driven through randomly arranged pinning sites (light blue dots) in a
plastic flow phase
\cite{Reichhardt15a}.
The trajectory of a single skyrmion is highlighted,
showing spiraling motions inside the pinning sites.
Reprinted with permission from C. Reichhardt {\it et al.}, Phys. Rev. Lett.
{\bf 114}, 217202 (2015). Copyright 2015 by the American Physical Society.
}
\label{fig:7}
\end{figure}

Particle-particle interactions of either the elastically coupled
or pairwise type
can be included by making further
modifications to the Thiele equation.
Lin {\it et al.} \cite{Lin13} proposed
a particle-based
model including
skyrmion-skyrmion, skyrmion-pinning, and skyrmion-driving force interactions
of the form:    
\begin{equation} 
\alpha_d {\bf v}_{i} + \alpha_m {\hat z} \times {\bf v}_{i} =
{\bf F}^{ss}_{i} + {\bf F}^{p}_{i} + {\bf F}^{D} .
\end{equation}
Here
${\bf v}_{i} = {d {\bf r}_{i}}/{dt}$
is the skyrmion velocity,
$\alpha_d$ is the damping constant which
aligns the skyrmion velocity in the direction of the external forces,
and $\alpha_m$ is the strength of the Magnus term
which aligns the skyrmion velocity in the direction perpendicular to
the external forces.
When both $\alpha_d$ and $\alpha_m$
are finite, the skyrmions
move at an angle
called the intrinsic skyrmion Hall
angle,
$\theta^{\rm int}_{\rm SkH} = \tan^{-1}(\alpha_{m}/\alpha_{d})$
with respect to an externally applied driving force.
In the work of Lin {\it et al.} \cite{Lin13},
the skyrmion-skyrmion interaction was modeled 
as a short range repulsive force of
the form
${\bf F}_{i}^{ss} = \sum^{N}_{j\neq i}K_{1}(r_{ij}){\hat {\bf r}_{ij}}$,
where $K_{1}$ is the modified Bessel function,
$r_{ij} = |{\bf r}_{i} - {\bf r}_{j}|$
is the distance between skyrmion $i$ and skyrmion $j$, and
${\hat {\bf r}}_{ij}=({\bf r}_{i}-{\bf r}_{j})/r_{ij}$.
In Fig.~\ref{fig:7}
we show a
snapshot from a 2D particle based
skyrmion simulation model 
illustrating the skyrmion locations, pinning
site locations, and the trajectory of one of the skyrmions, which
undergoes rotational motion due to the Magnus force as
it moves across the pinning sites \cite{Reichhardt15a}. 
The model proposed
by Lin {\it et al.} \cite{Lin13}
has both advantages and disadvantages.
It neglects inertial effects,
changes in the skyrmion shape,
Magnon generation, and  
possible many body interaction terms.
On the other hand, it allows
for greater computational efficiency compared to micromagnetic simulations,
permitting
many thousands of skyrmions
to be simulated over long periods of time.
In many cases the particle-based model
successfully captures the robust general features
of the system.

The particle based models can be substantially modified
based on insight gained either from micromagnetic simulations or
experiments.
For instance,
the skyrmion interactions are typically modeled
as a short range repulsion; however,
some micromagnetic simulations show evidence of
skyrmion clustering \cite{Loudon18,Rozsa16,Leonov19},
suggesting that the skyrmion interactions extend out to longer range.
Such effects could be captured by adding
some form of longer range attraction to
the particle based model.
In other samples,
the skyrmions exhibit
a transition from a square to a triangular
lattice, and this could
be modeled by including an additional
higher order symmetry
term in the pairwise potential of the form \cite{Olszewski18}
\begin{equation}
V(R,\theta) = K(r)(1 +  A\cos^2(n_{a}(\theta - \phi)/2)). 
\end{equation}
Here $\theta$ is the angle between
the two skyrmions, $\phi$ is the rotation
angle of the axis,
and $n_{a}$ is the number of symmetry directions
in the potential, where $n_{a} = 4$ would favor square ordering.  
In some skyrmion systems, the
skyrmion size can vary.
This could be modeled
by introducing a varying
screening length $\lambda_i$ in the skyrmion interaction potential,
$K_{1}(r/\lambda_i)$,
where $\lambda_i$ would have some nonuniform distribution.
It is also possible to add three-body and multi-body effects
to the interaction
by including
higher order potentials such as a three-body
$V_{i,j,k}$.
These potentials
could be extracted from micromagnetic simulations, similar to the techniques
used to model such effects
in colloidal systems \cite{Sengupta10}.
The skyrmion dynamics can also be modified.
For example, an
antiskyrmion could have
a four-fold modulation of its dissipative term
or different dissipation terms for different
directions of driving \cite{Kovalev18}.
Other studies have shown that trochoidal skyrmion motion
is possible, some types of
which can be modeled with particle based approaches \cite{Ritzmann18}.

A variety of potentials can be used to represent the pinning
term ${\bf F}_i^p$, such as
the short range attraction employed
in previous works \cite{Lin13}.
Other
possibilities
include short range repulsion, longer range pinning which could arise
from strain fields or magnetic interactions,
sites with competing attraction and repulsion of the type observed in
micromagnetic simulations \cite{Muller15}, 
or long range smoothly varying landscapes. 
It is also possible to add a thermal term to the skyrmion equation of motion
by introducing Langevin kicks
\cite{Brown18,Reichhardt19}. 

Although the particle model
does not capture all of the features found in micromagnetic simulations,
in some cases
it is possible to incorporate additional terms into the model in order
to mimic some of these effects, such as by including a time dependence of the
skyrmion interactions or the magnitude of the dissipative or Magnus forces in
order to represent breathing modes.
Similarly,
the shape changes that skyrmions can undergo
when they become compressed or elongated while
in pinning sites
can be modeled by modifying the particle-particle interactions
when at least one of the skyrmions is inside a pinning site.
Similar
modifications could be applied for systems in which the shape-changing
skyrmions are moving across a landscape.
Other rules could be
added to accommodate skyrmion creation or annihilation, such as by
defining
certain conditions
for the combination of the external force and the pinning force which,
when met, would cause the removal or addition of a skyrmion.
Magnon generation could be captured by introducing
retarded potentials, a dynamical pairwise skyrmion-skyrmion interaction term,
or multi-body interaction effects.

The particle description can be integrated
directly to examine the skyrmion dynamics;
however,
in order to identify ground state configurations such as
crystal, liquid or pinning-stabilized disordered structures,
Monte Carlo or other simulated annealing methods can be applied.
In simulated annealing, the system is initialized in a high
temperature state with rapidly diffusing particles, and the temperature is
gradually lowered to $T=0$ or to the desired final temperature.
The cooling must be performed sufficiently slowly that
the particles can explore phase space and find a configuration that is
either in or close to a ground state.
The cooling rate
can be tested by first considering a system containing no pinning in
order to determine whether the skyrmions are able to settle into a
triangular lattice.

\begin{figure}
\includegraphics[width=\columnwidth]{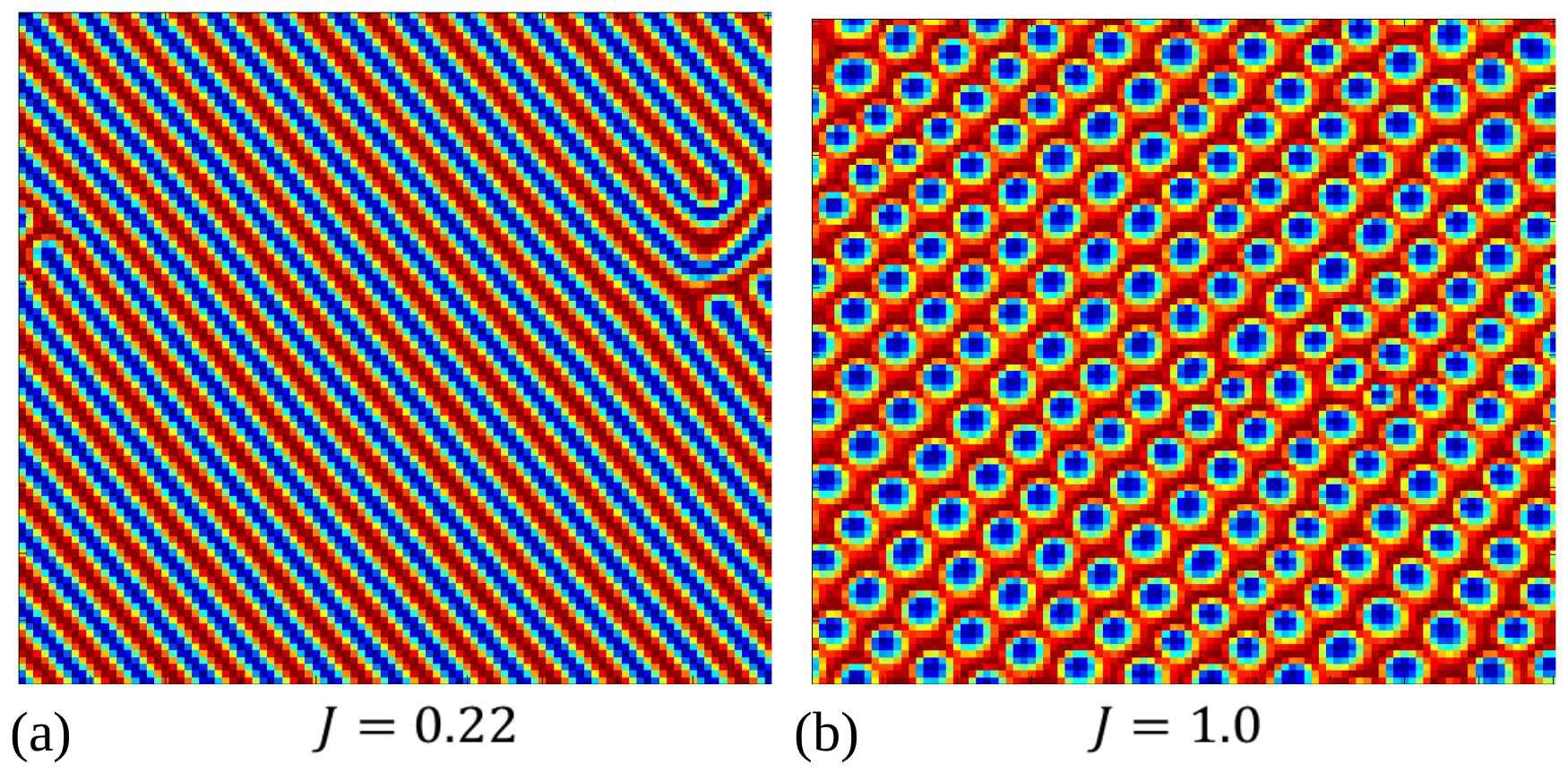}
\caption{ Images from 2D micromagnetic simulations \cite{Lin13a}.
  (a) The spiral state in the absence of a magnetic field, $H=0$.
  (b) A skyrmion lattice state for finite $H$. 
Reprinted with permission from S.-Z. Lin {\it et al.}, Phys. Rev. Lett.
{\bf 110}, 207202 (2013). Copyright 2013 by the American Physical Society.
 }
\label{fig:8}
\end{figure}

\section{Micromagnetic Models}

Since skyrmions are emergent objects composed of spins,
the other main modeling method that has been employed
is micromagnetic simulations in which
the
dynamics of the spin degrees of freedom are calculated directly
in the presence of different interaction terms including
exchange energy, DMI, anisotropy, 
and magnetic fields.
The starting point for these models
is \cite{Bogdanov89} 
\begin{equation} 
  \mathcal{H} = \int d{\bf r}^2\left[\frac{J_{ex}}{2}(\nabla {\bf n})^2
  + D{\bf n}\cdot \nabla \times {\bf n} -  {\bf H}_a \cdot {\bf n}\right], 
\end{equation} 
where $J_{ex}$ is the exchange term, $D$ is the DMI produced by spin-orbit coupling,
and ${\bf H}_{a}$ is the anisotropy term. 
Additional terms can be added to represent pinning, thermal forces,
gradient forces, and other effects.
This Hamiltonian can be integrated
using the Landau-Lifshitz-Gilbert equation \cite{Tatara08}, 
\begin{equation}
  \partial_t {\bf n} = ({\bf J}\cdot\nabla){\bf n} - \gamma{\bf n}\times {\bf H}_{\rm eff}
  + \alpha \partial_t {\bf n}\times {\bf n} .
\end{equation} 
In many cases, the skyrmions can be modeled
effectively using 2D micromagnetic simulations, 
but extensions to fully 3D simulations are possible.
In Fig.~\ref{fig:8}(a) we
show an image from a 2D micromagnetic simulation \cite{Lin13a}
of a spin system with DMI that has formed a helical state at
zero magnetic field, $H=0$.
The same system at finite $H$ forms a skyrmion lattice, as shown
in Fig.~\ref{fig:8}(b).

\begin{figure}
\includegraphics[width=\columnwidth]{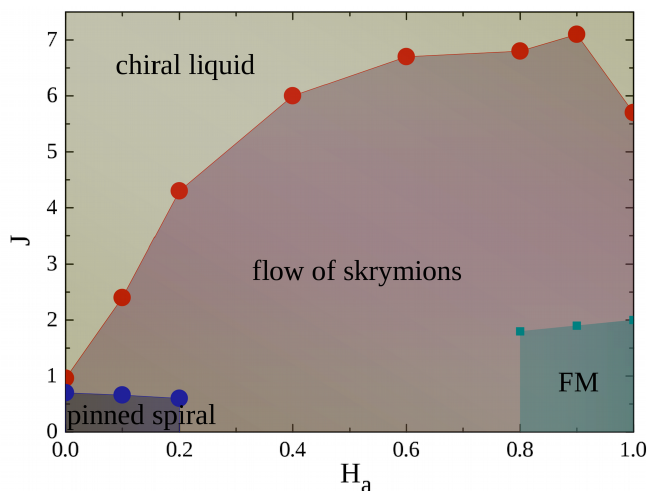}
\caption{ The dynamic phase diagram
  as a function of current $J$ vs magnetic field $H_a$
  from  2D micromagnetic simulations \cite{Lin13a}.
  In the absence of a current, $J = 0$,
  pinned spiral, skyrmion lattice, and ferromagnetic (FM) phases appear.
  At finite $J$, a moving skyrmion lattice and chiral liquid phase form at high drives.
  This indicates that it is possible to use a drive to
  nucleate skyrmions from a spiral or ferromagnetic state.
Reprinted with permission from S.-Z. Lin {\it et al.}, Phys. Rev. Lett.
{\bf 110}, 207202 (2013). Copyright 2013 by the American Physical Society.
 }
\label{fig:9}
\end{figure}

The advantage of micromagnetic models is that 
they allow skyrmion distortions and breathing modes
to occur along with skyrmion annihilation and creation.
The internal dynamics of a single skyrmion
can be studied in detail,
and it is possible to include additional terms
which can give
rise to remarkably rich behaviors.
Micromagnetic simulations can readily access the
basic phase diagram in the absence of drive
under an applied field,
showing the transition
from a zero field
helical state
to skyrmion lattices of varied density followed by
the emergence of a ferromagnetic
state at high fields.
When a driving force is applied,
even in the absence of pinning the
range of magnetic fields for which skyrmions
are stable changes.
An example of a dynamic phase diagram as a function of current
versus magnetic field for driven 
skyrmions in a pin-free system obtained through micromagnetic simulations
appears in Fig.~\ref{fig:9}.
A pinned spiral state forms at low fields,
and there are regions of flowing skyrmions,
a ferromagnetic state, and a high drive chiral state.
These simulations indicate
that application of a current can
cause skyrmions to emerge
from ferromagnetic or spiral states,
while strong driving can destroy the skyrmions \cite{Lin13a}.
Experiments have demonstrated the current-induced creation
and annihilation of 
skyrmions in weakly pinned systems \cite{Zhu17}.
Current-induced nucleation of skyrmions was also observed in
experiments in 
Co-based Heusler alloys \cite{Akhtar19};
however, these samples were strongly pinned, suggesting that 
pinning in combination with a drive can create skyrmions. 
This work also revealed that the current required to nucleate skyrmions increases
with increasing magnetic field. 

There are several
magnetic codes available
that can be used to simulate skyrmions interacting with pinning,
including MuMax \cite{Leliaert18}.
The disadvantage of micromagnetic simulations is that they are generally limited
in the number of skyrmions that can be simulated
and the time scales that can be studied.
As a result,
such simulations are unsuitable for examining
hundreds or thousands of skyrmions interacting
with pinning sites under a drive due to the relatively long transient
times that can occur before the system settles into a steady state.
It is also possible to use other types of
numerical models for skyrmions or skyrmion-defect interactions.
For example,
density functional theory can be particularly powerful
for extracting the energies of skyrmion-pin interactions
on the atomic scale.

\subsection{Pinning Mechanisms}

\begin{figure}
\includegraphics[width=\columnwidth]{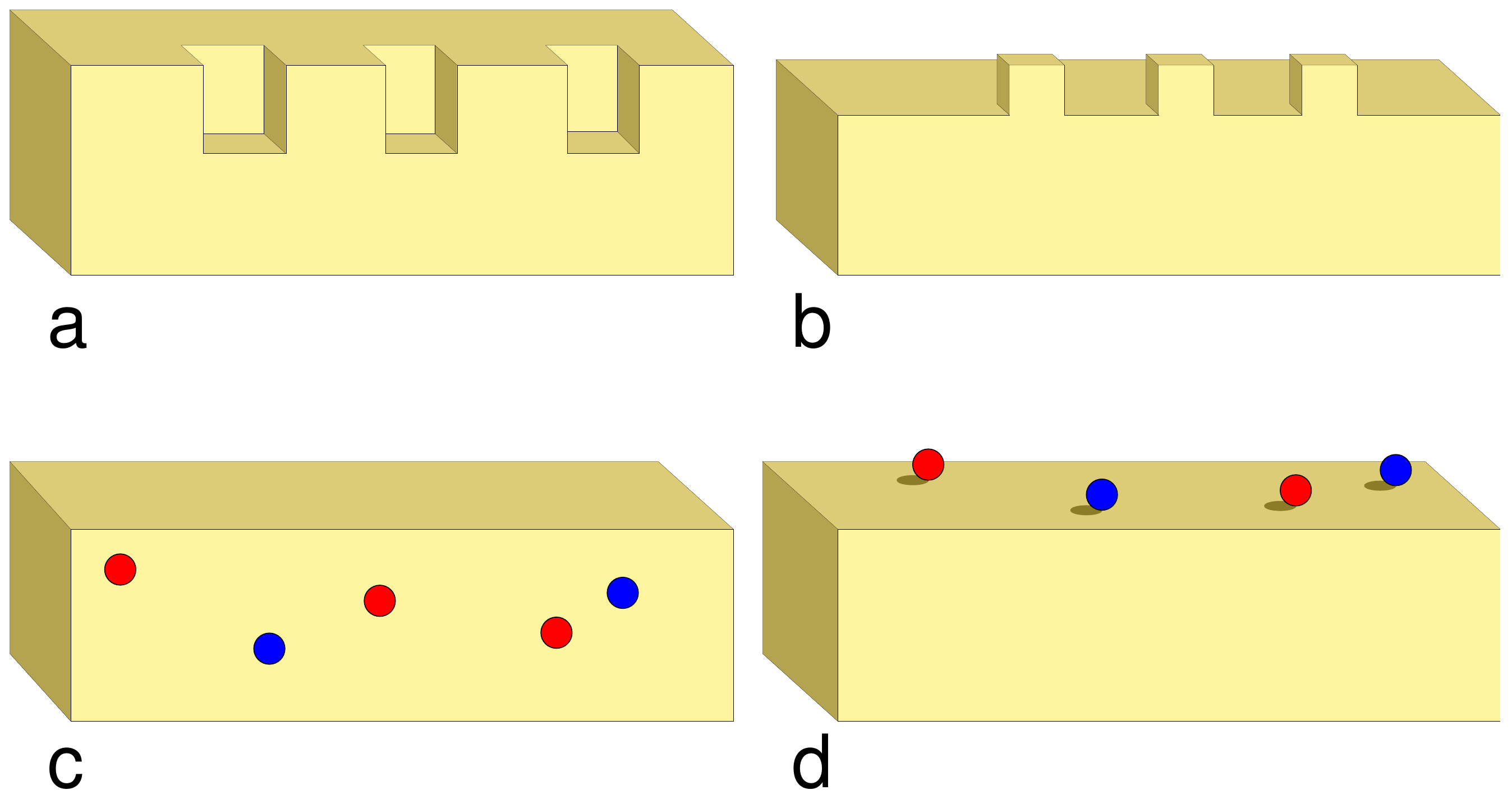}
\caption{Schematic illustrations of possible ways
 that pinning can arise in skyrmion systems. (a) Surface thickness modulations. 
 (b) The addition of nanodots to the surface.
 (c) Naturally occurring atomic defects or substitutions in the bulk of the sample. (d) 
Adatoms on the surface of the sample.  
 }
\label{fig:10}
\end{figure}

In the experiments by Schulz {\it et al.} \cite{Schulz12},
the motion of skyrmions was inferred from
observations
of changes in the topological Hall effect.
This technique provided evidence of
a finite depinning threshold for skyrmion motion,
and in many
subsequent imaging experiments,
a wide range of depinning thresholds has been observed 
ranging from $10^6$ to $10^{11}$ A/m$^2$.
In superconducting vortex systems, pinning arises at locations
where
the order parameter of the superconducting condensate is lowered.
The system can reduce its energy by placing a vortex at these locations
since the condensation energy is already suppressed to zero
at the vortex core
\cite{Blatter94}.
In the case of colloidal particles, pinning can be produced via optical 
trapping \cite{Reichhardt02} or simply by providing a substrate on which
the particles can be localized
\cite{Pertsinidis08,Tierno12}, while in 
Wigner crystals the pinning is produced by offset charges \cite{Reichhardt01}. 
For skyrmions, pinning effects can in principle arise in numerous ways.
These include
local changes in the DMI, missing spins,
holes in thin film samples, a local change in the anisotropy,
sample thickness modulations, localized
changes in the magnetic field,
impurity atoms embedded in the bulk, or adatoms adhering to the surface.  
Schematics of some of the possible pinning mechanisms appear in
Fig.~\ref{fig:10}.
It is possible to
introduce
a surface modulation by fabricating holes or antidots
as in Fig.~\ref{fig:10}(a),
place dots in the form of magnetic nanoparticles on the surface as in
Fig.~\ref{fig:10}(b),
take advantage of
naturally occurring atomic defects in the bulk such as missing atoms
or substitutions as shown in Fig.~\ref{fig:10}(c), or
place adatoms on the surface as in Fig.~\ref{fig:10}(d).
Grain boundaries, twin boundaries,
or dislocations
can also serve as pinning sites
in thin film systems.

There is no threshold current for skyrmion motion
in micromagnetic simulations of uniform samples without defects
\cite{Lin13a}. 
One of the first theoretical studies of skyrmion pinning was
performed by Iwasaki {\it et al.} \cite{Iwasaki13},
who used micromagnetic simulations with
parameters appropriate for MnSi and modeled the pinning
as small regions in which
the local anisotropy $A$ varied.
In this system, where the
ratio of the local anisotropy to the exchange term $J$ is
$A/J = 0.2$,
the depinning threshold is $j_{c} \approx 10^{10} - 10^{11}$A/m$^2$
and the skyrmion depins elastically.
Lin {\it et al.} used a combination of micromagnetic simulations
and particle based simulations
for 2D skyrmions and also found
finite depinning thresholds for both cases \cite{Lin13}.       

Liu and Li \cite{Liu13} considered a local exchange mechanism for
producing skyrmion pinning,
achieved by
varying the local density of itinerant electrons.
Using micromagnetics 
and a Thiele equation approach, they found
that the skyrmion
is pinned due to the
lowering of the skyrmion core energy.
They also showed that under perturbation by a small drive,
the skyrmion
performs a spiraling trajectory as it returns to the pinning site,
in contrast to an overdamped particle which moves linearly back to
its equilibrium position.
The spiraling motion is produced by the Magnus force.
When the current is large, the skyrmion 
is able to escape the trap and a depinning phenomenon occurs.

\begin{figure}
\includegraphics[width=\columnwidth]{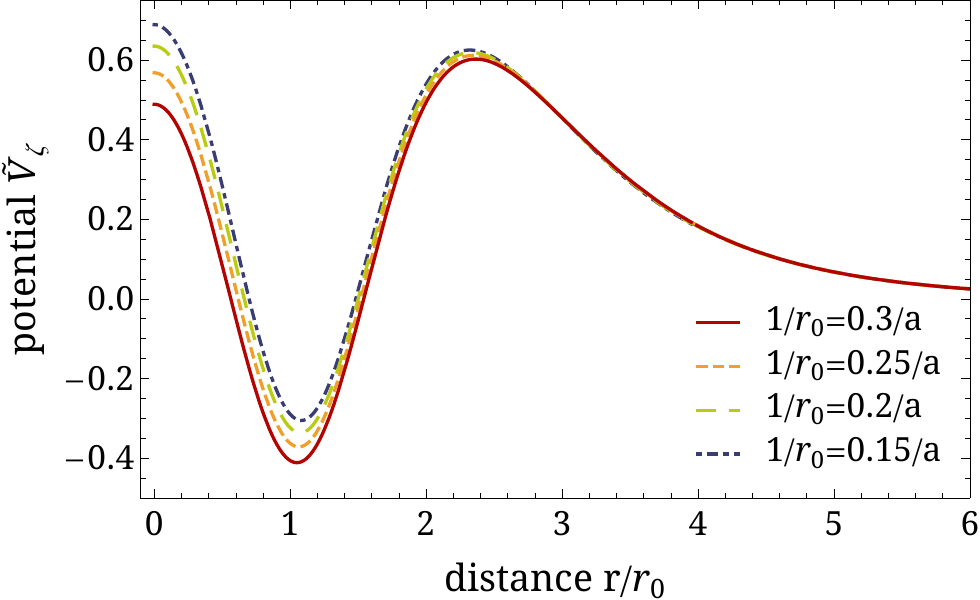}
\caption{
The shape of the pinning potential produced by a hole in the sample,  
which has longer range repulsion and a short range attraction \cite{Muller15}.
Reprinted with permission from J. M{\" u}ller and A. Rosch, Phys. Rev. B
{\bf 91}, 054410 (2015). Copyright 2015 by the American Physical Society.
}
\label{fig:11}
\end{figure}

Muller {\it et al.} \cite{Muller15}
considered the interaction of a skyrmion with a hole
or locally damaged region
both analytically and numerically
using continuum methods and the Thiele equation approach.
They found that the potential
generated by the hole has the interesting property of combining
a longer range repulsion with
a short range attraction.
The competition that results when a drive is applied produces
an unusual effect.
The skyrmion moves around the
pinning site at low drives due to the repulsion,
but at high drives 
it jumps over the longer range repulsive barrier and is captured by
the short range attraction.
At even higher drives, the skyrmion  
escapes from the attractive part
of the pinning site and the system enters a flow regime.
The competing attractive and repulsive
potential experienced by the skyrmion due to the hole is illustrated
in Fig.~\ref{fig:11}.

Choi {\it et al.} \cite{Choi16} used density functional
theory to study the effects of atomic defects on 
skyrmions in MnSi. They found that if
Si is substituted or if Mn is substituted
by Zn or Ir,
the resulting defect sites attract the skyrmions,
whereas if Mn is substituted with Co, the interaction with the defect sites
is repulsive.

\begin{figure}
\includegraphics[width=\columnwidth]{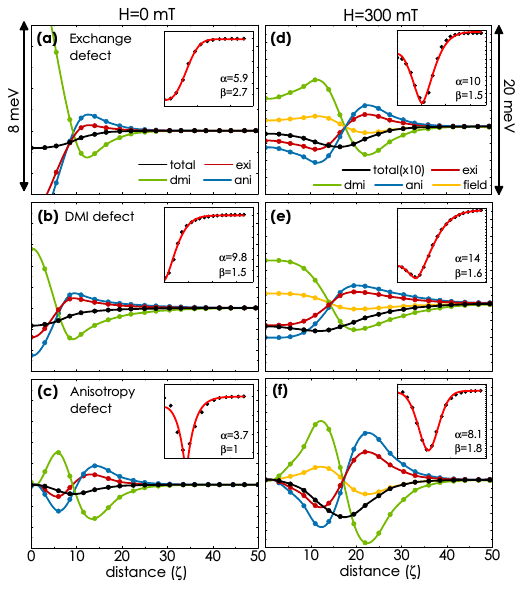}
\caption{
  The
  total (total),
  exchange  (exi),
  Dzyaloshinskii-Moriya (dmi), anisotropy (ani), and Zeeman (field)
  energies
  plotted as a function of distance $\zeta$
  along the minimum energy path for a skyrmion to escape from
  the defect
  for three different types of defects at
  external fields of $H=0$ mT [(a)-(c)]
  and $H=300$ mT [(d)-(f)] \cite{Stosic17}.
  In the insets, the total energy landscape or effective pinning potential
  is fit to an exponential power function.
Reprinted with permission from D. Stosic {\it et al.}, Phys. Rev. B
{\bf 96}, 214403 (2017). Copyright 2017 by the American Physical Society.
}
\label{fig:12}
\end{figure}

For Co monolayers on Pt,
Stosic {\it et al.} \cite{Stosic17} examined the interactions
of skyrmions 
with atomic defects
and studied the pinning potentials
at different locations
including on or between domain walls.
Figure~\ref{fig:12} shows the
total, exchange, DMI, anisotropy, and Zeeman energies
as a function of the distance $\zeta$ along the
minimum energy path for the skyrmion to escape from the pinning.
The insets indicate that the total energy $G$
can be fit to an
exponential power function
$G(\zeta) \propto -\exp[-(\zeta/\alpha)^\beta$], 
where $\alpha$ and $\beta$ are the scale and shape parameters.
Stosic {\it et al.} found that off-center pinning sites
are well described by a similar energy expression
with a radial shift.

Navau {\it et al.} \cite{Navau18}
used micromagnetic simulations to study
the properties of skyrmion-defect interactions in thin films containing DMI
modulations, and 
also obtained analytic expressions for the skyrmion-defect forces within
a rigid skyrmion model approximation.
They found that the pinning is enhanced when the defect increases
the DMI but weakened when the defect decreases the DMI.
Anisotropic defects
can be attractive, repulsive, or have a combination
of the two effects. 

\begin{figure}
\includegraphics[width=\columnwidth]{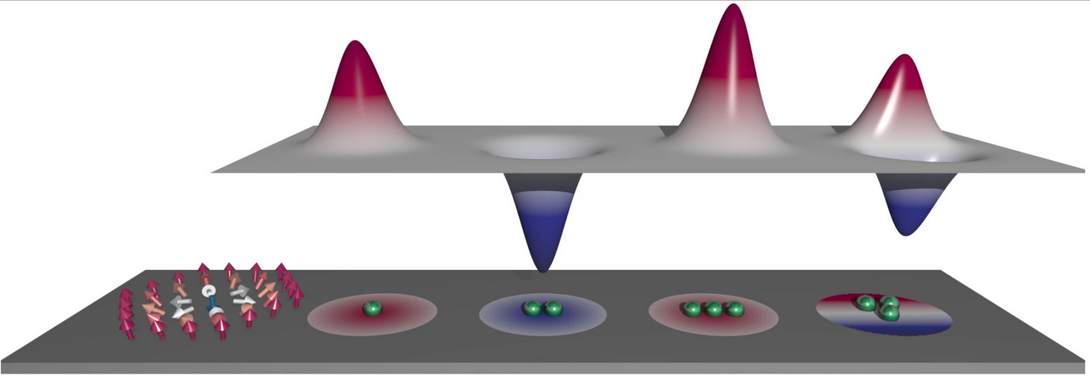}
\caption
{
Schematic of atom-by-atom construction of potential landscapes for
skyrmions. The leftmost cluster of arrows illustrates the size of a typical
skyrmion. Green spheres are atoms which have been placed so as to construct,
from left to right, a repulsive, attractive, strongly repulsive, or
combined attractive and repulsive pinning potential \cite{Arjana20}.
Reprinted under CC license from I. G. Arjana {\it et al.}, Sci. Rep.
{\bf 10}, 14655 (2020).
}
\label{fig:fig.new.2}
\end{figure}

From first principles calculations for skyrmions interacting with
single-atom impurities,
Fernandes {\it et al.} \cite{Fernandes18}
found that 
defects can be
both attractive and repulsive or purely attractive
depending on the impurity type.
They focused on PdFe bilayers on an Ir substrate and
considered a range of defect transition metal atoms
including 3d (Sc, Ti, V\ldots) and 4d (Y, Zr, Nb\ldots) atoms as well as
Cu and Ag atoms, with the defects either
located on the surface or embedded in the Pd surface layers.
By analyzing the 
binding energy to determine whether it is positive or negative,
they found that it is possible to have both attractive and repulsive
interactions with various strengths that depend strongly on the element used.
A key feature of this system
is that strongly magnetic defects locally stiffen the skyrmion,
leading to a repulsive skyrmion-defect interaction, while 
weakly interacting defects produce
attractive pinning due to the substrate contribution.
Since the pinning
originates from
surface atoms, it would be possible to use
scanning tunneling microscopy
to
add atoms in prescribed patterns in order to create
attractive and repulsive pinning sites that precisely
control the deviations of the skyrmions.
Arjana {\it et al.} \cite{Arjana20}
pursued this
idea by examining atom by atom crafting of skyrmion defect landscapes
using single,
double, and triple atom states to create repulsive, attractive,
and combined repulsive and attractive pinning sites,
as illustrated in
Fig.~\ref{fig:fig.new.2}.
They also
generated asymmetric landscapes and
demonstrated that atomic clusters
could be used to
construct reservoir computing devices.

Larger scale magnetic defects can be created using
a variety of nanoscale methods.
These include
irradiating
particular regions
of the sample
in order to
change the local magnetic properties
\cite{Fassbender09},
introducing large scale thickness modulations to change the
DMI \cite{Yang15}, or adding magnetic dots to the surface in a manner similar
to that used for introducing pinning in superconductors \cite{Martin97,Marchiori17}.  
In extensive 
micromagnetic simulations
of skyrmion trapping by larger scale magnetic defects,
Toscano {\it et al.} \cite{Toscano19}
found that
the defects can act either as attractive traps or
as repulsive scatterers
depending on the exchange stiffness, DMI, perpendicular anisotropy,
and saturation magnetization.
A defect that modifies the exchange stiffness
acts as a skyrmion trap when its exchange stiffness is smaller than
that of the surrounding material
but as a repulsive scattering site when the exchange stiffness
exceeds that of the surrounding material.
Additionally, the strength of the interaction with a pinning site increases
when the skyrmion becomes smaller than
the size of the defect.
In other micromagnetic simulations for skyrmions moving in 
nanostructured  materials,
a large region with altered local anisotropy
was shown to act as a repulsive area
for the skyrmions \cite{Wang18,Ding15}.

Wang {\it et al.} \cite{Wang17} introduced
the concept of pinning skyrmions with magnetic field gradients and
showed that the pinning strength depends on the intensity of the
gradient as well as on
the skyrmion size.
They
demonstrated that a skyrmion can be dragged 
and manipulated with a suitable magnetic field gradient,
suggesting a new way to move skyrmions by using
a magnetic tip.  

Beyond the evidence for skyrmion pinning obtained from transport studies, 
pinning effects can also be deduced via the manipulation of individual
skyrmions.
Hanneken {\it et al.}~\cite{Hanneken16}
explored
the interactions between nanometer-scale skyrmions
and atomic scale defects in PdFe
by measuring the force
needed to move a skyrmion, which revealed the presence of a range of
pinning strengths.
They also found that interlayer defects such as single Fe atoms interact
strongly with a
skyrmion
while single Co adatoms on the surface
are weak pinning centers;
however, clusters of such adatoms can serve as strong pinning sites.

\subsection{Skyrmion Pinning by Individual versus Extended Defects and the Role of the Magnus Force}

\begin{figure}
\includegraphics[width=\columnwidth]{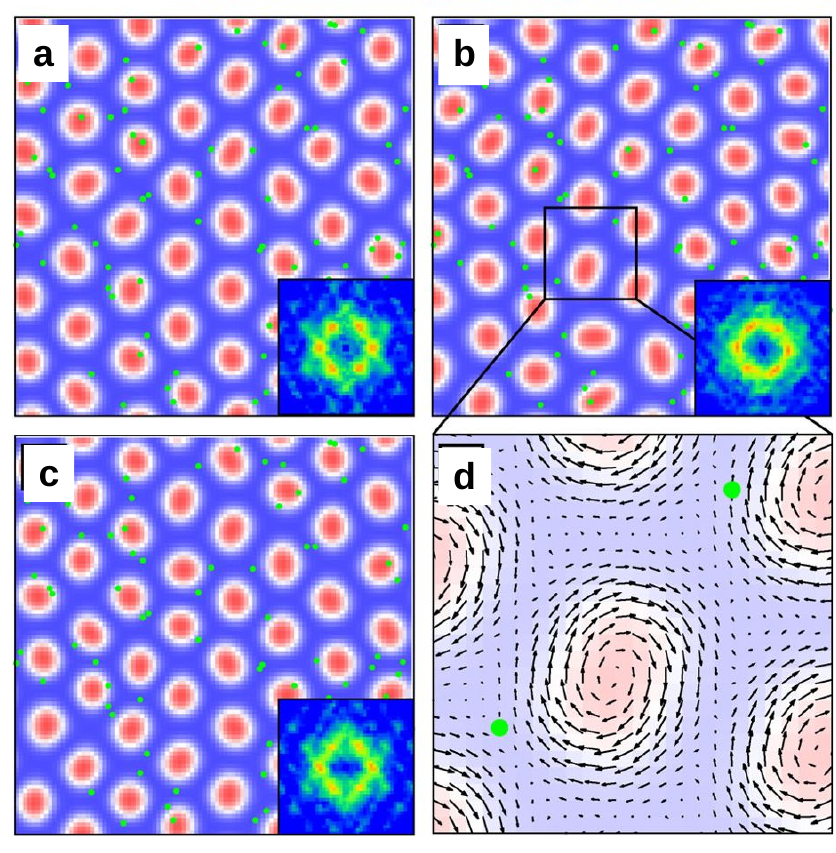}
\caption{
Micromagnetic simulation images of the evolution of
a skyrmion crystal
and the distortions of the skyrmions
for different times under the application
of a driving current \cite{Iwasaki13}. The times are
(a) $t=1.30 \times 10^{-8}$ s,
(b) $t=2.60 \times 10^{-8}$ s,
and (c) $t=8.47 \times 10^{-8}$ s.
Green dots are the defect sites and red regions are the skyrmion centers,
while the insets show the corresponding structure factor measurement.
Panel (d) shows a magnified view of the distortion of the circular
shapes of the skyrmions in panel (c).
Reprinted by permission from: Springer Nature,
``Universal current-velocity relation of skyrmion motion in chiral magnets,''
Nature Commun. {\bf 4}, 1463 (2013),
J. Iwasaki {\it et al.}, \copyright 2013.
}
\label{fig:14}
\end{figure}

\begin{figure}
\includegraphics[width=\columnwidth]{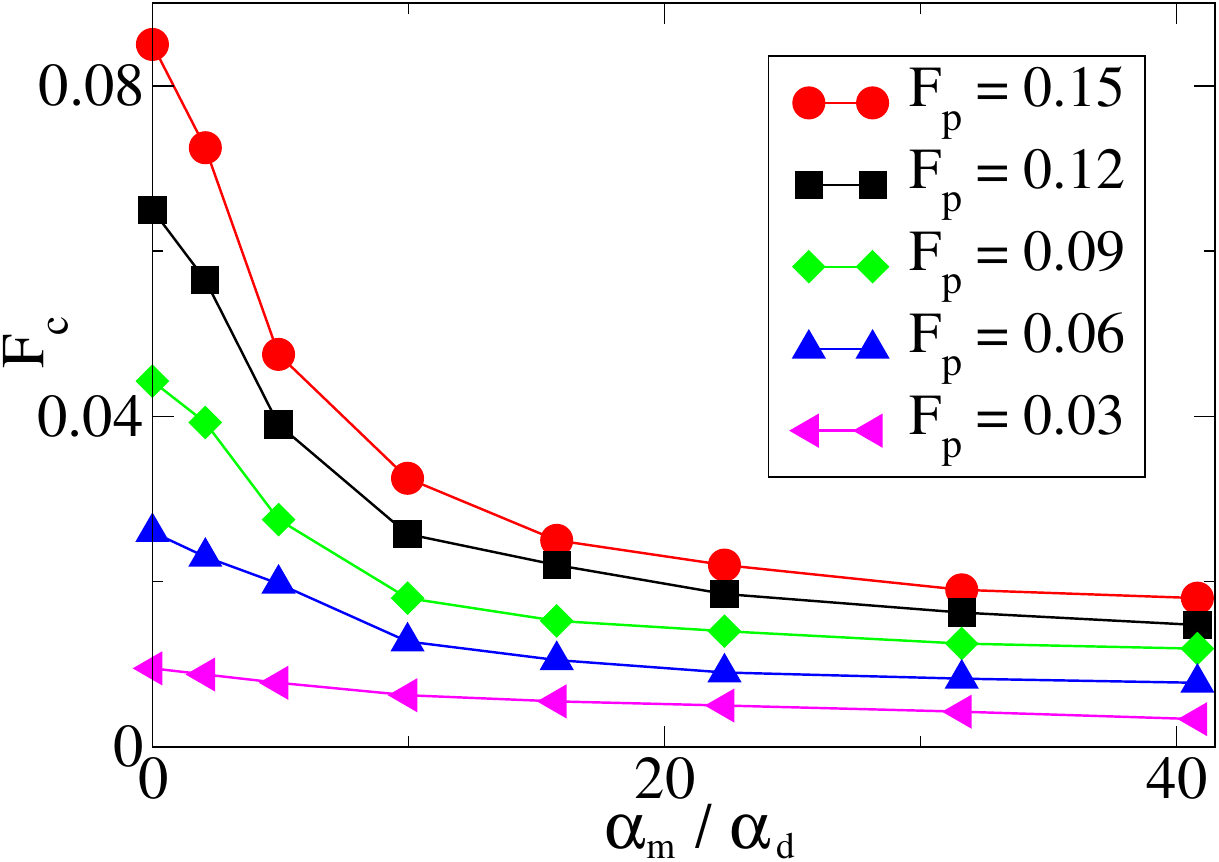}
\caption{The critical depinning force $F_{c}$ vs the ratio $\alpha_m/\alpha_d$
  of the Magnus
force to the dissipative
term
for 2D particle based simulations of skyrmions
moving over pointlike disorder sites for 
varied pinning strength $F_p$ \cite{Reichhardt15a}.
The depinning threshold decreases with increasing Magnus force.
Reprinted with permission from C. Reichhardt {\it et al.}, Phys. Rev. Lett.
{\bf 114}, 217202 (2015). Copyright 2015 by the American Physical Society.
}
\label{fig:15}
\end{figure}

In many systems such as vortices in type-II superconductors,
it is known that very different pinning 
effects can occur when the defects are extended or linelike instead of pointlike.
Such extended defects can form naturally,
as in the case of twin boundaries \cite{VlaskoVlasov94}, or they can be
introduced with nanoscale techniques \cite{Guillamon14}. 
In superconducting vortex systems,
a line defect can serve as a region of increased pinning
for motion across the lines
\cite{VlaskoVlasov94,Groth96},
but it can also produce guided or easy flow for motion
of vortices along the line \cite{Groth96,Duran92}. 
In skyrmion systems,
it was initially argued that a skyrmion can
move around a point pinning site due to the Magnus effect \cite{Nagaosa13}.
Micromagnetic simulations by Iwasaki {\it et al.}~\cite{Iwasaki13} 
showed that pining was reduced not only by this avoidance
motion but
also by the ability of the skyrmions
to change their shape, as shown in Fig.~\ref{fig:14}.

\begin{figure}
\includegraphics[width=\columnwidth]{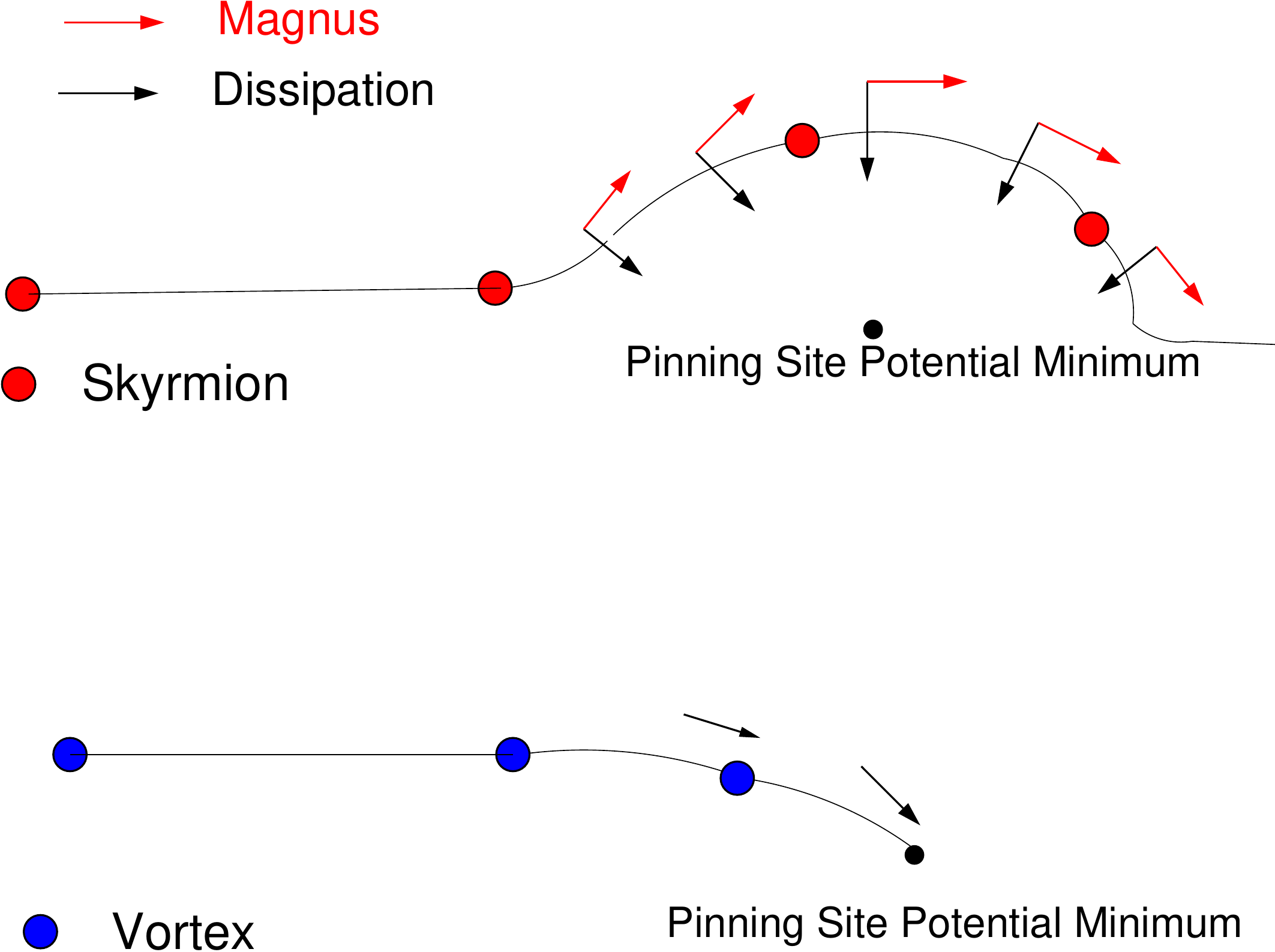}
\caption{ Schematic of a skyrmion (red dot) with both Magnus and dissipative terms 
and a vortex (blue dot) with only a dissipative term interacting with 
an attractive point pinning site (black dot)
to illustrate how the Magnus force can decrease the
effectiveness of the pinning.
The velocity component induced by the dissipation
is indicated by black arrows, and that produced by the 
Magnus force is shown as red arrows.
Since the Magnus force
velocity component is perpendicular to the
attractive force from the pinning site,
the skyrmion deflects around the pinning site.
In contrast, the vortex moves directly toward the
potential minimum and is more likely to be trapped by the pinning site.
}
\label{fig:16}
\end{figure}

Particle based simulations
of skyrmions interacting with
pointlike random pinning \cite{Reichhardt15a} in 2D systems
indicate
that the depinning threshold decreases as the ratio $\alpha_m/\alpha_d$ of the
Magnus force to the dissipative term increases
over a wide range of pinning strengths, as shown in Fig.~\ref{fig:15}.
A schematic illustration of
how the Magnus force reduces the effectiveness of the pinning
for a skyrmion interacting with a point pinning site appears in
Fig.~\ref{fig:16}.
The black arrows indicate the direction
of the attractive force from the pinning site, which always points toward the pin.
The red arrows represent the Magnus force component,
which is always perpendicular to the force from the pinning site.
The net effect is that, although the dissipative term favors the motion
of the skyrmion toward the pinning site,
the 
Magnus force causes the skyrmion to deflect around the pinning site.
In contrast, a purely overdamped
particle such as a superconducting vortex
moves directly toward the center of the pinning site and is likely to
be trapped.
The deflection of the skyrmion around the pinning site
depends strongly on the relative size and extent of the skyrmion compared to
that of the
pinning site.

\begin{figure}
\includegraphics[width=\columnwidth]{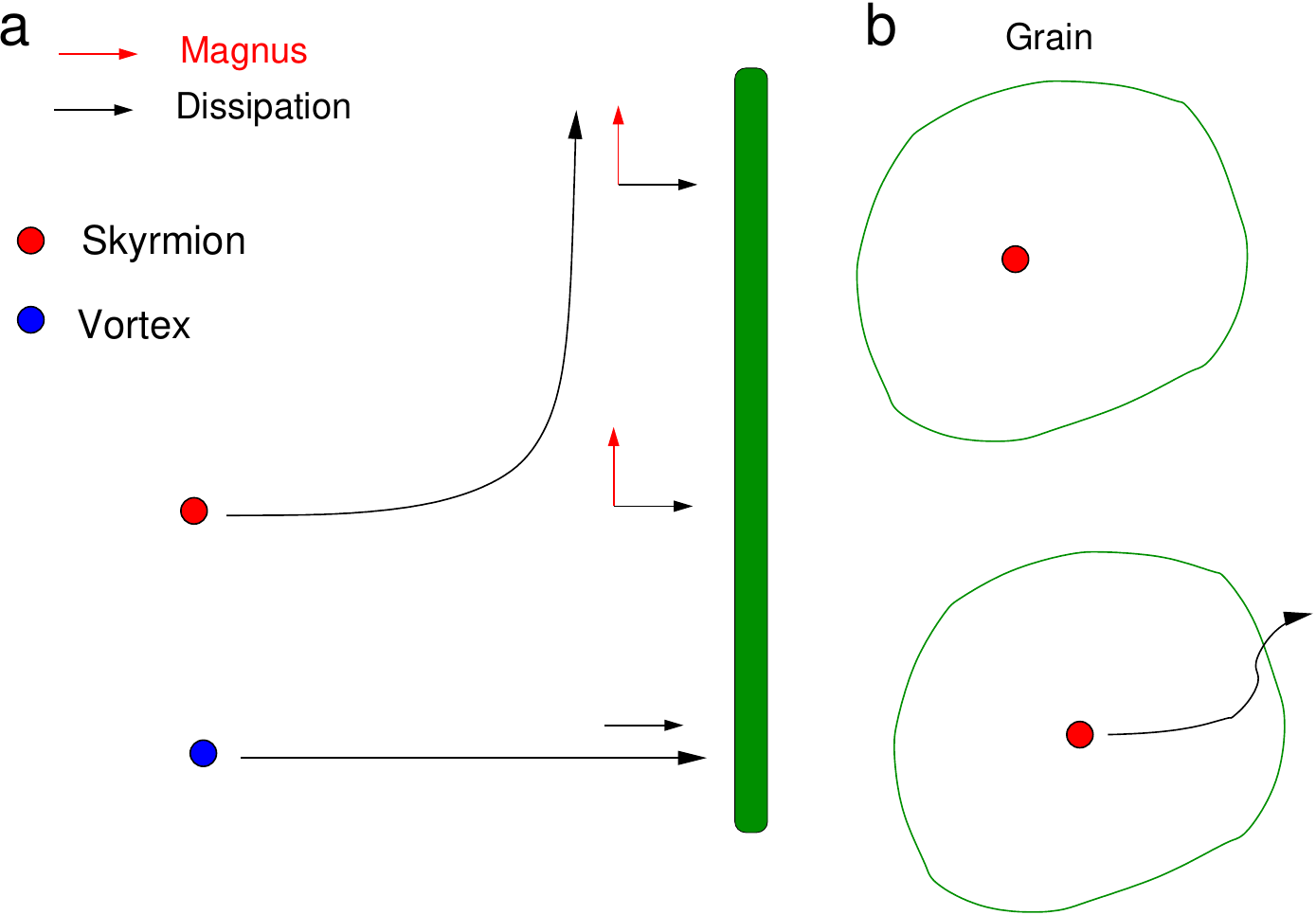}
\caption{ (a) Schematic showing the dissipative (black arrows) and
Magnus (red arrows) velocity components for a skyrmion (red dot) or
vortex (blue dot)
moving toward an attractive extended line defect (green). 
The overdamped vortex moves directly toward the line defect,
while the skyrmion is deflected but 
gradually approaches the defect.
Unlike the case for point pinning in Fig.~\ref{fig:16},
the skyrmion cannot simply move
around the line defect but eventually reaches the defect and
interacts with it.
(b) Schematic of a skyrmion (red dot) located inside a closed grain
boundary (green).  The skyrmion may
be deflected as
it moves toward the grain boundary;
however, it must cross the pinning potential minimum
in order to pass through the grain boundary.  
}
\label{fig:17}
\end{figure} 

Experiments by Woo {\it et al.}~\cite{Woo16}
on room temperature ultrathin systems
unexpectedly showed that the skyrmions experience strong pinning. 
The intrinsic pinning in these samples may not be pointlike
due to
the nature of the films,
which contain grain boundaries or extended defects.
Continuum based simulations of skyrmions
\cite{Legrand17} confirmed
that grain boundaries induce skyrmion pinning
and that the strength of this pinning increases for smaller grain sizes; however,
there is a minimum grain size below which pinning cannot occur.
One explanation for the stronger pinning by
extended defects
is that although the Magnus force can cause a skyrmion to skirt
a point defect, the same process
is not possible for an extended defect.
In Fig.~\ref{fig:17}(a)
we
schematically illustrate
the Magnus and dissipation induced velocity components
of a skyrmion moving toward
an extended line defect.
If the defect is attractive,
the dissipative term aligns the velocity toward the line defect,
but the Magnus force generates a velocity component
perpendicular to the line defect, causing the skyrmion trajectory to
bend sideways as the skyrmion approaches the defect.
If the defect line extends across the sample,
the skyrmion is unable to avoid the defect, but eventually reaches and
crosses it.
As a result, the skyrmion experiences the full pinning potential
of the line defect. This is
in contrast to the point pinning situation where the skyrmion can
completely avoid the pin.
If a driven skyrmion is inside an extended line defect such
as a grain boundary,
as shown schematically in Fig.~\ref{fig:17}(b),
thee skyrmion trajectory may bend upon
approaching the boundary due to the Magnus force,
but the skyrmion eventually must pass through the potential minimum
in order to exit the grain boundary,
as illustrated in the lower panel of Fig.~\ref{fig:17}(b).
As a result, extended defects
are always more effective
than point defects at exerting pinning forces on skyrmions.

\begin{figure}
\includegraphics[width=\columnwidth]{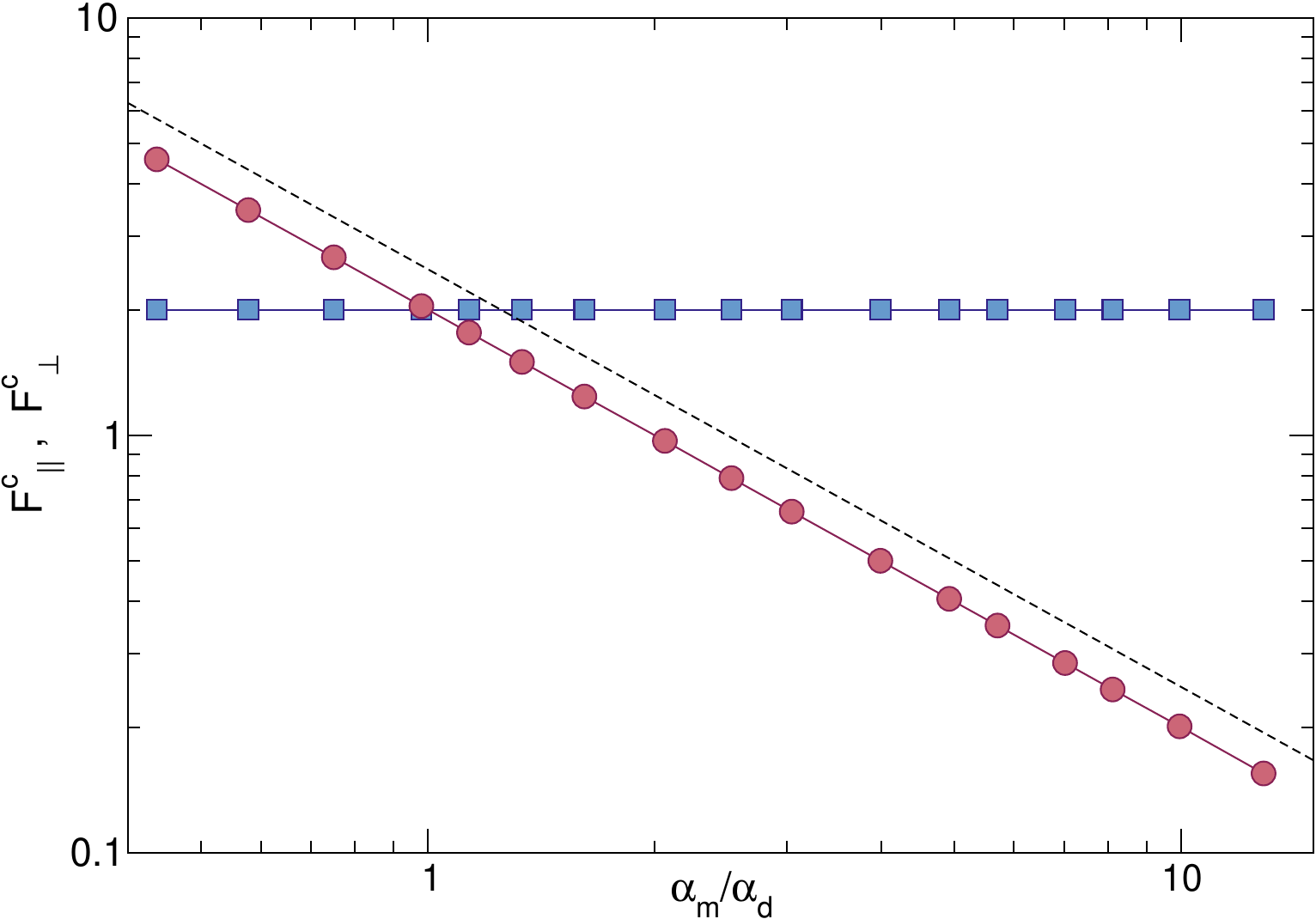}
\caption{
2D particle based numerical simulations of a skyrmion interacting
with a 1D defect line showing  
the critical depinning force
for driving applied parallel,
$F^c_{||}$
(blue squares),
and perpendicular, $F^c_{\perp}$ (red circles), to the line, vs
the relative strength $\alpha_m/\alpha_d$ of the Magnus force.
$F^c_{||}$ is insensitive to the Magnus force while
$F^c_{\perp}$ decreases with increasing Magnus force.
\cite{Reichhardt16a}.
Reprinted with permission from C. Reichhardt {\it et al.}, Phys. Rev. B
{\bf 94}, 094413 (2016). Copyright 2016 by the American Physical Society.
}
\label{fig:18}
\end{figure}

A numerical test of 
the effect of the Magnus force on skyrmions
moving perpendicular to a line defect
was performed 
by Reichhardt {\it et al.}~\cite{Reichhardt16a}
for a 2D skyrmion moving over a
1D pinning line.
As shown in Fig.~\ref{fig:18},
when the driving direction is parallel to the pinning line,
the critical current $F_{c}$ is independent of the size of the Magnus term,
in contrast to the decrease in critical current with increasing
Magnus term found for point pinning.
On the other hand, when the drive
is applied perpendicular to the line defect,
the depinning threshold
decreases with increasing Magnus force.
This effect would be most pronounced for skyrmions
moving over 1D pinning features such as twin boundaries,
but would likely be absent in a sample filled with closed grain boundaries.

The best model for the interaction between skyrmions and
extended defects
is dictated by the nature of the defects.
For example, if the defects arise due to  
thickness modulations, the pinning will be short range and attractive,
whereas if the defects are in the form of
magnetic stripes,
the pinning will be
longer range with a
dipolar form $A/r^3$ and
it could be either attractive or repulsive.
In some cases the extended defect could have a competing potential
in which
the interaction with the skyrmion
is repulsive at longer distances but becomes attractive close to the defect.

\begin{figure}
\includegraphics[width=\columnwidth]{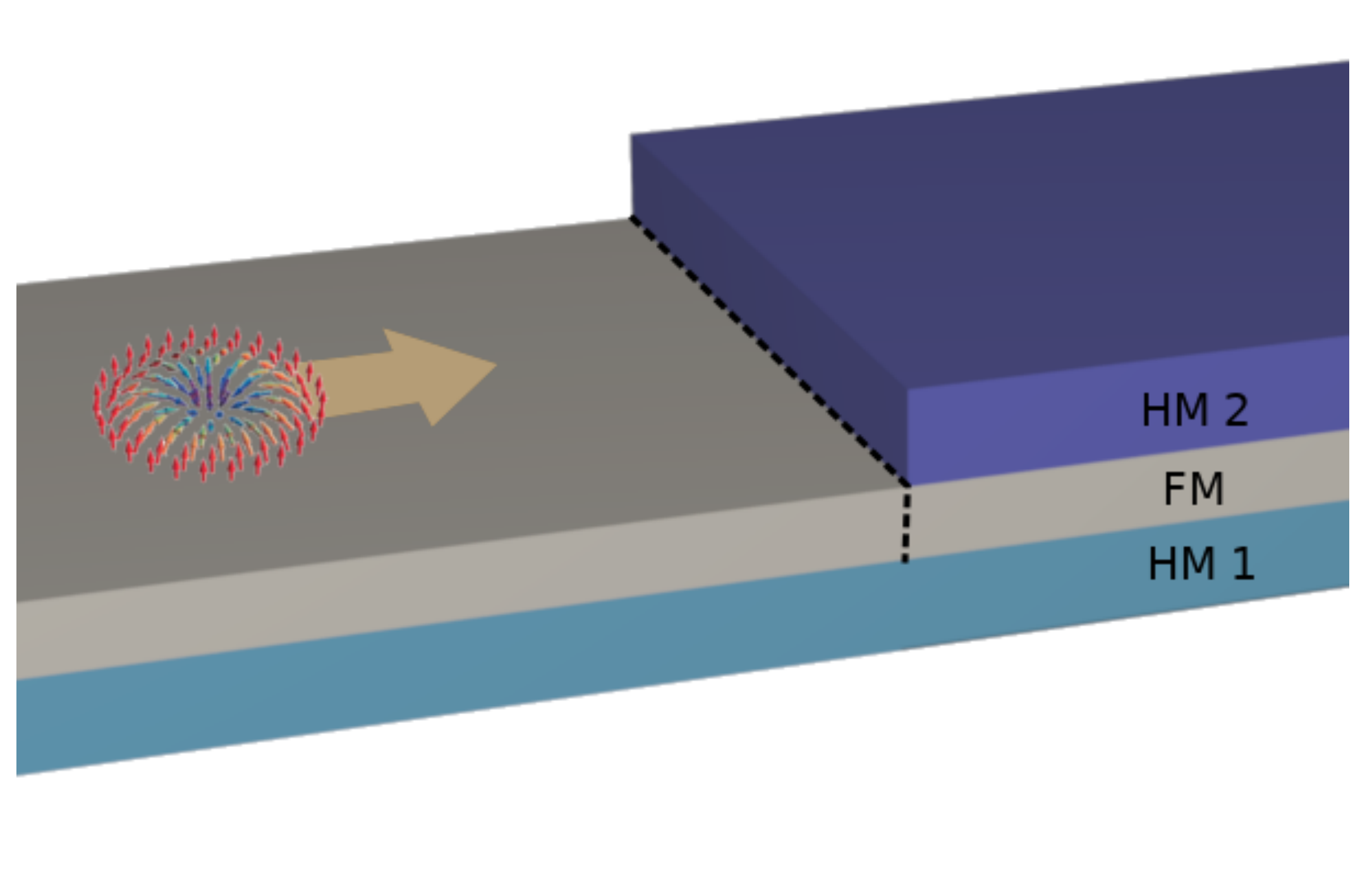}
\caption{
Schematic of a skyrmion moving toward an interface
composed of a 
ferromagnetic (FM) layer sandwiched between two
heavy metal (HM) layers.
The upper heavy metal layer is truncated in
order to create a heterochiral interface
\cite{Menezes19}.
The dashed line indicates the interface where the DMI changes.
Reprinted with permission from R. M.  Menezes {\it et al.}, Phys. Rev. B
{\bf 99}, 104409 (2019). Copyright 2019 by the American Physical Society.
}
\label{fig:19}
\end{figure}

\begin{figure}
\includegraphics[width=\columnwidth]{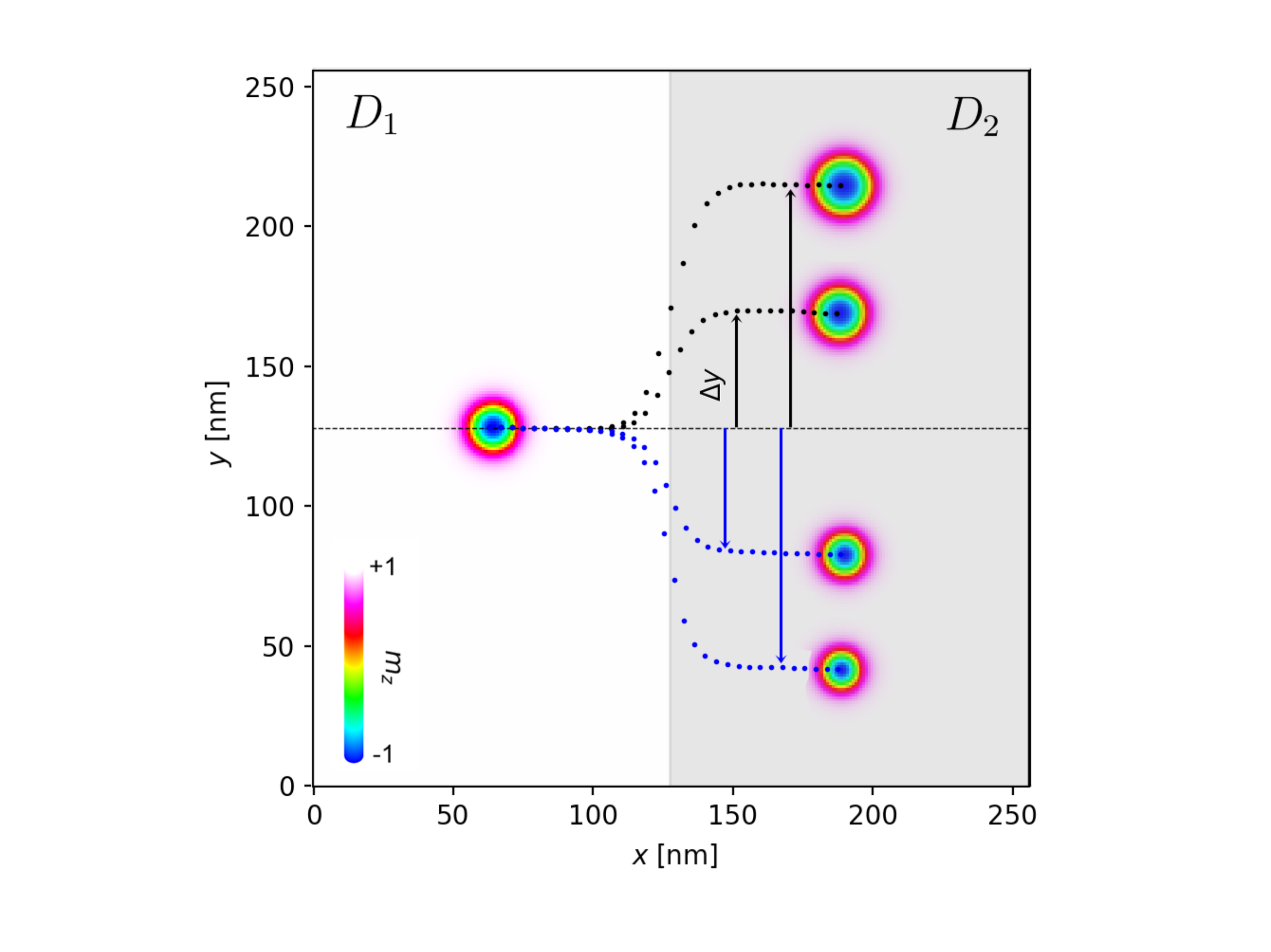}
\caption{
Illustration of skyrmion motion through the heterochiral interface
shown in Fig.~\ref{fig:19} \cite{Menezes19}.
The initial skyrmion position is on the
left side of the interface. As the relative DMI strengths $D_1$ and $D_2$
are varied with respect to each other, the skyrmion trajectory is
deflected by a distance $\delta y$ in the positive (black arrows)
or negative (blue arrows) $y$ direction.
Reprinted with permission from R. M.  Menezes {\it et al.}, Phys. Rev. B
{\bf 99}, 104409 (2019). Copyright 2019 by the American Physical Society.
}
\label{fig:20}
\end{figure}

Navau {\it et al.}
simulated the
Thiele model 
for a single skyrmion interacting with an extended defect
consisting of an edge \cite{Navau16}.
The edge exerts a force on the skyrmion of the
form ${\bf f} = -f_{0}e^{-d/d_{0}}{\bf {\hat n}}$,
where $d$ is
the distance between the skyrmion center of mass
and the edge, ${\bf {\hat n}}$ is a unit 
vector perpendicular to the edge,
and $d_{0}$ is approximately equal to the diameter of the skyrmion.
The skyrmion is strongly deflected by the edge.
Other work \cite{Navau18}
showed that extended defects can exert either repulsive or attractive forces
on a skyrmion.
The dynamics of a skyrmion interacting
with an extended defect depends on both the form of the defect
and the skyrmion type.
Menezes {\it et al.}~\cite{Menezes19} considered micromagnetic simulations
of a skyrmion moving toward a heterochiral interface created
with multilayers as illustrated in Fig.~\ref{fig:19}.
They found that a ferromagnetic skyrmion is deflected by
the interface, and that the deflection amplitude
can be tuned by changing the applied current or 
by modifying the difference in the DMI across the interface,
as shown in Fig.~\ref{fig:20}. On the other hand,
antiferromagnetic skyrmions experience no deflection at the interface.

\subsection{Further Directions}

There are numerous theoretical, computational, and 
experimental directions for further study
of the basic mechanisms of skyrmion pinning.
Simulations and theory have shown that there are many ways to create
attractive, repulsive, or both attractive and repulsive
pinning sites, 
so one of the next steps is to consider how to combine
these different types of pinning sites to produce novel 
dynamical phenomena, control the skyrmion motion, and
reduce or enhance pinning.
In many other systems
where pinning occurs,
such as for vortices in type-II superconductors,
the natural or artificial defects producing the pinning 
reduce the superconducting condensation energy,
so studies
have focused on
strictly attractive pinning sites. 
In colloidal systems, optical forces and most surface modifications
also create attractive pinning sites.
As a result,
systems with repulsive defect sites represent a relatively unexplored
regime of collective dynamics.
Many skyrmions in thin films seem
to show strong pinning effects from attractive pins;
however, there may be a way to
introduce additional
repulsive defect sites
that would effectively reduce the
overall pinning by competing with 
the attractive pinning centers.

Another question is the nature of the pinning process
for antiskyrmions or antiferromagnetic skyrmions.
The work of Menezes {\it et al.}
\cite{Menezes19} on a line defect separating two regions
with different DMI suggests that antiferromagnetic skyrmions
may not be very susceptible
to changes in the DMI, thus reducing the number of possible
methods available for pinning skyrmions of this type.
This could mean that
antiferromagnetic skyrmions would be more mobile
than ferromagnetic skyrmions
due to a reduction in the amount of pinning present; however,
the lack of a Magnus
effect in the antiferromagnetic skyrmions
could make any pinning sites that are present more effective.
It would also be interesting to explore the pinning of
biskyrmions, merons, and other related objects such
as skyrmionium \cite{Kolesnikov18},
as well as the role pinning plays 
in determining the direction
of current flow.
For example, Stier {\it et al.} \cite{Stier21}
showed in simulations that
although magnetic impurities do not interfere with
a uniform applied current, conducting impurities can change the current paths.
If defects could be introduced that
are able to move over time in response to a current,
they would create a pinning
landscape that can gradually be sculpted in a manner
similar to electromigration.
This could produce interesting memristor-like effects.  

\begin{figure}
\includegraphics[width=\columnwidth]{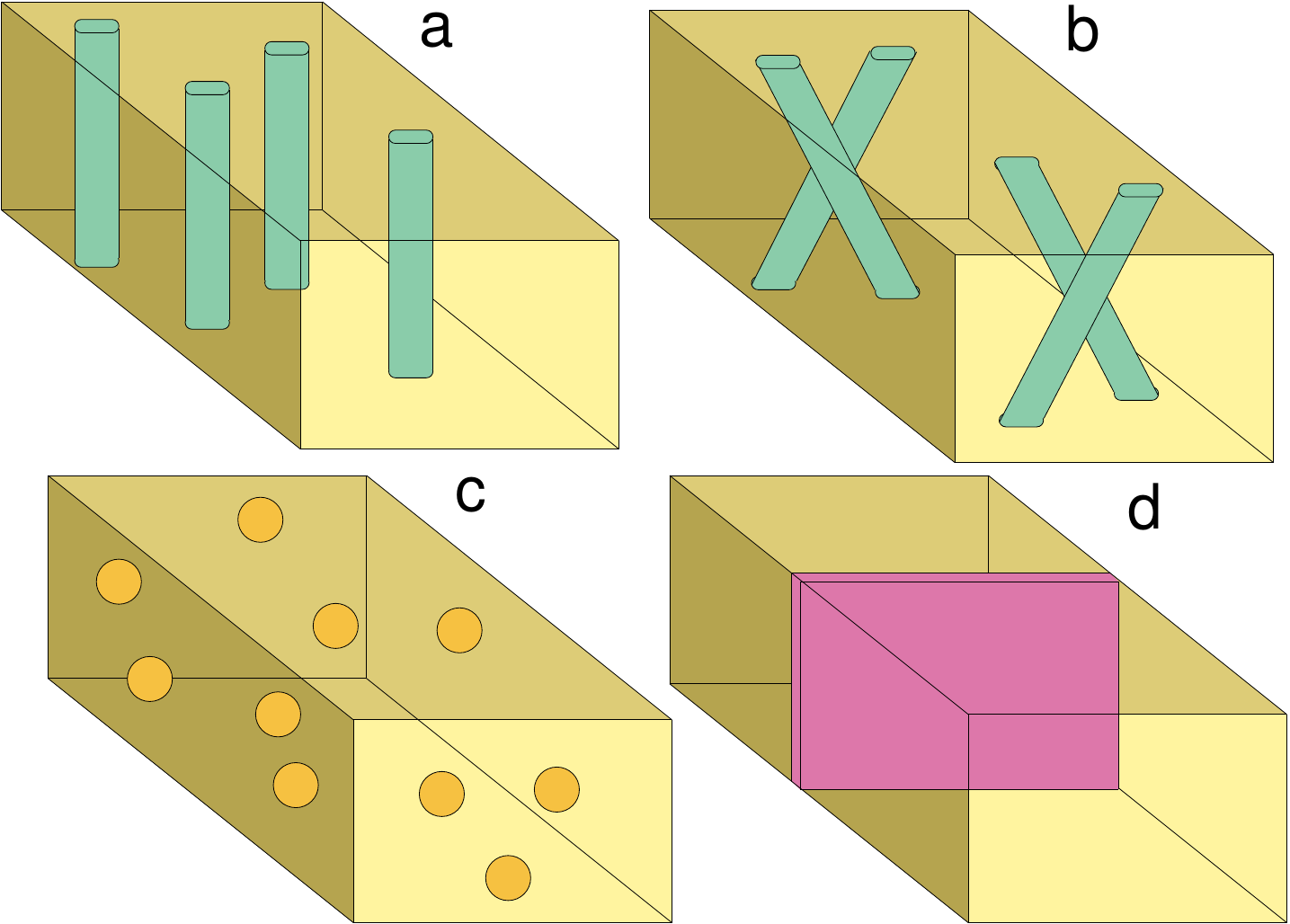}
\caption{ Schematic of possible different types of 3D defects
  that could be created for bulk skyrmions. 
  (a) Columnar defect tracks, which could induce the formation of a skyrmion Bose glass.
  (b) Splayed columnar defects, which could
  create a splayed skyrmion glass or promote skyrmion entanglement.
  (c) Random point defects, which could 
  generate a skyrmion glass.
  (d) 3D planar defects.  
 }
\label{fig:21}
\end{figure}

Most studies of pinning performed up until now have focused on
defects in 2D;
however, for   
3D line like skyrmions, entirely new types of pinning effects could arise along
with an array of new methods for creating 3D pinning.
In 3D superconducting vortex systems, it 
was shown that columnar pinning
enhances the critical depinning current
\cite{Civale97} by trapping
the vortex line along the entire length of the pinning site,
and a similar effect could occur for 3D skyrmions.
Splayed columnar defects \cite{Hwa93} could promote 
the entanglement of skyrmion lines, while proton irradiation could be used
to create
random point defects \cite{Haberkorn12}
or 3D line defects \cite{Kafri07}. 
In the schematic in Fig.~\ref{fig:21}
we show possible 3D pinning arrangements that could
be created in skyrmion
systems, including
columnar, splayed, 3D point-like, and 3D planar defects.
It would be interesting to learn whether 3D pinning is more
effective than 2D pinning or whether it can reduce skyrmion creep at
finite temperatures.
Some types of defects repel skyrmions rather than attracting them,
and adding
3D versions of such defects could increase the net
skyrmion mobility.
One possible experiment would be to irradiate bulk
samples and determine whether the depinning threshold changes
as measured by changes in the topological Hall effect.
If a sufficiently large density of 3D defects were added to the sample,
they could create percolation paths which could serve as easy flow channels
for skyrmion motion, leading to a net increase rather than
a net decrease in the
skyrmion mobility.

In 3D samples it would be possible
to place
one type of pinning on the top of the sample, such as through nanopatterning or
by adding adatoms,
while simultaneously placing a different type of pinning on the bottom of
the sample.
For example, if
antipinning sites are present on top of the sample
and pinning sites are present on the bottom,
a shear effect could arise under driving that would
promote skyrmion cutting 
or the creation of monopoles along the skyrmion lines.  

\section{Collective States and Skyrmion Lattices With Pinning}

\begin{figure}
\includegraphics[width=\columnwidth]{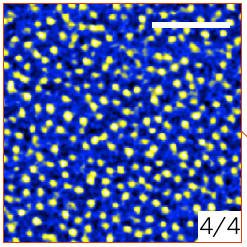}
\caption{Magnetic force microscope
  image of disordered skyrmions in Ir/Fe/Co/Pt multilayers
\cite{Soumyanarayanan17}.
Reprinted by permission from: Springer Nature,
``Tunable room-temperature magnetic skyrmions in Ir/Fe/Co/Pt multilayers,''
Nature Mater. {\bf 16}, 898 (2017),
A. Soumyanarayan {\it et al.}, \copyright 2017.
}
\label{fig:22}
\end{figure}

We next consider the effect of pinning on
the static configurations of collectively interacting skyrmions.
The very first experimental observation of skyrmions
was the imaging of a skyrmion lattice
with neutron scattering \cite{Muhlbauer09},
followed by direct visualization of the skyrmion lattice
with Lorentz microscopy \cite{Yu10}.
The fact that the skyrmions formed a lattice
suggests that in these initial experiments, the pinning was relatively weak;
however, it is
also possible that the thermal fluctuations were large enough to wash out the
effect of the pinning and create
what is known as a floating solid.
There are now many examples of skyrmion
systems, particularly in thin films,
that
form disordered states
\cite{Karube18,Hsu18,Zhang18b,Wang19}.   
Figure~\ref{fig:22}
shows an
image of disordered room temperature skyrmions in 
Ir/Fe/Co/Pt multilayers \cite{Soumyanarayanan17}.
The manner in which the system is prepared strongly impacts
whether the skyrmions form a lattice.
For example, consider a sample in which
the skyrmion ground state at temperature $T_1$ is disordered.
If the sample were prepared
at another temperature $T_2$ where the ground state is ordered and the
temperature was suddenly changed to $T_1$,
the skyrmions
could remain
in a metastable ordered lattice configuration.
The metastable state could be destroyed
by the application of a current
or drive 
which allows the skyrmions to reach
their disordered $T_1$ ground state configuration.

\begin{figure}
  \includegraphics[width=\columnwidth]{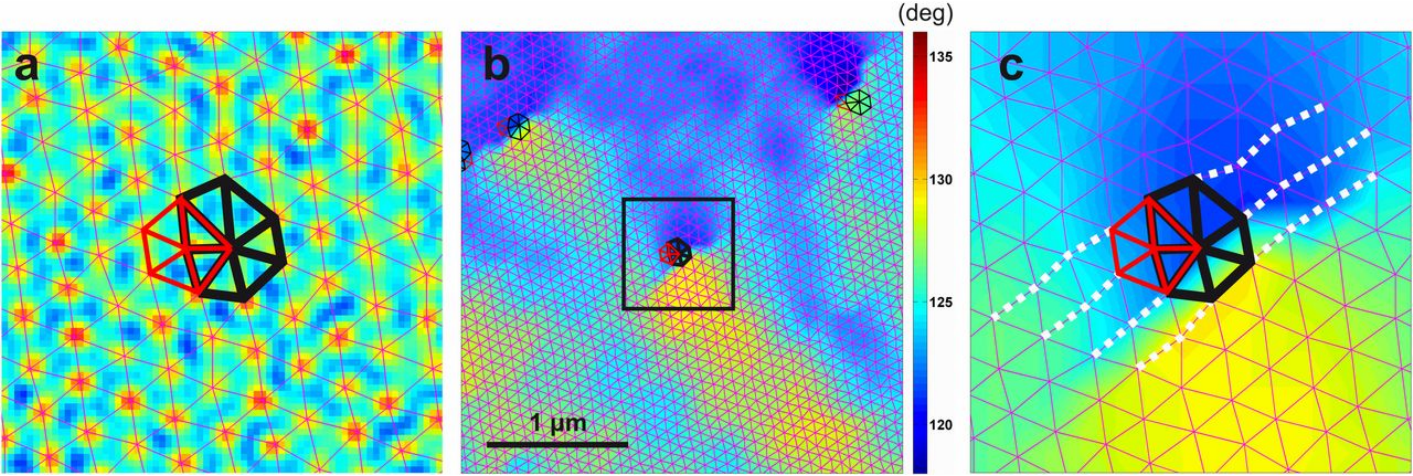}
  \caption{
Examples of Delaunay triangulations of skyrmion lattices \cite{Rajeswari15}.
(a) Image of a lattice defect consisting of sevenfold-coordinated (black)
and fivefold-coordinated (red) skyrmions adjacent to each other.
(b) A map of the local spatial angle superimposed on top of the Delaunay
triangulation with a defect at the center.
(c) A close-up view of the region marked with a square in panel (b),
showing the presence of a dislocation line at the domain boundary.
From J. Rajeswari {\it et al.}, Proc. Natl. Acad. Sci. (USA) {\bf 112}, 14212
(2015).
}
\label{fig:23}
\end{figure}

The structure of a skyrmion lattice can be
measured
using the structure factor,
\begin{equation}
S({\bf k}) =  \frac{1}{N}\left | \sum_{j=1}^N e^{-i{\bf k} \cdot {\bf R}_{j}}\right |^2
\end{equation}
where ${\bf R}_j$ is the position of skyrmion $j$ and $N$ is the total number
of skyrmions being sampled.
If the skyrmion forms a glass state, $S({\bf k})$ has a ring structure,
while if the skyrmions are in a triangular lattice,
$S({\bf k})$ has sixfold peaks.
The lattice structure can also be measured by using a
Voronoi or Delaunay construction to determine the fraction of
sixfold coordinated particles in the system,
as illustrated in Fig.~\ref{fig:23} \cite{Rajeswari15}.
These measures can also be
used to identify different types
of topological defects in the skyrmion lattice,
such as a skyrmion
with five neighbors that is connected to a skyrmion
with seven neighbors to form a dislocation pair, as shown 
in Fig.~\ref{fig:23}(c).
The dislocation pairs can glide or climb depending on the strength
of the driving. 
It is also possible that instead of completely disordering, the skyrmion lattice
can form domains defined by grain boundaries of a specific angle.
The grain boundaries are decorated by 5-7 dislocation pairs
and the spacing between the dislocations is determined
by the angular mismatch between the skyrmion lattices in the
adjacent grains \cite{Lavergne18}.

Disordered  skyrmion arrangements can be produced by
strong pinning, temperature, or polydispersity of the
skyrmion sizes or types.
In other systems containing random quenched disorder,
various types of equilibrium phases can arise.
For example, consider a 2D system with a triangular lattice ground state
or a 3D system of lines that form a
2D ordered triangular lattice in the absence of disorder.
Under increasing temperature, these systems generally
melt at a critical temperature $T_{c}$.
In the 3D system, the melting transition can be first or
second order, while in the 2D system it can be second order 
according to the
Kosterliz-Thouless-Halperin-Nelson-Young (KTHNY)
mechanism,
in which there is first
a proliferation of dislocations 
followed by the proliferation of free disclinations
\cite{Kosterlitz73,Young79,Nelson79,Strandburg88}. 
Evidence for 2D melting via intermediate hexatic phases
in the absence of a substrate
has been obtained
in numerous systems, including colloidal
assemblies \cite{Zahn99}
and it is even possible to
observe a first order transition into a hexatic 
phase \cite{Thorneywork17}.
The hexatic phase can be detected by measuring
the density-density correlation function
\begin{equation}
g_{G}(|{\bf r} - {\bf r^{\prime}}|) = \langle\exp(i{\bf G}\cdot[{\bf u}({\bf r}) - {\bf u}({\bf r}^{\prime})])\rangle
\end{equation}
and the bond-angular correlation function 
\begin{equation}
g_{6}(|{\bf r} - {\bf r}^{\prime}|) = \langle\exp(i6[\theta({\bf r}) - \theta({\bf r}^{\prime})])\rangle .
\end{equation}
Here ${\bf G}$ is the reciprocal lattice vector, ${\bf u}({\bf r})$ is the particle 
displacement field, and $\theta({\bf r})$ is the angle with respect to the $x$-axis. 
For a 2D crystal, $g_{6}({\bf r})$ is constant
and $g_{G}({\bf r})$ decays 
algebraically,
$g_{G}({\bf r}) \propto r^{-n(T)}$. In the hexatic phase,
$g_G({\bf r})$ decreases exponentially while
$g_{6}({\bf r})$ decays algebraically as
$g_6({\bf r}) \propto r^{-n_{6}(T)}$,
where $n_{6}$ approaches the value $1/4$. 
In the fluid phase, both correlation functions decay exponentially.
Several recent experiments have provided evidence for
a hexatic phase in skyrmion systems \cite{Huang20,Zazvorka20}.

\begin{figure}
\includegraphics[width=\columnwidth]{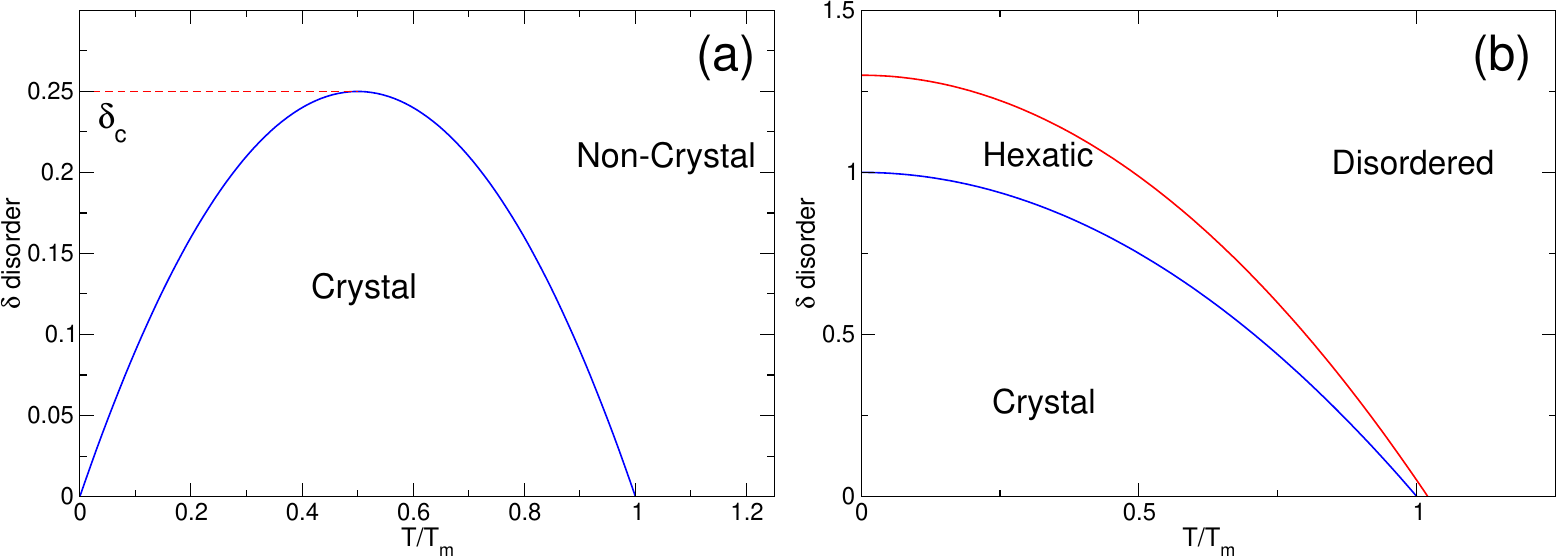}
\caption
{ (a) Schematic phase diagram as a function of quenched disorder
$\delta$ vs temperature $T/T_m$ for a 2D system, where $T_m$ is the
melting temperature.
The solid line indicates the transition
from a crystal to a disordered non-crystalline state 
predicted by Nelson \cite{Nelson89}.
The disordered state can be reentrant
when the temperature washes out the effect of the quenched disorder
before the crystal lattice melts.
The
dashed red line is
from the modified
phase digram proposed by Cha and Fertig \cite{Cha95}, where
the system is ordered at $T = 0$
and the low temperature disordered state does not appear until
a critical amount of disorder $\delta_{c}$
has been added
(b) A schematic phase diagram as a function of quenched disorder $\delta$
vs temperature  $T/T_m$ based on
2D colloidal experiments \cite{Deutschlander13},
where an intermediate hexatic phase
appears between the crystal and disordered phases. 
}
\label{fig:24}
\end{figure}

The more relevant situation for most skyrmion systems
is when disorder is present.
At $T = 0$ in a sample with random pinning, 
a lattice of interacting particles
takes advantage of the pinning energy $E_{p}$
at the cost of the elastic energy $E_{el}$.
If the pinning is weak, 
a small amount of elastic distortion can occur but the
triangular symmetry of the lattice is preserved.
When the disorder
is stronger,
the elasticity of the lattice can break down and
various types of topological defects can appear.
In 2D systems, a disordered KTHNY transition can occur
in which the system passes from a lattice to a hexatic phase,
while when the disorder is stronger,
a 2D glassy state appears. 
Nelson \cite{Nelson89} proposed the
phase diagram illustrated in Fig.~\ref{fig:24}(a)
as a function of disorder versus temperature.
In the absence of quenched disorder, the
lattice ordering begins
to disappear
at a finite value of $T$ when a transition occurs
into a hexatic state or
into a liquid.
When quenched disorder is present,
the lattice becomes disordered even for $T=0$;
however, the temperature can
wash out the effect of disorder,
producing a thermally induced transition from a disordered
non-crystalline state into a floating crystalline state
which then melts into a liquid at a higher temperature.
When the quenched disorder
is strong enough, the system is always in a disordered state.
Cha and Fertig \cite{Cha95} argued that at $T = 0$ the 
system remains in a crystalline state until
a critical amount of quenched disorder $\delta_{c}$  is added,
at which point the system disorders,
as indicated
by the horizontal dashed line in Fig.~\ref{fig:24}(a).
These different 
scenarios depend on the size scale of the quenched disorder,
since the thermal effects can only wash out the pinning
before the lattice melts
if the pinning sites are small.
Experiments in 2D colloidal systems \cite{Deutschlander13}
suggest the phase diagram shown in Fig.~\ref{fig:24}(b),
where an intermediate hexatic phase appears for zero quenched disorder and
increases in extent as quenched disorder is added to the sample.
In principle, a similar phase diagram could be constructed for 2D skyrmion systems in
which
the skyrmion size is roughly uniform.

\begin{figure}
\includegraphics[width=\columnwidth]{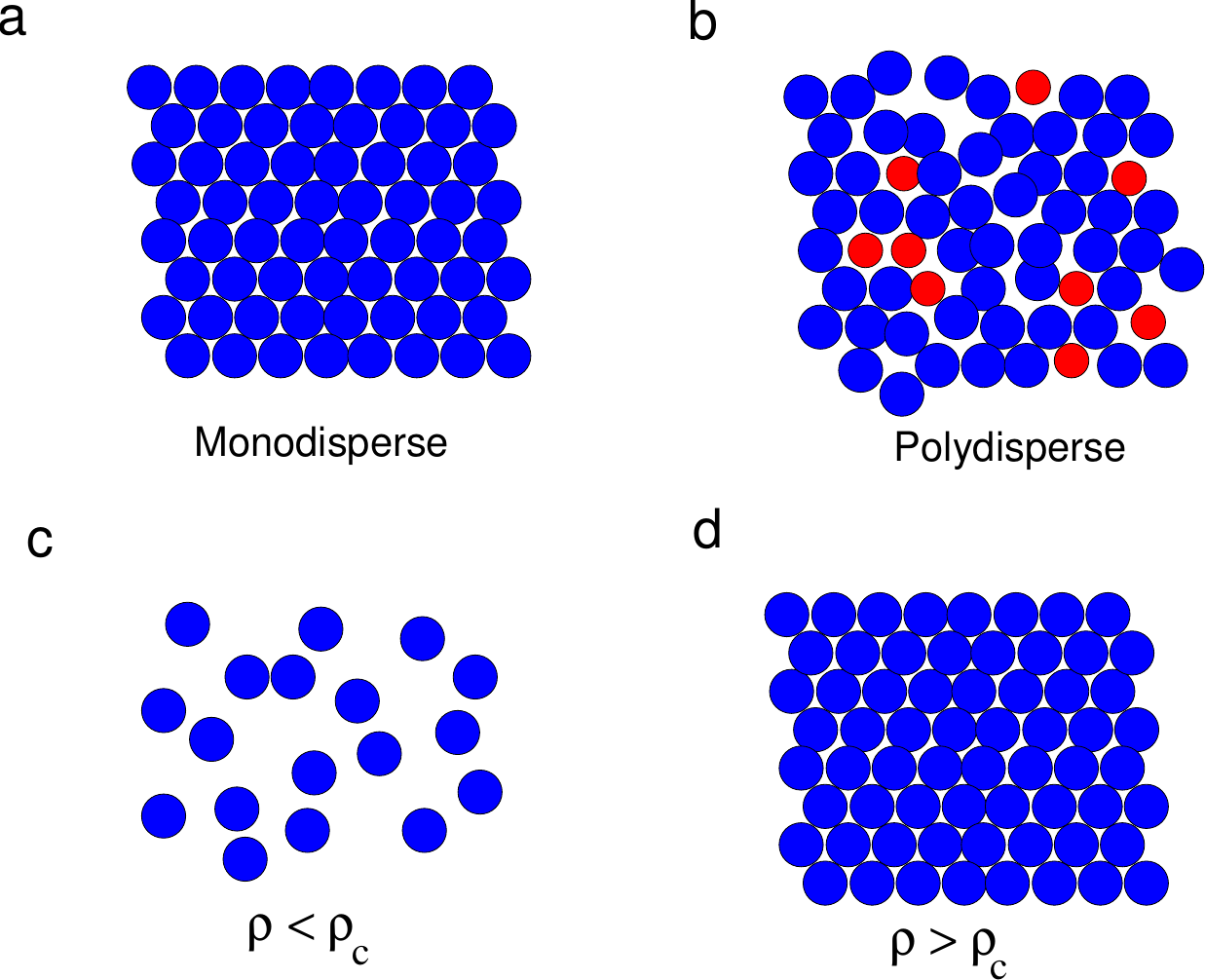}
\caption
{Schematic illustrations of
scenarios leading to disordered skyrmion structures without
quenched disorder or temperature. 
(a,b) Disordering induced by polydispersity in the skyrmion sizes.
(c,d) A jamming mechanism for skyrmions with short range contact forces.
(c) Below a critical density, $\rho < \rho_{c}$, the system is
liquidlike, while
(d) for $\rho > \rho_{c}$, the skyrmions are in contact and form a solid.    
}
\label{fig:25}
\end{figure}

Recent Monte Carlo simulations have shown
that a 2D skyrmion lattice
can melt without passing through
an intermediate hexatic phase \cite{Nishikawa19};
however, there are a wide
range of different 2D skyrmion systems,
and as suggested in Fig.~\ref{fig:24}, quenched disorder
could enhance the hexatic phase.
Skyrmions in 2D
are often already strongly disordered, but
in a dense regime
the skyrmion
interactions could become strong enough to favor
the formation of a
hexatic
phase.
In addition to quenched disorder, two other mechanisms could
be relevant in determining whether the
skyrmion arrangement is ordered or disordered.
In 2D assemblies of particles that form a triangular lattice in the absence of
quenched disorder,
there can be transitions to hexatic or disordered states
as a function of increasing particle polydispersity.
For example, polydispersity in the skyrmion sizes
could induce the formation of a hexatic state even when the quenched
disorder is weak.
Simulations of 2D Lennard-Jones systems
\cite{Sadrlahijany97} showed that, depending on the density of the sample,
a dispersity
in as few as 10\% of the particles
was sufficient to disorder the system.
In other work on bidisperse
Yukawa particles,
doubling the charge on 10\% to 20\%
of the particles
disordered the system \cite{Reichhardt08}.
Numerical evidence
by Zhang {\it et al.} for frustrated ferromagnetic films \cite{Zhang17} 
containing mixtures of skyrmions with different sizes
indicates that
polydispersity can produce disordered skyrmion states.
In Fig.~\ref{fig:25}(a,b) we schematically
illustrate
the disordering of
a monodisperse triangular solid
by the introduction of some dispersity in the particle size.
An open question
for skyrmion systems
is how much size dispersity is necessary to induce a transition
from a triangular solid to a disordered state.    

Another possible disordering mechanism is
an effective jamming transition.
In jamming, a system passes from
a fluid-like state where the particles
can move freely
to a solid state in which the particles are in contact and the assembly has
a finite response to a shear stress.
The jamming concept is typically used
to describe systems with short range or even hard sphere interactions,
such as grains and emulsions;
however, the interaction between larger skyrmions can be described as a
short range repulsion, making it possible for such skyrmions to
share features with emulsions.
In a hard disk jamming system,
there is a density or area coverage $\phi_J$, referred to as
``point J,'' at which
the disks first come into contact.
For a 50:50 mixture of 2D bidisperse hard disks with a radius
ratio of  $R_1/R_2=1.4$,
$\phi_J=0.84$ \cite{OHern03,Reichhardt14}.
At densities below point J,
the system is a disordered fluid,
while above point J,
a jammed amorphous solid appears. 
Monodisperse disks form a triangular solid
which is jammed but not disordered at
a critical density $\phi_{c} = 0.9$.
These results
suggest
that in skyrmion systems with very short range interactions,
even monodisperse skyrmions will
generally be disordered if
they are below the jamming or solidification density. 
The schematic in Fig.~\ref{fig:25}(c) illustrates particles with short range interactions
at a density well below the jamming or solidification density,
$\rho < \rho_{c}$,
while in Fig.~\ref{fig:25}(d),
the same particles, which could be skyrmions, at
$\rho > \rho_{c}$
are in contact and form a jammed crystalline solid. 
It is possible that for small skyrmions or for 3D skyrmions,
where a single skyrmion can interact with other skyrmions beyond nearest neighbor
interactions,
the skyrmions will form a lattice,
while for larger skyrmions or for 2D skyrmions where the interaction range
is short and reaches only nearest neighbors,
the system
can be described as exhibiting a jamming
transition to a disordered state.  

The density of skyrmions is nonmonotonic as a function of
magnetic field.  As a result, at intermediate fields
where the skyrmion density is high,
the skyrmions may form a triangular solid;
however, when the skyrmion density decreases for higher or lower fields,
the spacing between skyrmions could become large enough that the skyrmions
no longer interact, causing the system to
transition into a disordered state outside some critical window of
magnetic fields.
The skyrmions could exhibit
two glassy states associated with
the lower field low density limit, an intermediate field triangular
lattice, and a higher field disordered state. 

For 3D systems containing quenched disorder,
such as superconducting vortex lines,
a Bragg glass can form in which the system
has hexagonal order but also exhibits glassy features \cite{Giamarchi95,Klein01}.
If skyrmions in a bulk 3D sample 
form a Bragg glass, it could be detected through
measurements of the
in-plane correlation function $g({\bf r})$
or by finding 
a power law divergence of the Bragg peaks in a scattering measurement
\cite{Giamarchi95}.
In analogy to the transitions observed in superconducting vortex
systems,
3D skyrmions could also undergo a first order transition
from a Bragg glass to a liquid state or 
to a different type of more disordered glass.

When columnar disorder is present,
3D superconducting vortices
can form a disordered 
Bose glass, which suggests that skyrmions in linelike disorder
could form a skyrmion Bose glass.
Strong disorder in a 3D skyrmion system could also produce different
types of glasses
such as an entangled state in which the skyrmion lines wrap around each other.
These skyrmion glasses could have very
different properties than the vortex glasses since the skyrmions
can in principle break or merge to form monopole states.
In superconducting systems, glassy states can be detected through
magnetization or voltage measurements, while for skyrmion
systems, possible measurements that could reveal glassy features include
magnetization, slow changes in the topological Hall effect, or changes
in the structure factor $S(k)$ as a function of time.
The exploration of glassy states is an almost completely open field in skyrmions.

In samples with
intermediate disorder,
there are only a few strong pinning sites,
so the system may exhibit a polydisperse state in which
local ordering coexists with
a series of grain boundaries \cite{Moretti09}.
Conversely,
there could be locally disordered regions
coexisting with long range order.
There are numerous experimental
observations of domains and grain boundaries in skyrmion
lattices \cite{Rajeswari15,Matsumoto16,Matsumoto16a,Zhang16,Nakajima17a,Li17}.
In such systems the depinning could involve either
the motion of grain boundaries or the rotation of grains,
and the depinning dynamics could be very different from those of
purely ordered skyrmion lattices or completely disordered skyrmion states.
Since skyrmions can change shape, the grain boundary formation process
in a skyrmion lattice differs from that in a colloidal lattice or an atomic system,
and certain topological defects may be less costly in a skyrmion lattice than
in a system of rigid particles \cite{Matsumoto16}.

\begin{figure}
\includegraphics[width=\columnwidth]{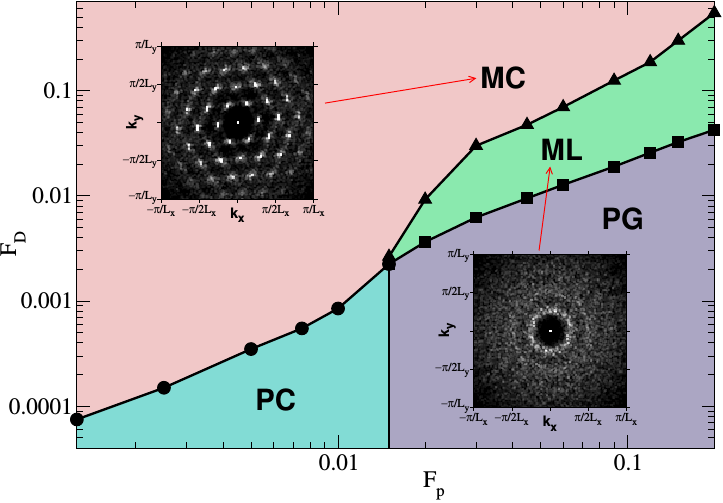}
\caption
{
Particle-based numerical simulations of a transition from a
pinned crystal (PC) to a pinned glass (PG) as a function of increasing
quenched disorder strength $F_p$ \cite{Reichhardt15a}.
When a driving force $F_D$ is applied,
transitions to a moving crystal (MC) or moving liquid (ML) state can
occur.
Upper inset: Peaks associated with
crystalline ordering appear in the structure factor $S({\bf k})$.
Lower inset: Rings associated with a liquid state appear in $S({\bf k})$.
Reprinted with permission from C. Reichhardt {\it et al.}, Phys. Rev. Lett.
{\bf 114}, 217202 (2015). Copyright 2015 by the American Physical Society.
}
\label{fig:26}
\end{figure}

The approach to a skyrmion glass at zero drive
was observed in particle based simulations by
Reichhardt {\it et al.} \cite{Reichhardt15a} for 
skyrmions interacting with attractive point pins.
At $T=0$ and low drives,
the skyrmions form a defect-free
lattice with six-fold 
peaks in $S({\bf k})$
and exhibit a finite elastic depinning threshold
when the quenched disorder is weak.
For stronger quenched disorder,
the skyrmions form a disordered
skyrmion glass as indicated by the ringlike structure of $S({\bf k})$,
illustrated in the lower inset of Fig.~\ref{fig:26}.
Although
the simulations in Fig.~\ref{fig:26} show
a transition from an ordered to a disordered state at $T = 0$
as a function of increasing quenched disorder strength,
in agreement with
the predictions of Cha and Fertig \cite{Cha95},
it was not determined whether or not
the transition from the pinned skyrmion crystal to
the pinned skyrmion glass was of KTHNY type.

\begin{figure}
\includegraphics[width=0.7\columnwidth]{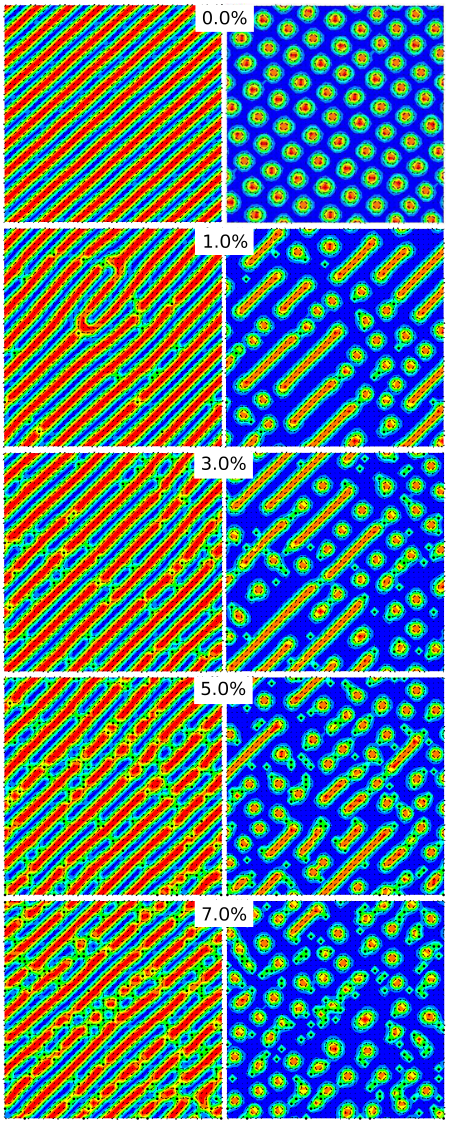}
\caption
{Images from Monte Carlo simulations \cite{Silva14} of
spiral (left column) and skyrmion (right column) states with increasing
magnetic spin vacancy densities $\rho$.
At $\rho=0$,
an ordered
spiral or triangular skyrmion lattice state forms.
As the density of magnetic spin vacancies increases,
skyrmions nucleate in the spiral state,
the skyrmion lattice
becomes disordered,
and bimerons appear.
Reprinted with permission from R. L. Silva {\it et al.}, Phys. Rev. B
{\bf 89}, 054434 (2014). Copyright 2014 by the American Physical Society.
}
\label{fig:27}
\end{figure}

Silva {\it et al.} \cite{Silva14} performed
Monte Carlo simulations of skyrmion formation
in the presence of 
pointlike nonmagnetic defects, and
found that a very small number of defects
could produce a disordered skyrmion structure.
They also observed the emergence of bimerons
for an increasing
density of
spin vacancies in both the spiral and
the skyrmion state,
as shown in Fig.~\ref{fig:27}.
Silva {\it et al.}
showed that inclusion of even 1\% of spin 
vacancies can strongly disorder the system, but did not identify whether
there is a critical amount of  disorder
that must be added to induce a transition
from a skyrmion crystal to a disordered state \cite{Silva14}. 
Hoshino and Nagaosa \cite{Hoshino18}
theoretically studied a collective skyrmion glass phase using various
methods such as replica theory
from the glass literature,
similar to
what has been employed for other pinned
systems \cite{Giamarchi95}.
They found
a number of scaling relations for the critical current and pinning frequencies,
along with the
key result that these quantities change sharply across the
transition from the helical state to
the skyrmion state. 
There have also been several studies demonstrating that quenched disorder
can generate skyrmions \cite{Mirebeau18,Chudnovsky18}.   

Transitions have been observed from square meron to hexagonal meron
to hexagonal skyrmion states
\cite{Yu18}.
Changes in the elastic constants can occur at these crossovers,
and if the elastic constants
drop below a certain level,
the system can disorder near the square to hexagonal transition.
It may also be possible to generate metastable
glassy skyrmion states by quenching rapidly from a
higher temperature liquid state to a lower temperature at which the equilibrium
state is an ordered solid.
In the presence of pinning, the resulting metastable supercooled liquid or glassy state
could be long lived.
Metastable disordered states
can be distinguished from equilibrium disordered states by
applying various
perturbations such as a changing magnetic field.
Experiments have shown that even in systems with large
intrinsic disorder,
an ordered skyrmion
lattice can be
produced by the judicious selection of field
application protocols
\cite{Gilbert19}.  

\subsection{Future Directions}

\begin{figure}
\includegraphics[width=\columnwidth]{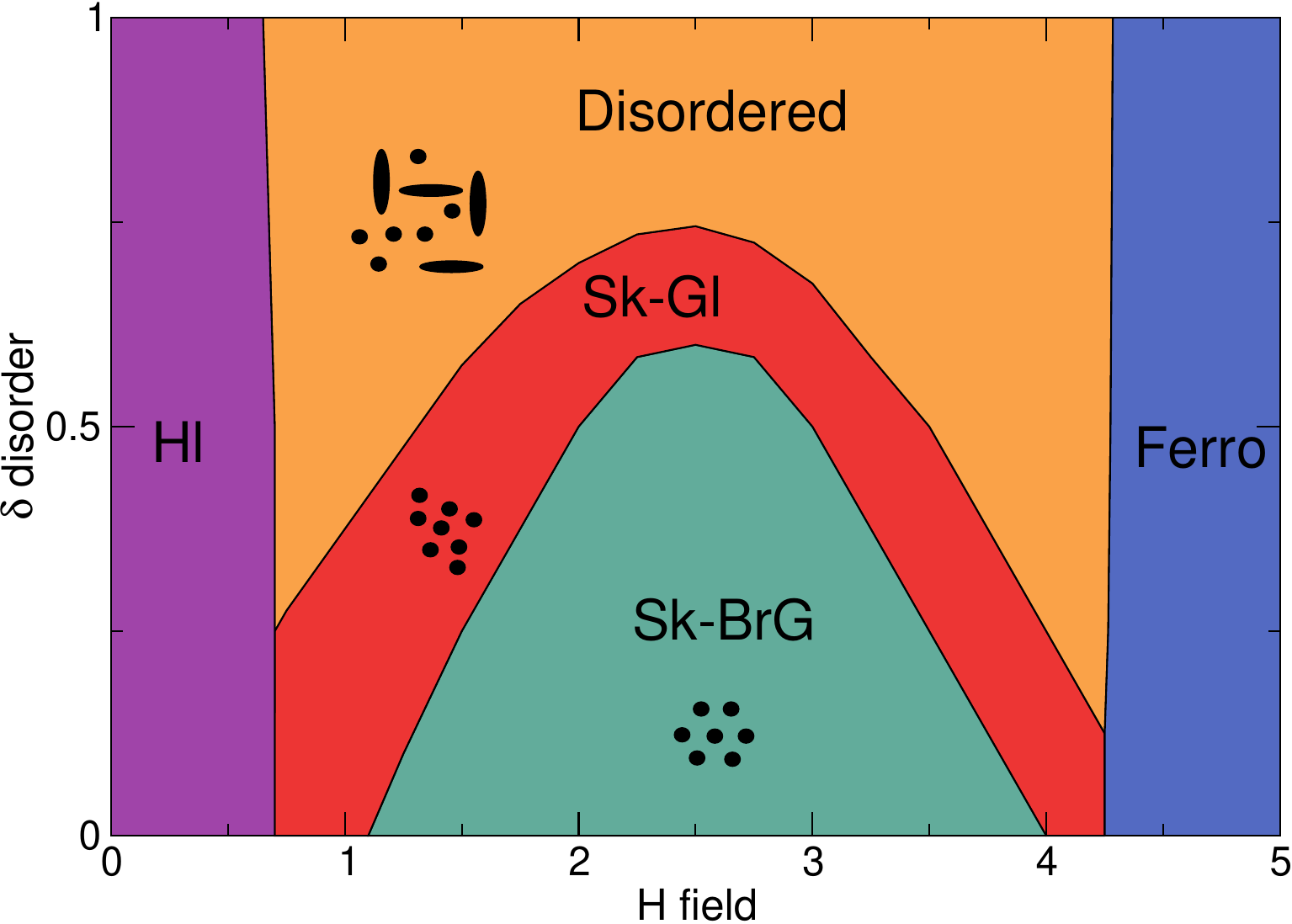}
\caption
 {Schematic of a possible phase diagram as a function of disorder
 strength $\delta$ versus magnetic field $H$ for a skyrmion 
 system.
 At low fields the system is in a helical state (Hl).
 At low disorder strength
 and high skyrmion density the system forms a skyrmion Bragg-Glass (Sk-BrG),
 and at low skyrmion densities and intermediate disorder,
 a skyrmion glass (Sk-Gl) appears.
 For strong disorder, the system forms a mixed skyrmion-meron state
 with skyrmion breaking (Disordered).
 At the highest fields, the system forms a ferromagnetic state (Ferro). 
}
\label{fig:28}
\end{figure}

There is a wide range of open issues for collective
skyrmion states with disorder,
such as
whether there are different types of glassy
states. These could include analogs to the
vortex glass with point pinning in type-II superconductors, 
a Bose glass, a splay glass, or even entirely new types of glassy phases
that
have not been observed previously.
For example,
for weak quenched disorder the skyrmions could form a Bragg glass,
while for stronger quenched disorder they could
form a skyrmion glass phase that is similar to
a vortex glass.
At even stronger disorder, the skyrmion lines
could break up to create something like a
monopole glass or a skyrmion-bimeron glass.
In Fig.~\ref{fig:28}
we show a schematic of a possible phase diagram as a function of disorder
strength versus magnetic field
for a skyrmion system.
At intermediate fields, where the skyrmion density is the highest, the
system could form a skyrmion Bragg glass,
while for larger disorder the skyrmions
could positionally disorder to form a skyrmion glass.
At the highest disorder strengths,
the skyrmion lines could break up into a disordered
configuration of coexisting skyrmions and bimerons.
Each of these states could show unique responses
to a drive, ac perturbations, retardation effects, or creep.
It is an open question whether a 
full phase diagram for static skyrmion states
can be measured
as a function of quenched disorder, field, and temperature.
Such a phase diagram could contain
skyrmion lattice, skyrmion glass, and 
skyrmion liquid states
similar to the vortex phase diagram
observed for type-II superconductors \cite{Crabtree97}. 
Another question is whether a 2D or 3D skyrmion liquid phase differs from a 2D
or 3D
skyrmion glass phase.
Since many materials now support skyrmions
at room temperature,
it is likely that systems will be identified in which
the thermal fluctuations are
strong enough to create a skyrmion liquid with diffusing skyrmions.
Already there is 
evidence for skyrmion thermal motion \cite{Zazvorka18,Nozaki19,Zhao19}
and liquid phases \cite{Chai18}.
The nature of the skyrmion liquid phase
could depend strongly
on quenched disorder and pinning effects.

Differences between a pinned liquid and
a pinned glass
are detectable in correlation functions
such as the density fluctuations or
$S({\bf k})$.
The same measures can be used to detect
the presence or absence 
of what is called
disordered hyperuniformity,
in which, unlike the behavior of a completely random system,
large scale density fluctuations are
suppressed
\cite{Torquato16}.
Hyperuniformity
can be used to distinguish jammed states from liquid states
\cite{Dreyfus15}, 
and it  has been observed in
simulations of interacting particles
with pinning \cite{LeThien17}.  
The key signature of hyperuniformity
is the behavior of the structure factor $S({\bf k})$ in the limit 
$|{\bf k}| \rightarrow 0$, given by
\begin{equation}
S({\bf k}) \propto |{\bf k}|^\alpha .
\end{equation} 
An exponent $\alpha > 0$ is indicative of hyperuniformity,
while in a random configuration, $S({\bf k})$ approaches a constant
value at small ${\bf k}$.
There are different
hyperuniform scaling
regimes with $\alpha > 1$, $\alpha = 1$, and $0 < \alpha < 1$.
In general, larger values of $\alpha$ indicate larger amounts of short range order. 
Hyperuniformity can also be characterized by
measuring the number variance 
$\sigma^2(R) = \langle N^2(R)\rangle - \langle N(R)\rangle^2$,
where $N(R)$ is the number of particles in a region of radius $R$.
For a random system, $\sigma^2(R) \propto R^2$,
while for
$d$-dimensional hyperuniform systems, $\sigma^2(R) \propto R^{d -\alpha}$ when
$\alpha < 1$ and $\sigma^2(R) \propto R^{d - 1}$ when $\alpha > 1$ \cite{Torquato16}.
Skyrmion assemblies are an ideal system
in which to test some of the hyperuniformity concepts
since skyrmions can easily be imaged over large scales.

It is an open question
how all of the disordered phases described above
would change for different species of skyrmions
such as an antiskyrmion lattice,
antiferromagnetic skyrmions,
or a 3D hedgehog lattice. 
Each variety of skyrmion
could exhibit different collective interactions
in the presence of disorder.

\section{Depinning Dynamics of Skyrmions with Pinning}

In this section we 
consider the dynamics of skyrmions in the presence of pinning.     
Skyrmions can be driven by various methods depending on
whether the host system is
a metal or an insulator.
In the case of a metallic system,
driving can be achieved through the
application of a current
by means of the spin torque effect
\cite{Nagaosa13,Woo16,Schulz12,Liang15,Yu12,Legrand17,Tolley18,Iwasaki13,Iwasaki13a}.
Other driving methods include
thermal gradients \cite{Kong13,Mochizuki14,Pollath17,Lin14,Kovalev14},
electric fields \cite{White14}, spin currents \cite{Shen18},
magnons \cite{Psaroudaki18}, magnetic field
gradients \cite{Zhang18a,Shen19}, and acoustic waves \cite{Nepal18}.
One of the first studies of skyrmion dynamics was performed by
Zang {\it et al.} \cite{Zang11}, who showed that
the skyrmion trajectories
are deflected from the direction of
the applied current
and generate a skyrmion Hall effect, which can be very large.
They also addressed
the effect of pinning and
identified
a weak pinning or
collective pinning regime along with a strong pinning regime.
Direct imaging of
skyrmion
dynamics has been achieved with a variety of experimental
techniques including Lorentz imaging, as will be discussed in more detail
below.

Skyrmions
produce the
topological Hall effect \cite{Nagaosa13,Neubauer09,Raju19}
which combines additively with the
other Hall effect terms
to give a measured resistivity of
\begin{equation}
\rho_{xy}(H) = R_{0}H + R_{s}M(H) + \rho_{TH}(H).
\end{equation}
Here $R_{0}H$ is the ordinary Hall effect and
$R_{S}M(H)$ is the anomalous Hall effect,
while $\rho_{TH}$ is the topological Hall effect, which is typically obtained
by accurately accounting for the contribution of the first two terms and
subtracting them from $\rho_{xy}$.
The topological Hall effect is linked to the skyrmion density according to
$\rho_{TH} = PR_{0}n_{T}\Phi_{0}$, where
$P$ is the density of mobile charges,
$R_{0}$ is an unknown Hall resistivity from the effective 
charge density
which is often taken to
be equal to the ordinary Hall coefficient, 
$n_{T}$ is the density of the total topological charge
from the skyrmions,
and $\Phi_{0} = h/e$ is the elementary flux quantum. 
According to this relation,
$\rho_{TH}$ is directly proportional to the number of skyrmions
in the sample \cite{Nagaosa13,Raju19}.

\begin{figure}
\includegraphics[width=\columnwidth]{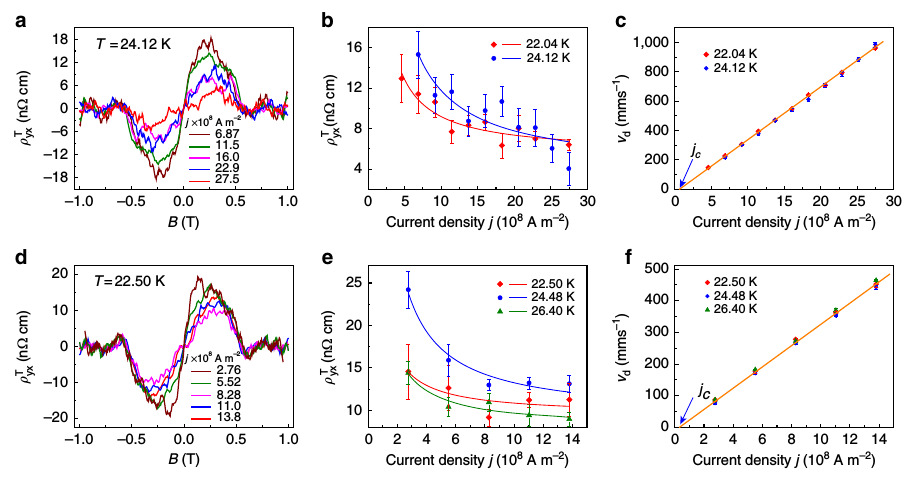}
\caption
{
Construction of skyrmion velocity-current curves based on
measurements of the topological Hall effect \cite{Liang15}.
Shown are transport results from two different devices,
one smaller (upper row) and one larger (lower row).
(a, d) The topological Hall effect $\rho^{T}_{xy}$ 
vs magnetic field $B$
measured at a range of applied current densities $j$.
(b, e) The average value of $\rho^T_{xy}$
over the range $B=0.2$ to 0.4 T vs current density $j$
at different temperatures.
(c, f) The estimated skyrmion drift velocity $v_d$ vs current density $j$.
Reprinted under CC license from D. Liang {\it et al.}, Nature Commun.
{\bf 6}, 8217 (2015).
}
\label{fig:30}
\end{figure}

Schulz {\it et al.} \cite{Schulz12} were able to construct
a skyrmion velocity-force curve based on changes 
observed in the topological Hall effect.
They argue that for constant $H$,
$\rho_{TH}$ remains
constant when the skyrmions are stationary
at zero current, $j = 0$, but that
when
the skyrmions begin to move under an applied current,
$\rho_{TH}$ decreases.
By measuring variations of $\rho_{xy}$ in the skyrmion phase as a function of
$j$,
they observed a drop at a specific value
of $j$ which was argued to correspond to the critical
depinning threshold, and
they were able to construct
an effective velocity-force curve.
In Fig.~\ref{fig:30},
a similar approach was
used to construct a skyrmion
velocity versus current curve for
MnSi nanowires of different sizes \cite{Liang15}.
The topological Hall effect $\rho^{T}_{xy}$,
which has a value different from zero only inside the
skyrmion phase,
is plotted versus $B$ for different 
currents in
Figure~\ref{fig:30}(a).
There is a decrease in
the average value of $\rho^T_{xy}$ with increasing $j$, as shown
in Fig.~\ref{fig:30}(b).
From this data, it is possible to extract an
estimated
skyrmion velocity $v_{d}$,
which is plotted versus $j$ in Fig.~\ref{fig:30}(c).
The value of the critical current
$j_{c}$ can be obtained from a linear fit of this curve.
Figure~\ref{fig:30}(d,e,f) shows that similar trends appear
in a larger device.
This work established
that $\rho^{T}_{xy} \propto 1/j$,
implying
a linear increase of the skyrmion velocity with drive
for drives well above $j_{c}$.
There is a nonlinear dependence of $v$ on $j$ near the threshold
current $j_c$.
When the depinning is elastic,
this nonlinear region
extends only as high as
currents below $1.1j_c$.
In contrast, for plastic depinning the
nonlinear regime can extend out to many multiples of $j_{c}$.

In principle, changes in the topological Hall effect as a function of current
could be measured carefully
as a function of drive, temperature, and magnetic field
in order to map
out the exact
behavior of $j_{c}$.
For example, a transition from elastic to plastic depinning
could be accompanied by a large increase
in $j_{c}$, similar to the peak effect phenomenon observed
at the transition from elastic to plastic
depinning in superconducting vortex systems \cite{Reichhardt17}.
Obtaining high precision measurements of
$\rho^{T}_{xy}$ down to the single skyrmion
level can be very difficult since all other
Hall contributions must be
carefully accounted for \cite{Zeissler18,Maccariello18},
so currently there are only a few studies that 
use changes in
$\rho^{T}_{xy}$ to deduce the critical depinning threshold \cite{Schulz12,Liang15}.
Other studies
in systems known to support skyrmions
have shown the presence of a topological Hall effect
that does not change with current \cite{Leroux18}.
Several issues
that can
complicate the picture include
sign changes
of the topological Hall effect or the existence of non-skyrmionic
topological Hall effect sources
\cite{Maccariello18,Denisov17,Denisov18}.
Recent experiments in which
$\rho^{T}_{xy}$ was measured simultaneously with the
number of skyrmions
confirm that $\rho^T_{xy}$
increases as the number of skyrmions increases;
however, there is not exact quantitative agreement with the
theory,
and the value of $\rho^{T}_{xy}$
is actually higher than would be expected from the
number of counted skyrmions \cite{Raju19}.  

\begin{figure}
\begin{minipage}{0.5\columnwidth}
  \includegraphics[width=\columnwidth]{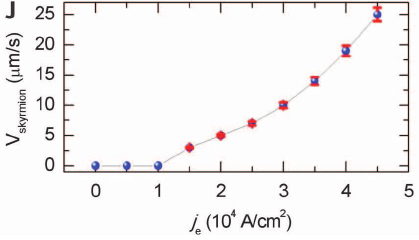}
\end{minipage}%
\begin{minipage}{0.5\columnwidth}
  \includegraphics[width=\columnwidth]{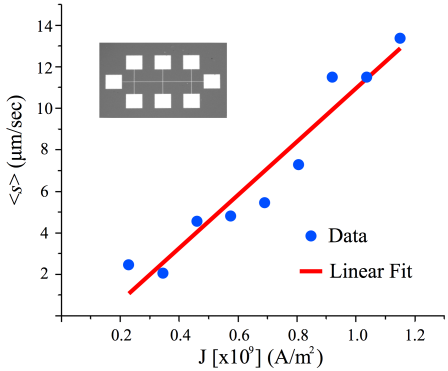}
 \end{minipage}
\caption
    { Left panel: The skyrmion velocity $V_{\rm skyrmion}$ vs current $j_e$
      for room temperature skyrmions deduced from direct imaging
      \cite{Jiang15}.
  From W. Jiang {\it et al.}, Science {\bf 349}, 283 (2015). Reprinted with
  permission from AAAS.
  Right panel: The skyrmion velocity $\langle s\rangle$ obtained from direct
  imaging
  versus current $J$ for Pt/Co/Os/Pt thin films
showing a linear fit \cite{Tolley18}.
Reprinted with permission from R. Tolley {\it et al.}, Phys. Rev. Mater.
{\bf 2}, 044404 (2018). Copyright 2018 by the American Physical Society.
}
\label{fig:31}
\end{figure}

Until now, the most common method
for generating skyrmion velocity-force or velocity-current
curves and obtaining the depinning threshold has 
been the use of direct imaging \cite{Jiang15,Woo16,Yu12,Jiang17,Litzius17,Woo18,Tolley18}.
An example of results obtained with this technique appears
in the left panel of Fig.~\ref{fig:31}
for room temperature skyrmions
with a critical depinning current near $j = 10^4$ A/cm$^2$ \cite{Jiang15}. 
The right panel of Fig.~\ref{fig:31}
shows the skyrmion velocity versus current
in room temperature Pt/Co/Os/Pt thin films
obtained from images taken with magneto-optic Kerr effect
(MOKE) microscopy \cite{Tolley18}.  
The
disadvantage of these
methods is the amount of time required for
imaging the skyrmion.
In many cases, the images
are obtained after applying a current pulse
rather than under a continuous current,
and the velocities must be deduced based on the skyrmion
displacements rather
than through a direct visualization of the skyrmion motion,
making it difficult to access
high frequency dynamics
or effects such as hysteresis
that can appear under a continuous current sweep.

\subsection{Elastic and Plastic Depinning}

\begin{figure}
\includegraphics[width=\columnwidth]{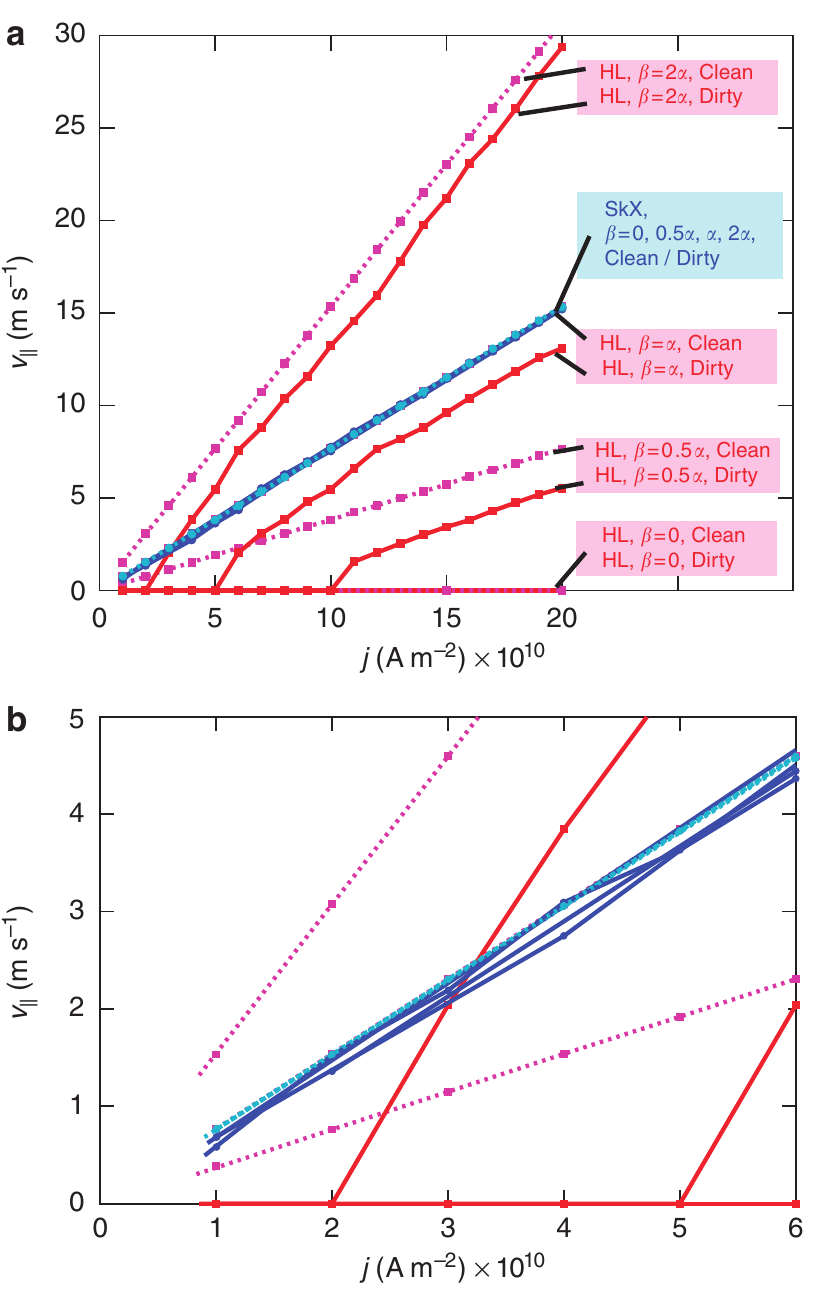}
\caption
{
Micromagnetic simulation measurements of the
current-induced longitudinal velocities $v_{||}$
of the helical (HL) and skyrmion crystal (SkX) phases vs
current density $j$ in both the clean (impurity-free) and dirty
limits for different 
values of the nonadiabatic term $\beta$
\cite{Iwasaki13}.
Blue lines: skyrmion phases; red and magenta lines: 
helical phases.
The skyrmions are much more weakly pinned than the helical
phases and show an elastic depinning transition.
(b) Magnification of panel (a) in the region of low current density.
Reprinted by permission from: Springer Nature,
``Universal current-velocity relation of skyrmion motion in chiral magnets,''
Nature Commun. {\bf 4}, 1463 (2013),
J. Iwasaki {\it et al.}, \copyright 2013.
}
\label{fig:32}
\end{figure}

Iwasaki {\it et al.} \cite{Iwasaki13} performed micromagnetic simulations 
of driven skyrmions interacting
with weak pinning and found that
the skyrmions form
a triangular lattice in
both the pinned and moving states.
Here
the pinning sites were much smaller than the skyrmion radius.
The depinning threshold is zero
in the absence of defects;
however, when pinning
is present, Iwasaki {\it et al.} observed elastic depinning in which
each skyrmion maintains its same neighbors over time.
They also found that as the ratio of the nonadiabatic
portion of the interaction decreases,
the critical current threshold $j_c$ increases.
The simulations revealed that
the skyrmions could not only move around the defects due to the Magnus force
but could also change shape.
Figure~\ref{fig:32} shows the
longitudinal skyrmion velocity $v_{||}$ versus
current $j$
from simulations for skyrmion
and helical phases
with and without disorder.
The helical phases are strongly pinned when disorder is present,
but
the skyrmion phases
are weakly pinned. 
Since the skyrmions form a triangular lattice,
Iwasaki {\it et al.}
also analyzed the Bragg peaks and
found that 
at lower drives the peaks were
somewhat weaker and showed strong fluctuations, while
at higher drives, the fluctuations were less pronounced and the Bragg peaks 
approached their pinning-free heights.
This result is similar to
the dynamical ordering found in superconducting
vortex systems \cite{Koshelev94,Olson98a}.
Although no
dislocations are generated at depinning,
the system
interacts more strongly with the pinning
at low drives and becomes less ordered.
Iwasaki {\it et al.} \cite{Iwasaki13}
argue that the particle-based Thiele equation approach can
be applied to understand both the depinning and
the skyrmion dynamics responsible for the behavior of the velocity-force curves.

In the simulations of Iwasaki {\it et al.} \cite{Iwasaki13},
the velocity-current curves were linear with
$v_{||} \propto F_{D}$; however,
the micromagnetic simulations could not
resolve the depinning threshold $F_c$ in the skyrmion regime.
Reichhardt and Reichhardt
examined a 2D particle based model for skyrmions interacting
with disordered pinning substrates of varied strength
\cite{Reichhardt15a,Reichhardt19a}, and
found that the velocity force curves are
consistent
with
$v \propto (F_{D} - F_{c})^\beta$ with $\beta < 1.0$.
For elastic depinning, $\beta < 1.0$, while for plastic depinning,
$\beta > 1.0$  \cite{Fisher98,Reichhardt17}; 
however, detailed finite size scaling
has not been performed in the skyrmion simulations in order
to confirm the exact values of the exponents
for either the elastic or plastic depinning case.
It is possible that the
presence of the Magnus force could change the
scaling compared to that
found in overdamped systems.
As for the behavior of the structure factor,
Reichhardt and Reichhardt 
 \cite{Reichhardt19a} examined the
magnitude $S(k_0)$ of one
of the six Bragg peaks
as a function of driving force $F_D$.
Although the skyrmions retain
their sixfold ordering for all drives,
there is a dip
in $S(k_{0})$ just at the depinning threshold,
indicating that during depinning,
the lattice becomes
more disordered or roughened,
as also observed by Iwasaki {\it et al.} \cite{Iwasaki13}.

\begin{figure}
\includegraphics[width=\columnwidth]{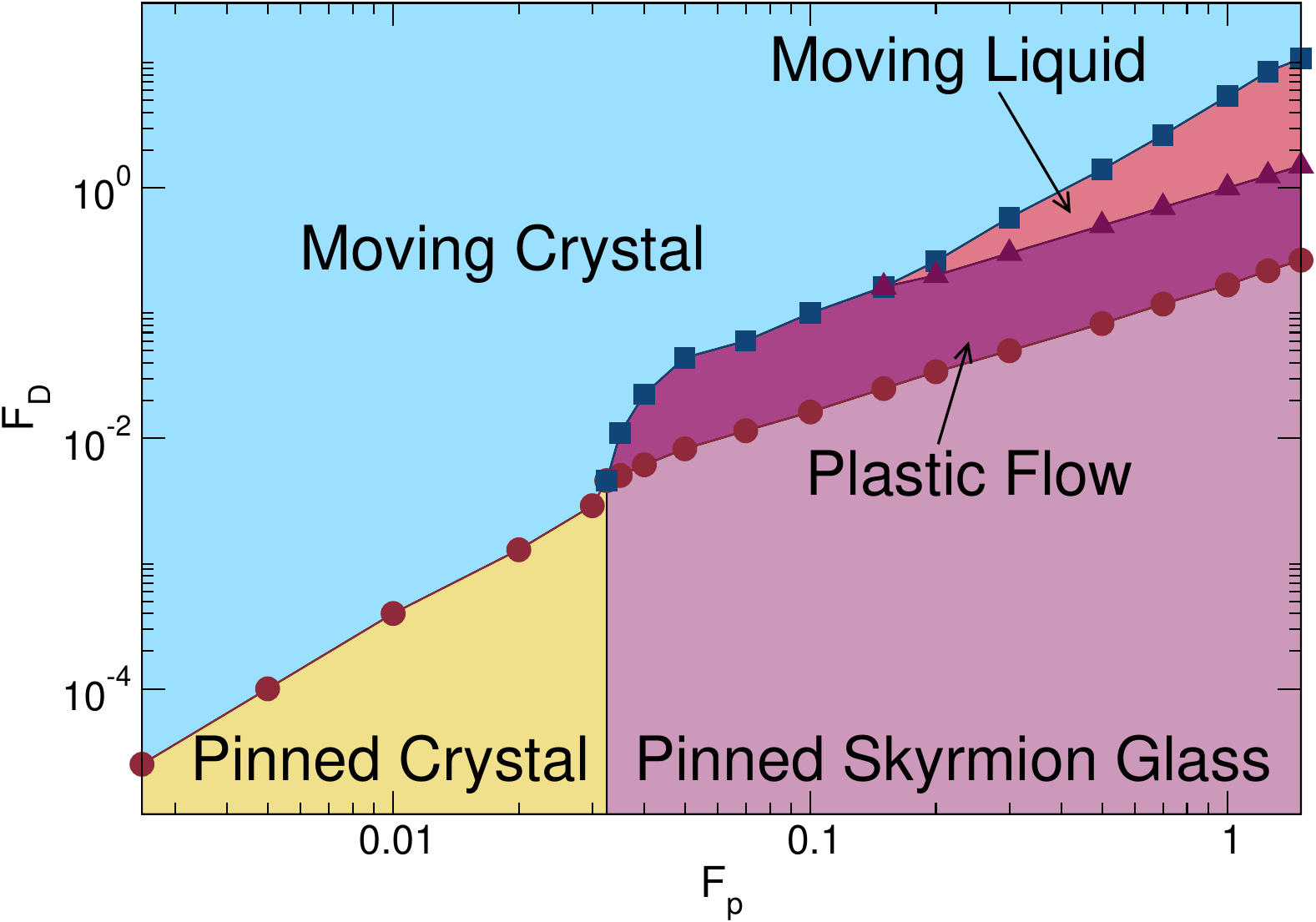}
\caption
    {
Particle-based simulations of a      
  dynamic phase diagram as a function of driving force $F_D$ vs
  pinning strength $F_p$
  highlighting the transition
  from a pinned crystal to a pinned glass
  \cite{Reichhardt19a}. 
  The pinned crystal depins elastically into a moving crystal phase,
  and the pinned skyrmion glass phase 
  depins plastically into a plastic flow regime which transitions into
  the moving liquid phase.
  At high drives the system forms a moving crystal. 
The pinned to moving crystal line behaves as $F_{c} \propto F^{2}_{p}$, while
the pinned glass to plastic flow line behaves as $F_{c} \propto F_{p}$.
Reprinted with permission from C. Reichhardt {\it et al.}, Phys. Rev. B
{\bf 99}, 104418 (2019). Copyright 2019 by the American Physical Society.
}
\label{fig:34}
\end{figure}

\begin{figure}
\includegraphics[width=\columnwidth]{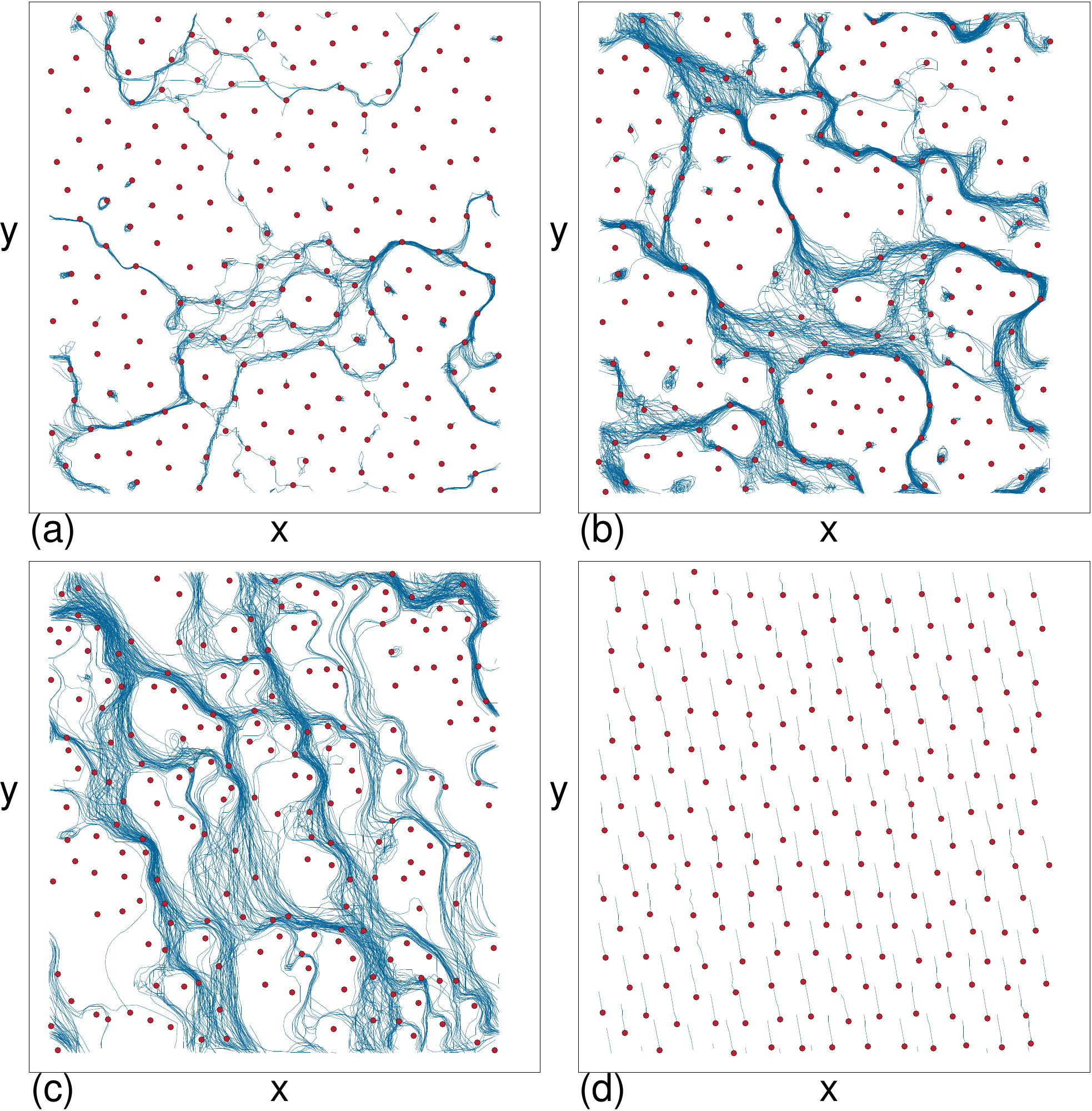}
\caption
    { The plastic flow phase just above depinning from particle-based
      simulations of
      skyrmions in strong random pinning \cite{Reichhardt16}.
      Skyrmion positions (dots) and trajectories (lines)
      are obtained over a fixed time period, and 
      the drive is in the positive $x$-direction.
      (a) At a drive close to depinning, channels of flow coexist
      with pinned skyrmions.
      (b) As the drive increases, the number of pinned skyrmions decreases.
      (c) For higher drives, plastic flow continues to persist and the
      direction of motion has rotated away from the driving direction.
      (d) Trajectories obtained over a shorter time period at
      a high drive where
      the skyrmions are dynamically ordered and move at an angle
      of $-79.8^\circ $ to the drive.
Reprinted under CC license from C. Reichhardt and C. J. O. Reichhardt,
New J. Phys. {\bf 18}, 095005 (2016).
}
\label{fig:35}
\end{figure}

\begin{figure}
\includegraphics[width=\columnwidth]{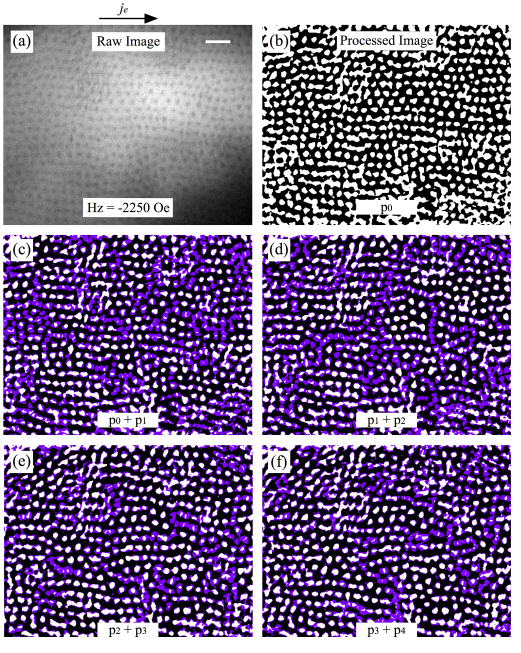}
\caption
    {
Images showing current-induced plastic motion of dipole skyrmions
from room temperature experiments on Ta (5 nm)/[Fe (0.34 nm)/Gd(0.4 nm)]
$\times$ 100/Pt (3 nm) \cite{Montoya18}.
(a) Original soft x-ray microscopy image of a close-packed skyrmion lattice.
(b) Postprocessed binary image of (a) where the background has been
subtracted.
(c,d,e,f) Skyrmion dynamics obtained by summing images of the domain
morphology before and after a current pulse is injected, where purple
regions indicate places where the domain morphology has changed.
Reprinted with permission from S. Montoya {\it et al.}, Phys. Rev. B
{\bf 98}, 104432 (2018). Copyright 2018 by the American Physical Society.
}
\label{fig:36}
\end{figure}

At stronger pinning,
Reichhardt and Reichhardt \cite{Reichhardt19a} found a transition to
a state in which, even for $F_D=0$,
dislocations proliferate and
the skyrmions are in a glassy configuration.
At higher drives, however, the
skyrmions can dynamically order into a moving crystal phase.
Figure~\ref{fig:34} shows the dynamical phase diagram
as a function of driving force versus pinning strength $F_{p}$
obtained from a series of
simulations.
Here there are two pinned phases:  the pinned crystal for weak disorder
and the pinned glass for stronger disorder.
In the pinned crystal phase, the critical driving force
$F_{c} \propto F^{2}_{p}$, as expected for elastic depinning
from collective pinning theory \cite{Blatter94},
while in
the pinned glass phase,
$F_{c} \propto F_{p}$ as expected for plastic depinning.
There is a sudden increase
in $F_{c}$ across the crystal to glass transition.
This is similar to the peak 
effect phenomenon
found for superconducting vortices,
where the particles in the plastic or disordered phase
have more local freedom to move and are better able to
optimize their
interactions with the randomly located pinning sites,
increasing the value of $F_c$
\cite{Reichhardt01,ToftPetersen18,Banerjee00,Bhattacharya93}.
In general, when the pinning is weaker,
the relative magnitude of the jump
in $F_c$ at the transition from elastic to 
plastic depinning becomes more pronounced \cite{Reichhardt17}. 
Just at depinning in the elastic state,
the motion can be jerky or intermittent but the
particles maintain the same neighbors.
On the other hand, for plastic depinning,
numerous dislocations and topological objects appear
in the system and there is a coexistence
of pinned and flowing skyrmions in the plastic flow state,
as illustrated
in Fig.~\ref{fig:35}
\cite{Reichhardt16}.
The moving liquid state in
Fig.~\ref{fig:34} is distinguished from the plastic flow state
by the fact that all of the skyrmions are moving simultaneously
but remain disordered.
At higher drives, within the particle model
the skyrmions dynamically reorder
into 
a moving crystal phase
and regain
their mostly sixfold ordering \cite{Reichhardt15a,Reichhardt16,Reichhardt19a}, 
\cite{Reichhardt19a}. 

Evidence for collective plastic flow as a function of drive has
been obtained with direct imaging for room temperature skyrmions in thin 
films.
The skyrmion trajectories
show a coexistence of moving and pinned
regions along with specific channels or rivers
in which flow is occurring, as illustrated in Fig.~\ref{fig:36} \cite{Montoya18}.
The images
closely resemble the motion observed experimentally near the depinning
transitions
of superconducting vortices \cite{Fisher98,Reichhardt17,Matsuda96} and 
colloidal particles \cite{Pertsinidis08,Tierno12} moving on random substrates.
Small angle neutron scattering experiments on MnSi under current flow showed
a broadening of the peaks above depinning,
which could be evidence of dynamical disordering
close to depinning;
however, it was also argued that the broadening could arise from
edge effects which produce
counter-rotating domains \cite{Okuyama18,Okuyama19}.  

An important difference between the experimental skyrmion system and
the particle based superconducting vortex and colloidal
simulations is that the skyrmions
have internal degrees of freedom which can become excited.
For example,
one end of skyrmion (a meron) could be pinned
while the meron in the other half of the skyrmion continues
to move.  This could be viewed
as
the motion of an elongated skyrmion or the emergence of
a helical stripe phase.

\begin{figure}
\includegraphics[width=\columnwidth]{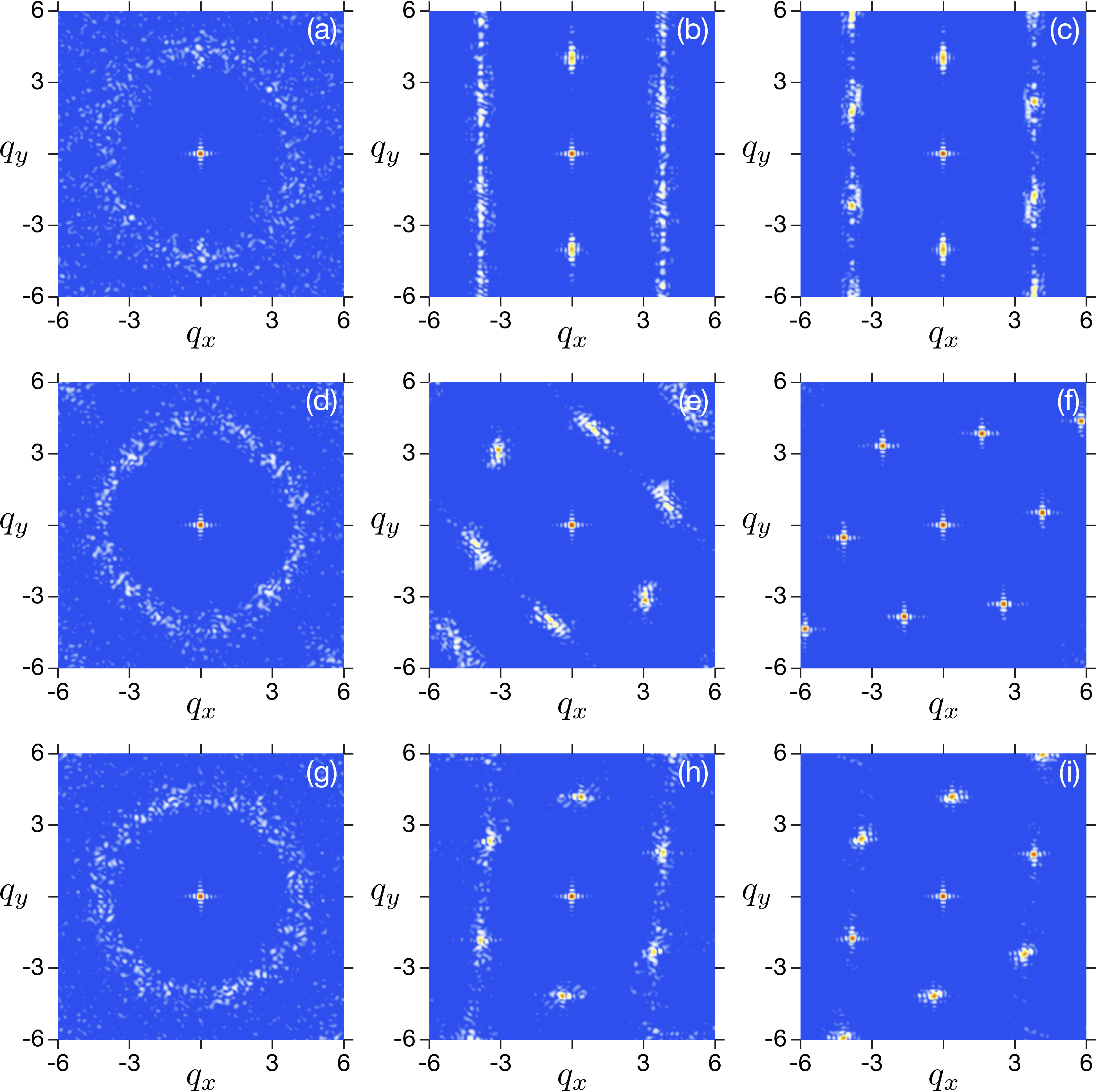}
\caption{
  Static structure factor $S({\bf q})$ from particle-based skyrmion
  simulations \cite{Diaz17}. The driving force increases from left
  to right in each row.
  (a-c) An overdamped
  system with an intrinsic Hall angle of $\theta^{\rm int}_{\rm SkH}=0$
  in (a) the plastic flow state, (b) the moving smectic state, and
  (c) the moving anisotropic crystal state.
  (d-f) A system with $\theta^{\rm int}_{\rm SkH}=45^\circ$ in
  (d) the moving liquid state, (e) a slightly anisotropic moving
  crystal state, and (f) the moving crystal state.
  (g-i) A system with $\theta^{\rm int}_{\rm SkH}=70^\circ$ in
  (g) the moving liquid state, (h) a slightly anisotropic moving crystal
  state, and (i) the moving crystal state.
Reprinted with permission from S. A. D{\' \i}az {\it et al.}, Phys. Rev. B
{\bf 96}, 085106 (2017). Copyright 2017 by the American Physical Society.
}
\label{fig:37}
\end{figure}

The dynamical ordering from a plastic flow state to a
more ordered state illustrated in
Fig.~\ref{fig:34}
is similar to that found for
superconducting vortices \cite{Reichhardt17,Koshelev94,Olson98a},
colloidal particles \cite{Reichhardt02},
Wigner crystals \cite{Reichhardt01}, pattern
forming systems \cite{Reichhardt03,Xu11,Zhao13},
and driven charge density waves \cite{Danneau02,Du06,Pinsolle12}.
There
are, however,
several differences in the moving states of skyrmions with a Magnus force
compared to the
previously studied overdamped systems. 
In the 2D vortex system as well as overdamped systems in general, 
the moving state is typically a moving smectic  
in which the vortices form rows that slide past one another.
Figure~\ref{fig:37}(a,b,c) shows $S({\bf k})$ 
at fixed drives for an
overdamped particle system
that could represent vortices moving over random disorder \cite{Diaz17}.
At lower drives in Fig.~\ref{fig:37}(a),
the structure factor has a ring shape
indicative of a liquid or glass and the particles
are in a disordered configuration.
At higher drives in Fig.~\ref{fig:37}(b), the system starts to dynamically
reorder into a moving smectic state where there are well defined
chains of particles moving past each other.  This creates a series of
aligned dislocations
and the structure factor contains two dominant peaks.
For even  higher drives
in Fig.~\ref{fig:37}(c),
the system is still in a moving smectic
state but some additional sixfold ordering is beginning to emerge,
producing additional smeared peaks in $S({\bf k})$.
At still higher drives, the structure factor
remains the same since the dislocations are dynamically trapped.
The approach to a moving crystal state
for overdamped particles such as vortices moving 
over random disorder has been predicted in theoretical calculations
\cite{Balents98,Giamarchi96} and observed in
numerous simulations \cite{Olson98a,Giamarchi96,Moon96,Kolton99,Fangohr01,Gotcheva04} and experiments \cite{Pardo98}. 

Simulations of driven skyrmions
have shown that when the Magnus force is present,
the dynamically reordered state has six strong peaks, indicating
the presence of a higher degree of
isotropic order than what is observed in overdamped systems \cite{Diaz17}.
This effect is
attributed specifically to the
Magnus force.
Viewed from a co-moving frame,
overdamped particles
experience 
force perturbations from the substrate that are strongest
in the direction of motion.
It can be argued that the
resulting fluctuations are equivalent 
to a shaking
temperature
$T_{sh} \propto 1/F_D$ \cite{Koshelev94}.
For sufficiently large drives,
the system freezes into a solid, but
because the shaking temperature is 
anisotropic with $T^{||}_{sh} > T^{\perp}_{sh}$
\cite{Balents98,Giamarchi96},
the direction perpendicular to the drive freezes first,
locking dislocations into the sample,
while the
direction parallel to the drive
remains liquidlike.
In the case of skyrmions,
the Magnus force mixes the fluctuations from the driving direction
into the perpendicular direction,
resulting in a more isotropic shaking temperature that
prevents the trapping of smectic defects and
allows the system to freeze in both directions simultaneously.
The isotropic nature of $T_{sh}$
was
confirmed in simulations
through direct measurements of the fluctuations in
both the transverse and longitudinal directions for
skyrmions moving through random pinning
\cite{Diaz17}.
It may be possible that for very large Magnus forces,
the system could form a moving smectic structure
that is aligned perpendicular, rather than parallel, to the drive.

Figure~\ref{fig:37}(d,e,f) shows
$S({\bf k})$ for three different drives in
simulations of a 2D driven skyrmion system  
with random pinning
where the intrinsic
skyrmion Hall angle is $\theta^{\rm int}_{\rm SkH} = 45^\circ$ \cite{Diaz17}.
At a low drive
in Fig.~\ref{fig:37}(d), the skyrmions are disordered
and $S({\bf k})$ has
a ringlike structure.
At a higher drive
in Fig.~\ref{fig:37}(e),
sixfold peaks  begin to emerge that are
much more isotropic than the peaks observed
in Fig.~\ref{fig:37}(b) at
the same drive in the overdamped
system,
although there is still a small amount of
smearing of the four side peaks.
For high drives,
illustrated in
Fig.~\ref{fig:37}(f),
there are now six sharp peaks of equal size and the skyrmions have
organized into a crystal.
A similar
evolution of the structure factor
with drive for skyrmions with
$\theta^{\rm int}_{\rm SkH} = 70^\circ$ appears in
Fig.~\ref{fig:37}(g,h,i).
Compared to the overdamped system, where a somewhat disordered crystalline
state emerges that is aligned with the driving direction,
the skyrmion crystalline state is very well ordered and it is {\it not} aligned
with the driving direction.
Instead, the orientation of the crystal rotates slightly with increasing drive.
This is another consequence of the
Magnus force, which causes the lattice to tend to align in the direction of motion
rather than in the direction of the applied driving force.
In an overdamped system,
these two directions are the same, but in the skyrmion system, they are
separated by the
intrinsic skyrmion Hall angle.

The moving smectic state can also be distinguished from the moving crystal
by measuring the relative motion of the particles in the co-moving frame,
where the center of mass motion has been subtracted.
Skyrmions exhibit
a long time diffusive motion in the driving direction,
but
subdiffusive motion or no diffusion perpendicular to the driving direction
\cite{Diaz17}.
The displacements in the moving frame are given by
$\Delta_{||}(t) = N^{-1}\sum_{i=1}^N[\tilde{r}_{i,||}(t) - \tilde{r}_{i,||}(0)]^2$, where 
$\tilde{r}_{i,||} = r_{i,||}(t) - R^{CM}_{||}(t)$, 
and  
$\Delta_{\perp}(t) = N^{-1}\sum_{i=1}^N[\tilde{r}_{i,\perp}(t) - \tilde{r}_{i,\perp}(0)]^2$, 
where $\tilde{r}_{i,\perp} = r_{i,\perp}(t) - R^{CM}_{\perp}(t)$. Here
${\bf R}^{CM}$ is the center of mass in the moving frame
and $N$ is the number of skyrmions. 
The different phases can be
identified through the power law behavior
\begin{equation}
\Delta(t)_{||,\perp} \propto t^{\alpha_{||,\perp}}
\end{equation}
For isotropic regular diffusion, $\alpha_{||} = \alpha_{\perp}= 1$;
for a smectic state,
$\alpha_{||} \geq 1$ and $\alpha_{\perp} = 0$;
for a moving crystal, $\alpha_{||} =\alpha_{\perp} = 0$; and for a moving liquid,
$\alpha_{||} \geq 1$ and $\alpha_{\perp} \geq 1$.
Other regimes are also possible.  For example,
at short times there can be
subdiffusive behavior with $0 < \alpha < 1$ in either direction,
but at long times a
crossover to regular diffusion occurs.
Within the smectic phase,
$\alpha_{||} = 2$, indicating
superdiffusive or ballistic motion in
the driving direction, while
$\alpha_{\perp} = 0$.
The ballistic behavior that appears even after the
center of mass motion
has been
removed arises because
the different rows in the smectic state are
moving at different speeds relative to one another.
In general, the moving smectic state in overdamped 2D systems
always shows regular diffusion or superdiffusion in the direction
parallel to the drive but no diffusion in the direction perpendicular
to the drive.
This is in contrast to the skyrmion system
which
exhibits no diffusion in either direction,
indicating the emergence of
a truly crystalline state as a function of drive.

\subsection{Noise} 

Noise fluctuations are a useful method for
characterizing condensed matter states
\cite{Weissman88,Sethna01}. 
For skyrmion systems,
transitions between plastic flow and
moving crystalline regimes
can be distinguished
with the power spectrum
\begin{equation}
S(\omega) = \left|\int \sigma(t)e^{-i2\pi\omega t}dt\right|^2 
\end{equation} 
of various time dependent quantities $\sigma(t)$,
which could include
the topological Hall resistance $\rho^{T}_{xy}$,
the local magnetization, or the fluctuations in $S({\bf k})$
at a particular value of ${\bf k}$.
Separate time series $\sigma(t)$ can be obtained
for different values of an applied drive
in order to
detect changes in the spectral response.
Such measures have been used to
study superconducting vortices
\cite{Olson98a,Kolton99,Marley95,DAnna95,Merithew96,Kolton02}, 
sliding charge density waves \cite{Gruner81,Bloom93}, and the
motion of magnetic domain walls \cite{Sethna01},
and they could prove to be a similarly
powerful technique for skyrmion systems.
Particle-based simulations of vortices showed
that in the plastic flow phase,
the velocity noise has a
broad band
$1/f^{\alpha}$ signature, where $f = \omega/2\pi$ \cite{Olson98a,Marley95}.
The value of the exponent $\alpha$ determines the type of the noise.
When $\alpha =0$,
the noise is white and has
equal power in all frequencies,
while $\alpha = 1$
or a $1/f$ signature is called pink noise
and $\alpha = 2$ or a $1/f^2$ signature is known as
brown noise or Brownian noise.
Brownian noise
can be produced by the trajectories
of a random walk,
whereas white noise has no correlations. 
In overdamped systems that undergo depinning,
values of $0.75 < \alpha < 1.8$
are associated with collective dynamics,
and
in some cases the presence of a critical point
produces a distinct spectral response
\cite{Travesset02}.
This implies that if depinning is a critical 
phenomenon,
it may be possible to use
the noise power
to determine the university class of the depinning.
In addition to broad band noise, there may be a knee at a specific
frequency
of the form
$S(f) \propto \tau/(1 + (2\pi\tau f)^2)$,
which approaches a constant value as $f$ goes to zero.
This type of spectrum is often
associated with telegraph noise,
where $\tau$ is the characteristic time of jumps between the two
states of the signal.
A narrow band noise signal
produces one or more peaks at characteristic frequencies that are
related to a length scale in the system.
For example, 
a random arrangement
of particles moving over random disorder can have a time of flight
narrow band noise peak
in which the characteristic frequency
is the
inverse of the time required to traverse
the entire sample \cite{Olson98, DAnna95}.
Alternatively,
if the particles are in a moving lattice, they can produce a washboard signal
corresponding to the time required for a particle to move one lattice constant
\cite{Olson98a,Harris95,Togawa00,Okuma07,Klongcheongsan10}.

\begin{figure}
\includegraphics[width=\columnwidth]{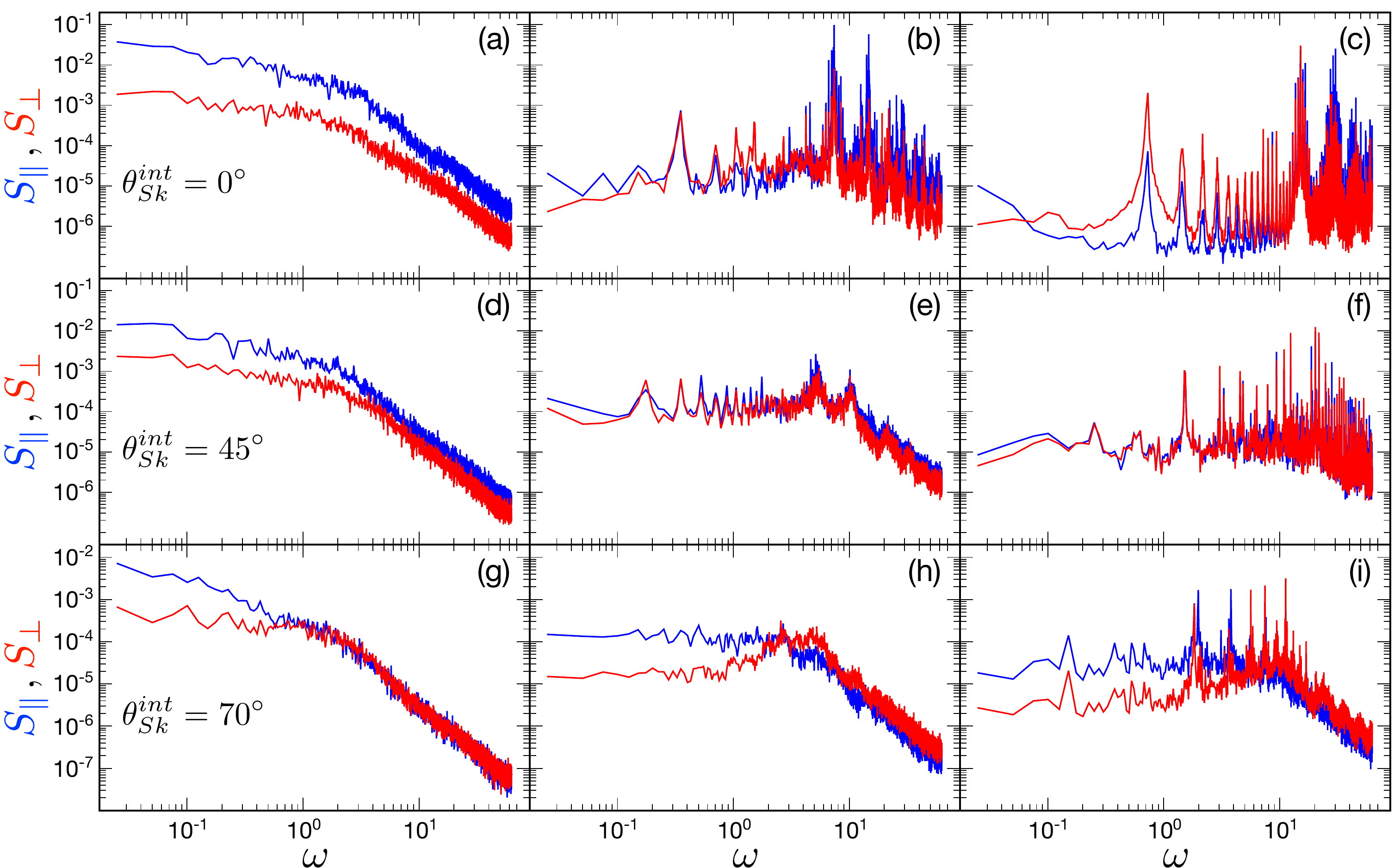}
\caption{
  Spectral density plots from
  particle based skyrmion simulations \cite{Diaz17}
  showing
  $S_{||}(\omega)$ (blue) for velocity fluctuations parallel
  to $\theta_{\rm SkH}$ and $S_{\perp}(\omega)$ (red)
  for velocity fluctuations perpendicular to $\theta_{\rm SkH}$.
  The driving force increases from left to right in each row.
  (a)-(c) An overdamped sample with $\theta_{\rm SkH}^{\rm int}=0^\circ$
  in (a) the disordered flow state and (b,c) two drives in the moving
  smectic phase.
  (d)-(f) A sample with $\theta_{\rm SkH}^{\rm int}=45^\circ$ in
  (d) the disordered flow state, (e) the moving liquid phase, and
  (f) the moving crystal phase.
  (g)-(i) A sample with $\theta_{\rm SkH}^{\rm int}=70^\circ$ in
  (g) the disordered flow state, (h) the moving liquid phase, and
  (i) the moving crystal phase.
Reprinted with permission from S. A. D{\' \i}az {\it et al.}, Phys. Rev. B
{\bf 96}, 085106 (2017). Copyright 2017 by the American Physical Society.
}
\label{fig:38}
\end{figure}

In simulations, a time of flight signal can arise from the motion
of a large scale structure, such as
a grain boundary in a skyrmion lattice,
through the periodic boundary conditions.
A signal of this type typically appears at relatively low frequencies.
In skyrmion experiments,
narrow band noise could be produced by
the periodic nucleation of skyrmions at the edge of the sample,
where the time of flight would correspond to the time required
for the skyrmion to cross to the other side of the sample
and be annihilated.
The washboard frequency of an elastic lattice moving over disorder is given by
$\omega = v/a$ \cite{Harris95}, where $v$ is the time averaged
dc velocity and $a$ is the lattice constant.
A measurement of
the washboard frequency thus provides a method for determining
the lattice constant.
Both the time of flight and washboard
narrow band
signals
are generated when the particles are in steady continuous motion,
rather than intermittently alternating between being pinned and moving.
If the system forms a continuously moving
liquid, 
the sharp narrow band peaks in the noise spectrum would be lost,
but
a smoother peak could still appear that is
associated with the
average time between collisions of a particle with a pinning site.
Figure~\ref{fig:38} shows power spectra
$S_{||}$ and $S_{\perp}$ of the longitudinal
and transverse velocity signals
from
a particle based simulation
of skyrmions moving over random disorder at various drives
\cite{Diaz17}.
In Fig.~\ref{fig:38}(a),
an overdamped system in the plastic flow  regime
has higher noise power parallel to the drive than perpendicular to the drive,
consistent with the
idea that the shaking temperature
from the fluctuations is largest
in the direction of drive
for overdamped systems moving over quenched disorder.
There is also a $1/f^{\alpha}$ tail with $\alpha \approx 1.5$,
similar to the noise observed in simulations of other overdamped
systems.
Figure~\ref{fig:38}(b,c) shows that at higher drives,
the broad band noise signal is lost and a series of high frequency peaks appear
at multiples of the washboard frequency.
At much lower frequencies,
the time of flight signal produces
a second series of peaks
which are the most pronounced in
Fig.~\ref{fig:38}(c).
Figures~\ref{fig:38}(d) and (g)
show the lower drive
plastic flow regime for skyrmion
systems with a Magnus force giving
$\theta^{\rm int}_{\rm SkH} = 45^\circ$ and $70^\circ$, respectively.
Here
the magnitude of the noise at higher frequencies is nearly identical in both
directions,
in agreement with the argument that
the shaking temperature for skyrmions is
more isotropic than that for overdamped particles.
At higher drives in Fig.~\ref{fig:38}(e, f, h, i),
peaks once again appear at both the time of flight and washboard frequencies.
Additional noise features can be deduced by
analyzing the evolution of these peaks as a function of current.
A sudden
switch in the peak frequency
would indicate
the reorientation of the lattice
or the annihilation
of dislocations in the lattice.    

Although skyrmions exhibit a number of dynamical features
similar to those found in overdamped vortex systems,
they also have
some unique
behaviors.
For example, if a current were used to create skyrmions,
this process
could be detected via changes in the narrow band noise signature.
In a sample where skyrmions
coexist with different species
of topological defects such as large ferromagnetic domains,
the low frequency noise generated by
the density fluctuations
could be used to determine the size of
 the domains \cite{Mohan09}.  
 It may also be possible to
 detect increases
 of noise as a function of increasing temperature
 near a 2D melting transition, where fluctuations are expected
 to increase strongly \cite{Koushik13}.
 In addition to the power spectrum,
 higher order measures such as the second
 spectrum or the noise of the noise
 can be analyzed to examine
 the persistence times of metastable processes \cite{Merithew96}.
 Noise has been used to measure various nonequilibrium effects
 such as negative velocity fluctuations \cite{Bag17},
 and similar
 studies could be performed
for driven skyrmions where the
 nonconservative
 Magnus force could produce novel effects.
Skyrmion systems in which
the dynamics of small numbers of skyrmions can be accessed
could be ideal for studying the routes to chaos
using techniques
similar to previous work performed with noise in
charge density wave \cite{Levy91} and superconducting vortex systems \cite{Olive06}.  

Up until this point, studies of skyrmion noise 
have been limited to particle based models \cite{Reichhardt16,Diaz17};  
however, continuum based
approaches
could
permit the
exploration of additional 
contributions to noise
from shape fluctuations or
the breathing modes of the skyrmions.
For example,
a moving skyrmion lattice would exhibit
a washboard frequency associated with the lattice spacing,
but a second much higher frequency
signal could appear as a result of
collective breathing modes
that are excited by the motion over random disorder.
In addition to noise generated purely by
the internal modes of the skyrmion, other noise signatures could
arise due to coupling of the internal modes with the skyrmion lattice.
Experimental  noise measurements in skyrmion systems are just
beginning,
with a recent experiment on
skyrmion motion in a narrow channel showing
a transition from $1/f$ noise to 
narrow band noise
similar to what has been seen in simulations \cite{Sato19}.   

\subsection{Avalanches}

Intermittent noise often takes the form of
time windows of little or no activity interspersed with 
windows of large activity or avalanches.
Avalanche-like behavior is a ubiquitous
phenomenon in 
driven systems with quenched disorder
\cite{Fisher98,Reichhardt17,Carlson94,Sethna01,Bak88},
and one of the best known examples is  Barkhausen noise
in magnetic systems \cite{Barkhausen19,Cote91,Bertotti94,Zapperi98}.
Avalanches are often most clearly resolvable
at low driving, where distinct jumps
can be distinguished from one another.

Numerous methods exist for analyzing the avalanches.
Construction of the probability distribution function of the magnitude of the
velocity or other avalanche signal as a function of time
can show whether the avalanches are all close to the same size,
are exponentially distributed,
have a specific range of sizes,
or are power law distributed.
A power law distribution of avalanche events is
often associated with critical behavior \cite{Bak88,Perkovic95}.
For example,
if depinning 
in systems driven over quenched disorder
is a critical phenomenon,
then avalanche behavior could appear at low drives close to the depinning transition.
It has been argued theoretically that
the avalanches
are critical only
for a critical
disorder strength $R_{c}$,
with large avalanches that are close to the same size occurring for
disorder strengths $R<R_c$,
and exponentially distributed avalanches
appearing for $R>R_c$;
however, it is possible to be fairly far from
$R_{c}$ and still observe a regime of power law distributed avalanche sizes
\cite{Sethna01,Perkovic95,Sethna93}.
Avalanches can occur in driven systems without thermal fluctuations;
however, there
are cases in which thermal effects can trigger avalanches.
Both
elastic and plastic systems exhibit
avalanches,
and
in principle the avalanche distributions would change
across an elastic-plastic transition.
Since avalanches occur so routinely in magnetic systems,
the skyrmion system is ideal for examining
avalanche effects.

Skyrmion avalanches remain largely unexplored, but were studied
by
D{\' \i}az {\it et al.}
\cite{Diaz18} using  a 2D  particle based model in which skyrmions
entered the edge of the sample under a low driving force through a series
of jumps.
For  zero or weak Magnus forces,
the avalanche sizes $S$ and durations $T$ are power law distributed,
$P(T) \propto T^{\alpha}$ and $P(S) \propto S^\tau$, with
$\alpha = 1.5$
and $\tau = 1.33$.
Near a critical point there
should be an additional scaling relation
$\langle S\rangle \propto T^{1/\sigma\nu z}$
between the avalanche sizes and durations 
\cite{Sethna01},
so that in this case,
$1/\sigma\nu z = 1.63$. 
The exponents should also obey
\begin{equation}
\frac{\alpha -1}{\tau -1} = \frac{1}{\sigma\nu z}
\end{equation}
near the critical point.
In the work of D{\' \i}az {\it et al.},
this equality was satisfied,
indicating that near depinning,
the system is critical.
Interestingly,
for large values of $\theta^{\rm int}_{\rm SkH}$,
the scaling exponents for the avalanches change
but equality (15) holds,
suggesting that the
nature of the criticality changes with increasing Magnus force.
The avalanches can also be characterized
by scaling the shape of
avalanches that have the same duration.
In certain universality classes such as the random field Ising
model, such scaling will produce a symmetric curve
\cite{Sethna01,Mehta02}.
D{\' \i}az {\it et al.} found that the avalanches in the overdamped system
and in samples with weaker Magnus forces
were symmetric in shape, while those for strong Magnus forces were strongly
skewed.
This is also correlated with the change in the avalanche exponents at
strong Magnus forces.
Skewed avalanche shapes can result from nondissipative effects,
such as inertia which tends to speed up the avalanche at later times
and produce a leftward skew, or
negative mass effects
which have the opposite effect and
give a rightward skew \cite{Zapperi05}.
Skyrmions have a tendency to be more strongly deflected at later times, which
is similar to
a negative mass effect.
Experimental studies of
avalanches or cascades in stripe and skyrmion phases that
focused on
jumps or changes in the pairwise correlation functions showed evidence
for power law distributions of jump sizes in the skyrmion regime, as well as
different avalanche exponents in
the skyrmion and stripe phases \cite{Singh19}.  

\subsection{Continuum Based Simulations of the Dynamic Phase Diagram}

\begin{figure}
\includegraphics[width=\columnwidth]{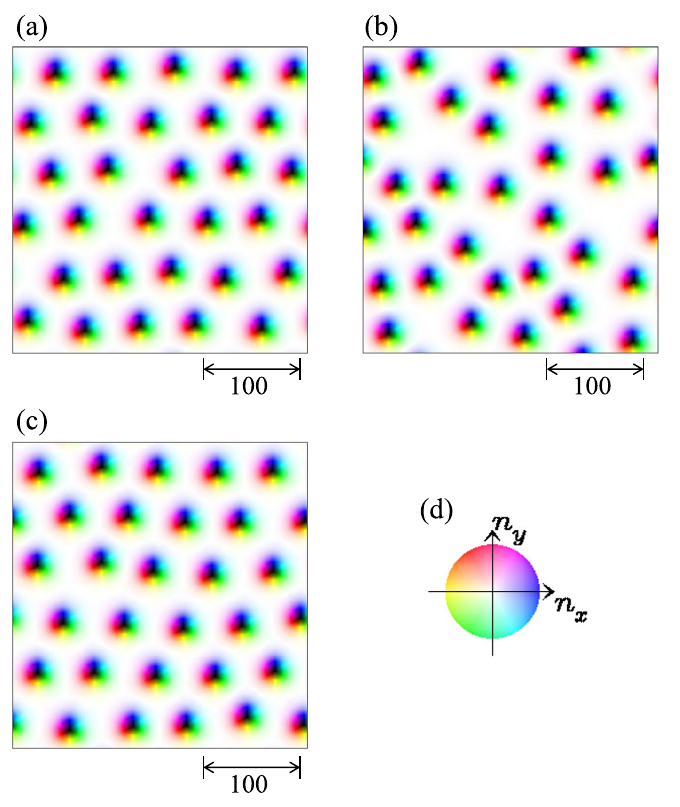}
\caption
{ Continuum simulations of current driven skyrmions interacting with
  weak impurities \cite{Koshibae18}.
  The magnetic field is $h=0.025$ and the impurity
strength is $K_{\rm imp}=0.01$.
(a) The initial configuration at zero drive, $j=0$.
(b) A glassy configuration at $j=0.001$. (c) A moving skyrmion crystal at
$j=1.0$. (d) The color code used in panels (a-c) to represent the in-plane
magnetic moment.
Reprinted under CC license from W. Koshibae and N. Nagaosa, Sci. Rep.
{\bf 8}, 6328 (2018).
}
\label{fig:41}
\end{figure}

\begin{figure}
\includegraphics[width=\columnwidth]{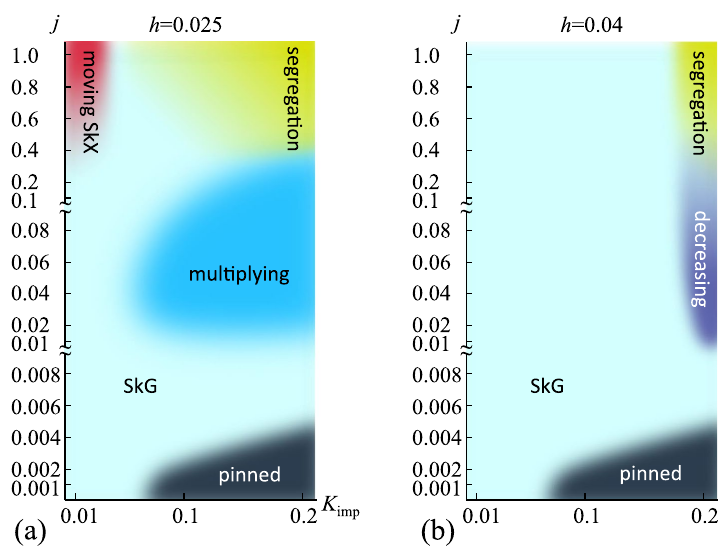}
\caption{
  Dynamic phase diagrams as a function of applied current $j$ vs
  impurity strength $K_{\rm imp}$ from the continuum simulations illustrated
  in Fig.~\ref{fig:41} \cite{Koshibae18}.
  The applied magnetic field is
  (a) $h=0.025$ and (b) $h=0.04$.
The different dynamic phases are labeled, including the
skyrmion glass (SkG) and moving skyrmion crystal (SkX) states.  
Reprinted under CC license from W. Koshibae and N. Nagaosa, Sci. Rep.
{\bf 8}, 6328 (2018).
}
\label{fig:42}
\end{figure}

A variety of continuum simulation studies
have explored the dynamical ordering of 
driven skyrmions in the presence of quenched disorder.
Koshibae and Nagaosa  \cite{Koshibae18} used
a 2D continuum model for skyrmions interacting with random 
point pinning to construct a driving force versus disorder
strength phase diagram.
They 
initialized the system in a skyrmion lattice
at a drive of $j = 0$
as shown in Fig.~\ref{fig:41}(a).
When a finite drive is applied,
the skyrmions move plastically,
creating a disordered
structure as shown in Fig.~\ref{fig:41}(b) at $j = 0.001$. 
At higher drives, the system transitions into  a moving skyrmion lattice
as illustrated in Fig.~\ref{fig:41}(c) for $j = 1.0$.
The resulting phase diagram as a function of the driving current $j$
versus the impurity strength $K_{\rm imp}$ appears in Fig.~\ref{fig:42}(a,b)
for two different magnetic field strengths of $h = 0.025$
and $h = 0.04$, respectively.
At $h = 0.025$, there is a pinned phase which 
grows in extent with increasing impurity strength, as
well as large regions
of moving skyrmion glass or disordered 
moving phases.
For $K_{\rm imp} < 0.1$, the moving skyrmion glass  
orders into a moving crystal as shown in Fig.~\ref{fig:41}.
The first of two new phases that appear is
a multiplication phase in which skyrmions can be created dynamically
by the combination of the current and pinning effects.
The second is a segregated or clustered state.
For $h = 0.04$, there is also a decreasing phase
in which skyrmions are annihilated.
The segregated phase  
was argued to result from the
modification of the skyrmion-skyrmion
interactions by the emission of spin excitations,
which produce an effective attractive interaction between the skyrmions.
In subsequent 2D particle 
based simulations of skyrmions  
moving over strong disorder,
a segregated phase was also observed that was  
argued to be due to
a Magnus-force induced effective attraction between
skyrmions that are moving at different skyrmion Hall angles
\cite{Reichhardt19a}.

The different phases in Fig.~\ref{fig:42}
could be detected using imaging and neutron scattering techniques.
They could also in principle be identified
by analyzing the noise
fluctuations
since, 
as was shown previously, a change in the noise power occurs across
the transition from the
moving glass to the moving lattice state.  The
multiplying, decreasing and segregated phases
shown in Fig.~\ref{fig:42} could each have their
own distinct noise signatures or changes in the topological Hall 
effect.   

\subsection{3D Skyrmion Dynamics}

Although stiff 3D skyrmions can be treated with 2D models,
in a fully 3D system there are numerous new
effects that can appear
such as skyrmion line wandering, skyrmion breaking,
and skyrmion cutting or entanglement. 
In 3D driven superconducting vortex systems with random disorder,
a variety of phases
that are distinct from those found in 2D systems arise
depending on the
material anisotropy  
and the pinning strength \cite{Reichhardt17,Olson00,Chen03,Zhao16}. 
In particular, the 3D vortex system often shows signatures of dynamical first order
phase transitions \cite{Reichhardt17,Olson00,Chen03}, observable as
sharp jumps and hysteresis in the
velocity-force curves.  Similar effects could occur in 
skyrmion systems.
Driven 3D skyrmions moving over quenched disorder could also exhibit
unusual behavior
such as the proliferation of monopoles in driven phases when the skyrmions
break or cut \cite{Milde13,Schutte14b,Lin16,Zhang16a}.

\begin{figure}
\includegraphics[width=\columnwidth]{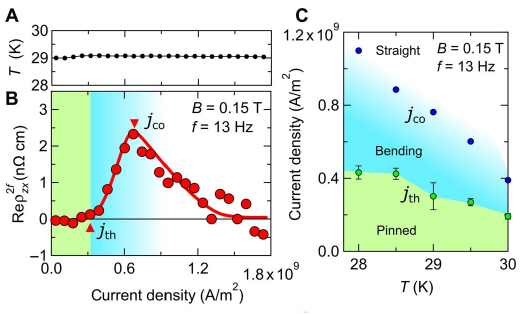}
\caption{
  Results from Hall measurements
  of 3D skyrmions in MnSi thin-plate samples \cite{Yokouchi18}.
(a) Sample temperature $T$ and (b) the real part of the second-harmonic
Hall resistivity, Re $\rho_{zx}^{2f}$, vs driving current density measured at
a frequency of $f=13$ Hz.
(c) Dynamic phase diagram as a function of current density vs temperature $T$
showing regions where the skyrmions are pinned (green), bending (blue), and
straight (white).
Reprinted under CC license from T. Yokouchi {\it et al.}, Science Adv.
{\bf 4}, eaat1115 (2018).
}
\label{fig:43}
\end{figure}

In transport experiments, Yokouchi {\it et al.}
\cite{Yokouchi18} examined the current-induced 
skyrmion motion in MnSi and found strong nonlinear signatures
above the threshold current.
These effects are reduced at higher drives.
Figure~\ref{fig:43}(b) shows the real part of the second-harmonic Hall
resistivity Re $\rho_{zx}^{2f}$ versus 
current density at a fixed magnetic field.
It was argued that the peak in Re $\rho_{zx}^{2f}$
arises from the bending 
of the skyrmion strings just above
the depinning threshold.
Such bending occurs in an asymmetric manner 
due to the creation of a nonequilibrium or nonlinear Hall response by the DMI.
At higher drives,
the skyrmions become straighter and
the effect is reduced.
The features in
Re $\rho^{2f}_{zx}$
can be used to construct the dynamical phase diagram
shown in Fig.~\ref{fig:43}(c).
A pinned phase appears below the threshold
current $j_{\rm th}$, while the current at which the skyrmion string
transitions from bent to 
straight is labeled $j_{\rm co}$.
As the temperature increases, $j_{\rm th}$ decreases
since
thermal activation makes it easier for the skyrmions to jump out of the pinning 
sites.
There is also some experimental evidence for the unwinding of skyrmion
strings in 3D systems under repeated drive pulses \cite{Kagawa17}.
Pinning could be playing a
role in this process
since a partially unwound string can become trapped by the disorder
during the intervals between driving pulses.

\begin{figure}
\includegraphics[width=\columnwidth]{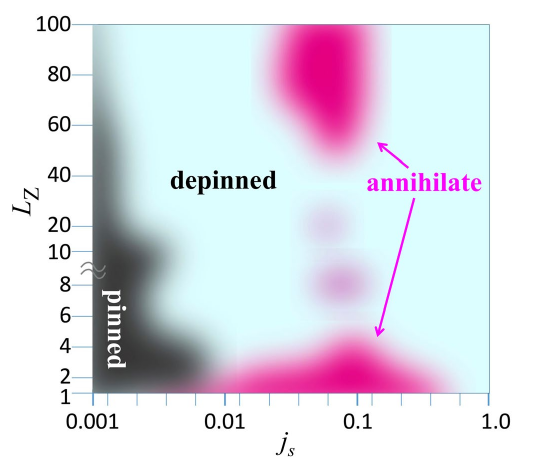}
\caption{
  Dynamic phase diagram from numerical simulations of skyrmion strings
  as a function of $L_Z$, the thickness of the 3D system,
  vs $j_s$, the applied current density \cite{Koshibae19}.
Reprinted under CC license from W. Koshibae and N. Nagaosa, Sci. Rep.
{\bf 9}, 5111 (2019).
}
\label{fig:44}
\end{figure}

Koshibae and Nagaosa \cite{Koshibae19}
numerically studied a skyrmion string driven through random disorder 
in a 3D system for varying sample thicknesses and
identified a pinned regime,
a moving skyrmion regime, and regions of skyrmion string annihilation.
Interestingly, they 
found
that current-induced skyrmion annihilation occurs at a finite current for
thin and thick samples, but 
not for samples of intermediate thicknesses,
indicating that there is an optimal sample length for
skyrmion stability.
Figure~\ref{fig:44} shows a dynamic phase diagram for the
skyrmion string as a function of 
sample thickness $L_{z}$ versus applied current. 
The extent of the pinned regime decreases with increasing $L_{z}$,
meaning that
it is more difficult to pin long 3D skyrmion strings than 2D skyrmions.
This is in agreement
with experimental observations in which
the depinning threshold is low in bulk samples but high in thin films.
In the regime where
skyrmion annihilation does not occur, the skyrmions show pronounced roughening
at low currents but become straighter at higher drives,
similar to the dynamic ordering transition
observed in 2D driven skyrmion assemblies with disorder \cite{Koshibae18}. 

\subsection{Further Directions for Dynamic Skyrmion Phases with Random Disorder} 
There are many future directions for studying the collective dynamics of skyrmions
with random disorder, including noise analysis, imaging, neutron scattering, or 
other experimental probes.
Of highest priority is developing a method to
obtain
clear transport 
measures using a topological Hall effect or other signal that would allow
access to the dynamics on
size and time scales beyond what
is possible in imaging
measurements.
A straightforward
analysis of such transport curves
could be used to determine
when the skyrmions have depinned,
whether they are undergoing elastic or plastic flow, and whether they
pass through drive-induced transitions
such as dynamical reordering, skyrmion annihilation,
or skyrmion creation. This is in analogy to the fact that for
driven vortices in type-II superconductors,
the boundaries between different
dynamic phases can generally be deduced from transport
measurements alone.
It would also be interesting to explore
the time required for a skyrmion system to relax after a single
or repeated driving pulse is applied.
For example, if the skyrmions are subjected to a small ac drive,
they will undergo some spiraling motion,
and there could be a crossover in the response depending on whether
the size of this spiral is larger or smaller than the effective dimension
of the pinning or disorder sites in the sample.
The addition of ac driving could also strongly influence
the dc depinning threshold,
and
Jin {\it et al.} \cite{Jin20}
have found numerical evidence that
an ac drive can substantially lower the dc threshold for the motion of
antiferromagnetic skyrmions.

The role of boundaries is also of interest, since
sample edges can be associated with
the injection or annihilation of skyrmions
or the presence of nonuniform edge currents.
One method for eliminating edges is to
consider a Corbino geometry in which
the skyrmions
circulate around the sample
rather then entering from the edges.
This geometry has been
used to study vortex dynamics in type-II superconductors
in the absence of edge contamination.
Most work on skyrmion dynamics has been
performed using dc driving; however, there
should be a range of unique dynamics that
can be accessed with ac driving, which could 
also substantially reduce the role of the sample edges.
Measurements of the
ac susceptibility 
could
detect different types of dynamical response that are associated 
with specific frequencies,
such as a pinning frequency in which
each skyrmion is only oscillating  within a pinning site instead of entering or
exiting the pinning, or a characteristic washboard frequency that can
be excited when the skyrmions are flowing elastically.
There should be many different ways to observe different types
of avalanche behavior in skyrmions
as well.
For example, if the direction of the magnetic field is changed,
3D skyrmion lines would have to 
reorient,
and if pinning is present,
this process could occur in a series of jumps instead of the smooth behavior
expected in the absence of pinning.
If temperature is relevant,
a finite drive could be applied below the depinning threshold in order 
to observe thermally activated avalanches.
It would also be
interesting
to apply a global current simultaneously with  local excitations
such as local heating or a local probe
to see whether large scale rearrangements of the skyrmions could
be induced by a local perturbation.

Beyond 2D and 3D line-like skyrmions, unique dynamics should appear for 
3D skyrmion hedgehog lattices \cite{Fujishiro19,Lin18}, which could provide 
one of the first realizations of the depinning of a 3D particle-like lattice.
This type of 3D system could be used to study
inhomogeneous pinning,
such as a sample in which pinning is present at the top but not at the bottom,
making it possible to create
a transformer geometry in a uniform field.
The effects of temperature
can also be explored.
It is possible that near the skyrmion
melting transition,
there could a be divergence in the amplitude of
the drive required to dynamically order the skyrmion lattice,
similar to what is found in vortex systems \cite{Koshelev94}.
Both 3D skyrmion lines and point particle skyrmions could exhibit a peak effect
\cite{ToftPetersen18,Banerjee00,Bhattacharya93,Cha98}
in which the depinning threshold current strongly 
increases
when the skyrmions
transition from 3D lines to broken lines or
from a 3D point particle lattice 
to a 3D glass.
Such a peak effect could also 
occur as a function of drive
in the form of a reentrant pinning effect,
where the skyrmions form 
mobile straight lines at low drives,
but break apart or disorder at higher drives 
and become pinned again.

Metastability and memory effects
associated with dynamical phases
are common features in other
systems that  exhibit depinning \cite{Henderson96,Paltiel00,Olson03,Xiao99}.
Such effects
can produce
hysteresis in the
velocity force curves or
persistent memory between driving pulses that generates an
increasing or decreasing response depending on the pulse duration.
Memory effects could be observed
by initializing
skyrmions in either a
metastable ordered or metastable disordered state,
applying a series of drive pulses,
and determining whether the metastable state gradually transitions to
a stable state,
similar to what has been observed  
for metastable states in type-II superconducting vortices  
\cite{Paltiel00,Olson03,Pasquini08}.
Even if one skyrmion phase has a much lower equilibrium energy,
when pinning is present
the skyrmions may be trapped in a different phase due to the pinning barriers,
so that only
the application of a current gives the skyrmions access to the dynamics that will permit
them 
to reach the low energy state.
In this case,
there could be a critical threshold drive that is required to destabilize the
metastable state.

\section{Pinning and the Skyrmion Hall Angle}

\begin{figure}
\includegraphics[width=\columnwidth]{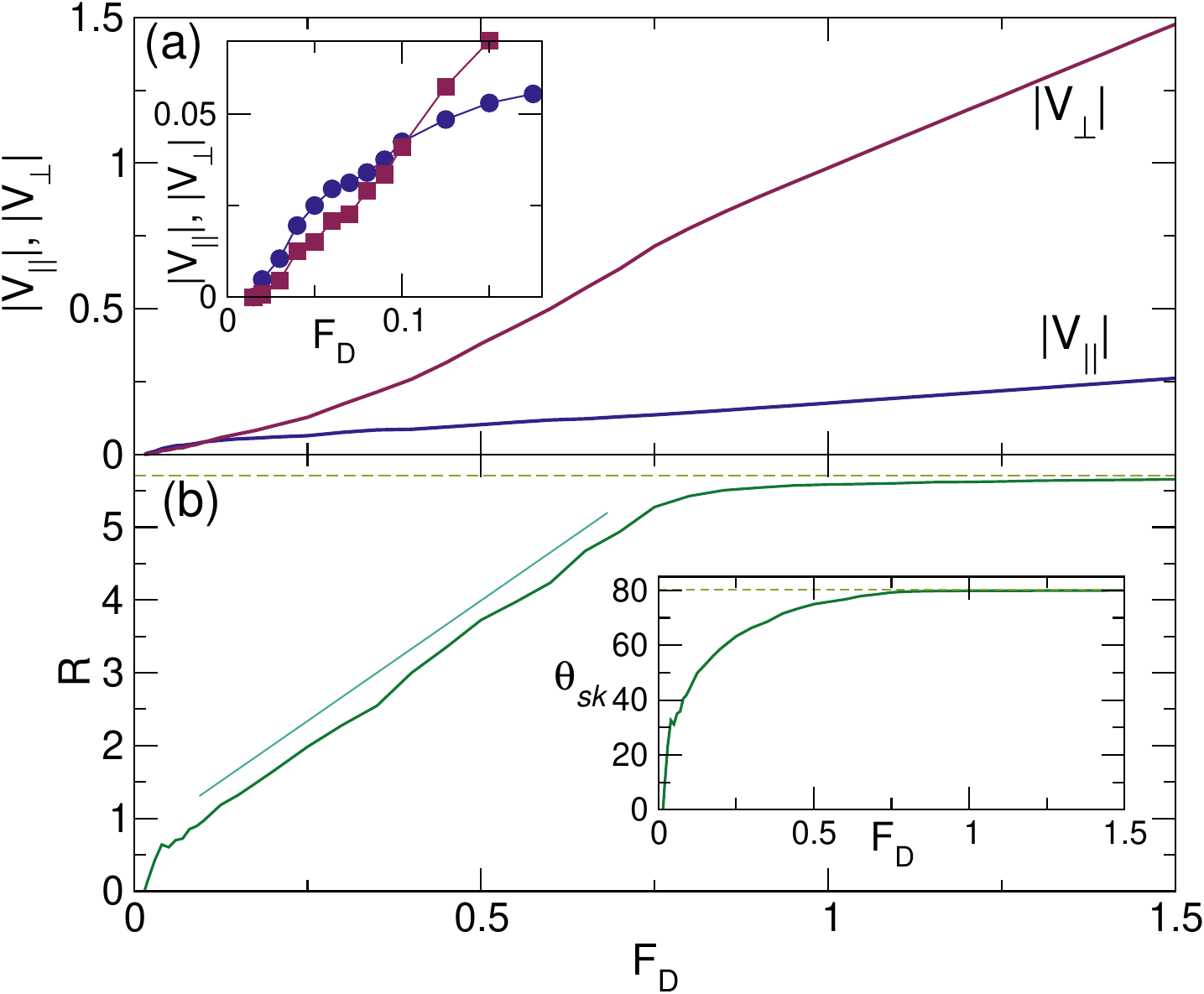}
\caption{
Particle-based simulation measurements of the behavior of the skyrmion  
Hall angle $\theta_{sk}$
for skyrmions driven over random disorder \cite{Reichhardt16}.
(a) The skyrmion velocities in the directions parallel
($|V_{||}|$, blue) and perpendicular
($|V_{\perp}|$, red) to the driving force
vs $F_D$.
Inset: a blowup of the
main panel in the region just above depinning where there is
a crossing of the velocity-force curves. (b) The corresponding
$R=|V_{\perp}/V_{||}|$ vs $F_D$.
The solid straight line is a linear fit and the dashed line is the clean
limit value of
$R \approx 6.0$.
Inset: $\theta_{sk}=\tan^{-1}(R)$ vs $F_D$.
The dashed line is the clean limit value of
$\theta_{sk}$.
Reprinted under CC license from C. Reichhardt and C. J. O. Reichhardt,
New J. Phys. {\bf 18}, 095005 (2016).
}
\label{fig:45}
\end{figure}

A skyrmion under an applied drive moves at an
angle called the skyrmion Hall angle
$\theta_{\rm SkH}$ with respect to the drive.
This angle is proportional
to the Magnus force,
and in the absence of pinning, it is independent of the magnitude of the driving force 
\cite{Nagaosa13,Zang11}.   
Particle based simulations for skyrmions moving over random and
periodic pinning showed that
$\theta_{\rm SkH}$ is
not constant but is nearly zero just at depinning and then increases with
increasing drive before saturating at a value
close to the intrinsic or pin-free value $\theta^{\rm int}_{\rm SkH}$ at higher drives
\cite{Reichhardt15,Reichhardt15a,Reichhardt16,Diaz17}.  
For
a particle based simulation of a collection of skyrmions
with $\theta^{\rm int}_{\rm SkH}=80.06^{\circ}$
driven over random pinning,
Fig.~\ref{fig:45}(a) shows the average velocity
in the directions parallel,
$|V_{||}|$, and perpendicular, $|V_{\perp}|$, to the
drive versus driving force $F_{D}$
\cite{Reichhardt16}.
The corresponding ratio
$R = |V_{\perp}/V_{||}|$
along with $\theta_{\rm SkH}=\tan^{-1}(R)$
appear in Fig.~\ref{fig:45}(b),
where the dashed lines are the expected values
of each quantity in the pin-free limit.
The inset of Fig.~\ref{fig:45}(a) shows that there is  a finite
depinning threshold as well as a range of drives
for which $|V_{||}| > |V_{\perp}|$;
however, as the drive increases,
$|V_{\perp}|$ grows more rapidly than $|V_{||}|$, 
since the intrinsic Hall angle in the clean limit would give
$R=|V_{\perp}/V_{||}| \approx 6$.
Figure~\ref{fig:45}(b) shows that
over a wide range of drives,  $R$ increases
roughly linearly with $F_{D}$ up to $F_{D} = 0.75$,
and then saturates close to its intrinsic
value.
As a result,
at small drives the skyrmions
move in the direction of 
the drive,
but as the drive increases, they gradually
develop a greater component of motion perpendicular to the drive,
until at large drives
they are moving along the direction of the
intrinsic Hall angle. 
Over the 
drive interval in which $R$ is linearly increasing with $F_D$,
the skyrmions are moving plastically,
while at higher drives when the skyrmions
begin to move in a more coherent fashion,
$R$ starts to saturate.
The same general behaviors are robust over
a range of different
intrinsic Hall angles, disorder strengths,
and pinning densities,
while when the Magnus force
is zero,
$|V_{\perp}| = 0$ and
$\theta_{\rm SkH} = 0$ for all values of $F_{D}$ \cite{Reichhardt16}.
For $\theta^{\rm int}_{\rm SkH} < 50^\circ$, the skyrmion Hall angle generally increases
linearly with $F_{D}$
since $\tan^{-1}(x)$ can be expanded as $\tan^{-1}(x) = x - x^3/3 + x^5/5...$. 
For small $R$, the first term dominates,
while for $\theta^{\rm int}_{\rm SkH} > 50^\circ$,
nonlinear effects appear in
$\theta_{\rm SkH}$ with increasing $F_{D}$. 
In particle based simulation
studies of skyrmion noise,
the measured skyrmion Hall angle
begins to saturate when the skyrmions
begin to order, which is correlated
with a reduction 
in the noise power \cite{Diaz17}.

\begin{figure}
\includegraphics[width=\columnwidth]{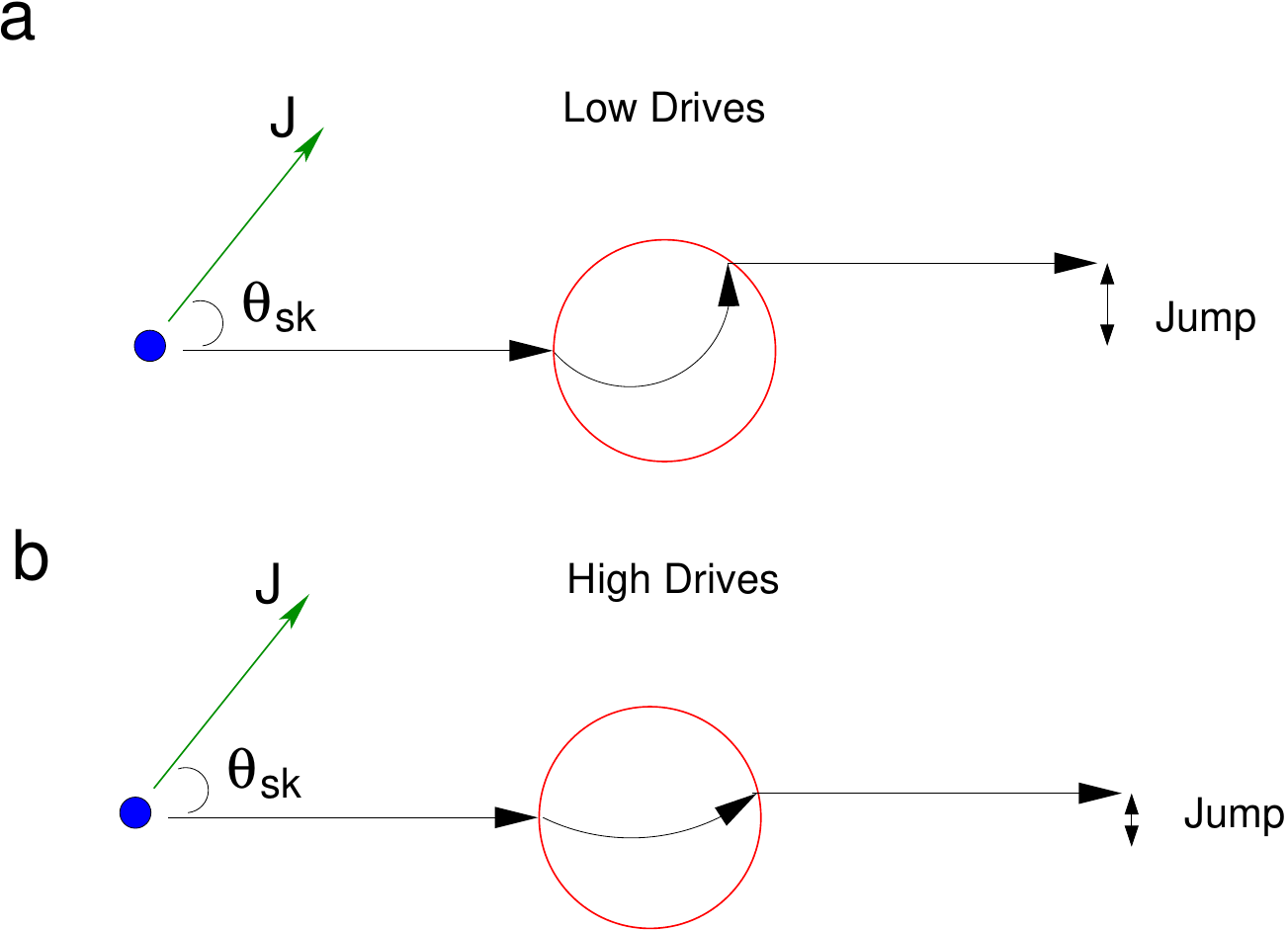}
\caption{
A schematic illustration of how pinning can change the
effective skyrmion Hall angle.
The blue dot is the skyrmion and the
red circle is the pinning site,
while $J$ is 
the direction of the applied current and
$\theta_{sk}$ is the
intrinsic skyrmion Hall angle determining the direction in which
the skyrmion moves with respect to the drive.
(a) At low drives, the skyrmion executes a Magnus-induced orbital motion
as it moves through the pinning site,
leading to a side jump as the skyrmion exits the pinning site.
The jump is in the direction of the current and
therefore reduces the
effective skyrmion Hall angle. (b) At higher drives, the skyrmion moves
rapidly through the pinning site, so the
magnitude of the side jump is strongly reduced. 
}
\label{fig:46}
\end{figure}

The appearance of a drive dependent skyrmion
Hall angle was also partially observed in continuum and Thiele equation 
based work by M{\" u}ller {\it et al.} \cite{Muller15} for a single skyrmion
interacting with a single defect. A more extensive study of the
evolution of $\theta_{\rm SkH}$ with drive was subsequently conducted using particle
based simulations of skyrmion motion 
in a periodic pinning array \cite{Reichhardt15}
and in random pinning \cite{Reichhardt15a}. 
In the work of M{\" u}ller {\it et al.} \cite{Muller15} and
Reichhardt {\it et al.} \cite{Reichhardt15a}, 
the microscopic origin of the drive dependence of $\theta_{\rm SkH}$
was argued to be a side jump effect, 
as illustrated in Fig.~\ref{fig:46}.
A skyrmion executes a Magnus-induced orbit as it moves
through a pinning site, 
so that when the skyrmion 
leaves the pinning site, it has effectively jumped in the direction of the
applied drive.  Repeated jumps
lower the
effective skyrmion Hall angle compared to
the pin-free situation.
The motion of a skyrmion is similar to that of a charged particle in a
magnetic field \cite{Nagaosa13},
and the skewed scattering of the skyrmion by a pinning site
is similar to what is known as a side jump effect for
electron scattering off magnetic defects,  
where an electron undergoes a sideways displacement when interacting with 
a potential as a result of 
spin-orbit interactions \cite{Berger70}.   
As illustrated in Fig.~\ref{fig:46}(a),
a more slowly moving skyrmion 
spends more time in the pinning site,
resulting in a larger jump.
At higher drives,
when the skyrmion is moving faster,
the jump is smaller and
the measured skyrmion Hall angle is closer to the defect-free
value, while at the highest drives, the skyrmions
move so rapidly through the pinning sites that there is
hardly any jump.
This is illustrated in Fig.~\ref{fig:46}(b),
which corresponds to the saturation of $\theta_{\rm SkH}$
at higher drives as
observed in simulation \cite{Reichhardt15,Reichhardt15a,Reichhardt16,Diaz17}.
The size of the jump
is also determined by which side of the pinning site the skyrmion
approaches,
so that for an ensemble of different impact parameters,
strongly asymmetric jumps appear
as demonstrated for a single skyrmion 
moving through a pinning site \cite{Reichhardt15}.
This same work showed
that when the Magnus force is zero, the particle
can still experience a jump as it moves through the pinning site, 
but the jump is symmetric as a function of impact
parameter, so that in the ensemble average,
there is no net jump.

\begin{figure}
\includegraphics[width=\columnwidth]{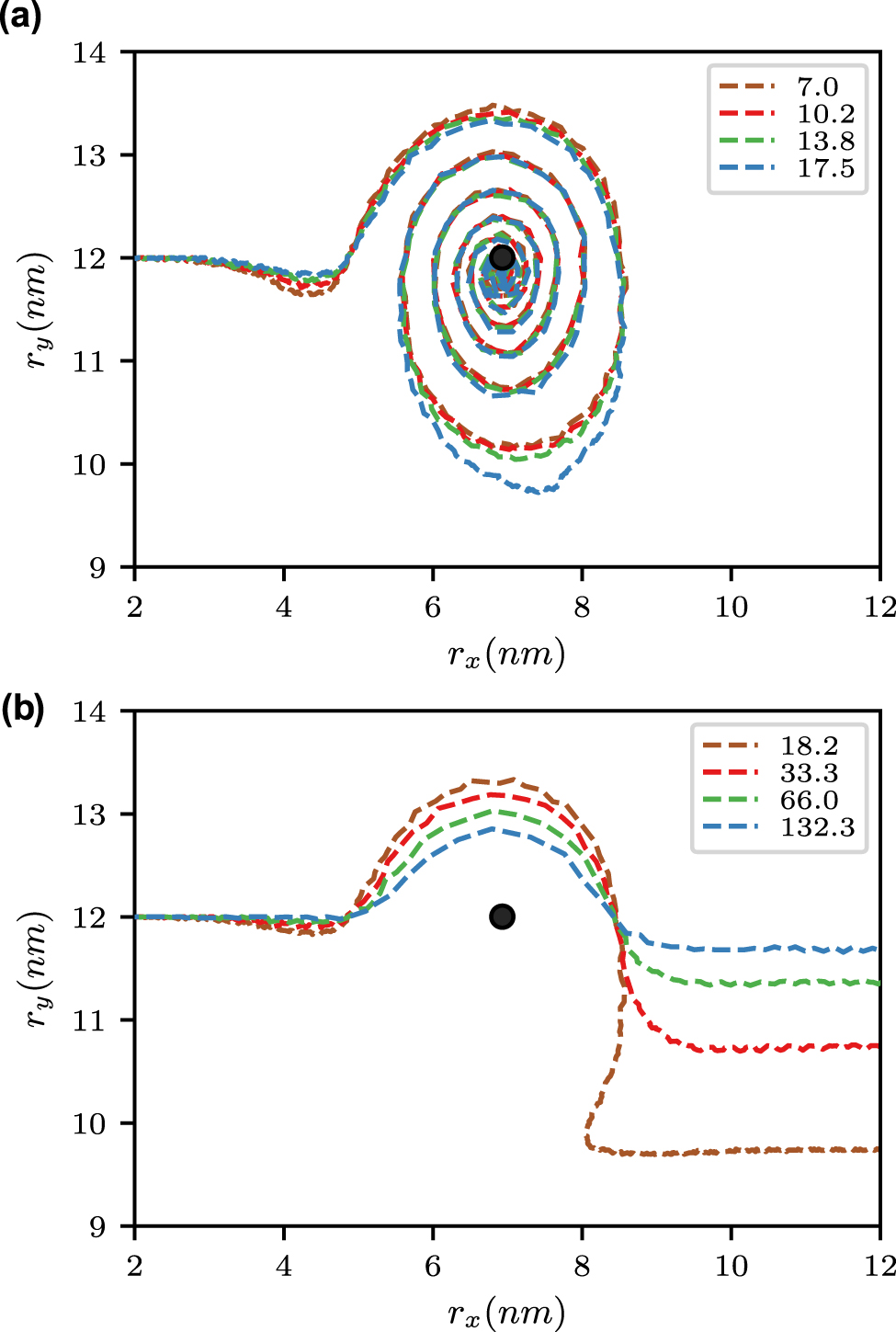}
\caption{
  Multiscale simulations of the trajectories of skyrmions scattering
  from a defect site consisting of a single atom \cite{Fernandes20}.
  The black circle indicates the position of the defect. (a) For
  low driving currents, the skyrmions are pinned. (b) For higher driving
  currents, the skyrmions can escape from the defect but their trajectories
  are deflected. The magnitude of the deflection decreases as the driving
  current increases.
Reprinted under CC license from I. L. Fernandes {\it et al.},
J. Phys.: Condens. Matter {\bf 32}, 425802 (2020).
}
\label{fig:fig.new.7}
\end{figure}

Fernandes {\it et al.} \cite{Fernandes20} used multi-scale simulations
to examine the deflections of skyrmions interacting with single atom defects
consisting of a Pd layer deposited on an Fe/Ir(111) surface.
At low driving currents in
Fig.~\ref{fig:fig.new.7}(a),
the trajectories indicate that the skyrmions become
trapped at the defect site, while in
Fig.~\ref{fig:fig.new.7}(b), the driving currents are high enough that the
skyrmions can escape from the defect site but experience a trajectory
deflection.
The magnitude of the deflection decreases as the skyrmion velocity increases.
If the disorder site is repulsive rather than attractive,
skyrmions are deflected in the opposite direction as they pass the defect.
This work also showed
that the Thiele equation approach is a reasonable approximation
for capturing the skyrmion dynamics. 

\begin{figure}
\includegraphics[width=\columnwidth]{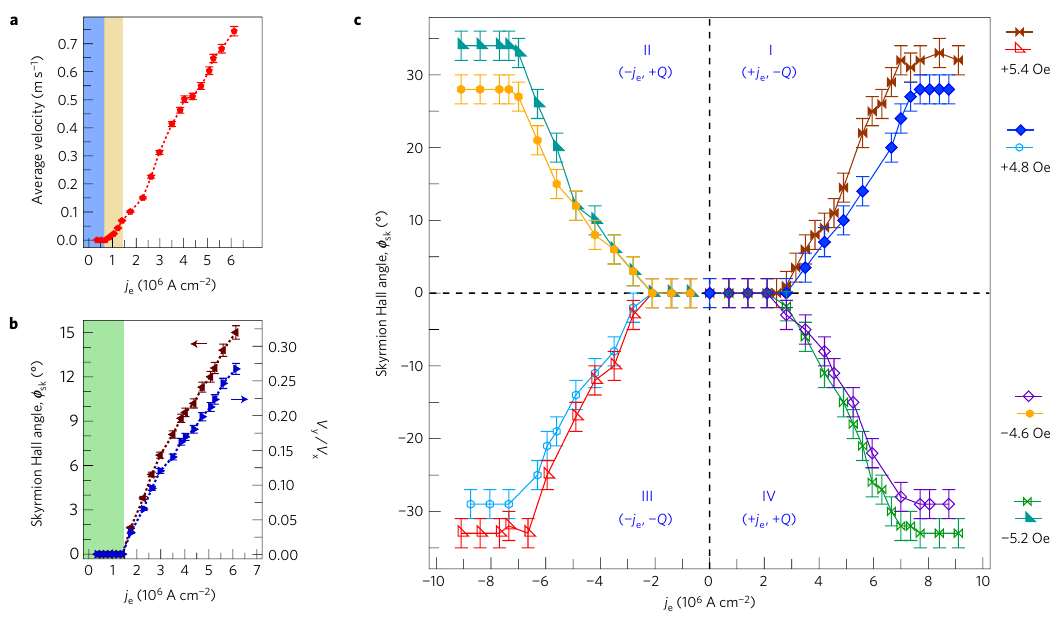}
\caption{
Skyrmion velocity and skyrmion Hall angle obtained from direct imaging of
the skyrmion motion \cite{Jiang17}.
(a) Average skyrmion velocity vs current density $j_e$ showing a pinned
regime of no motion (blue) and a region in which the skyrmions are moving
at zero skyrmion Hall angle (orange).
(b) The corresponding skyrmion Hall angle vs $j_e$.
(c) The skyrmion Hall angle for both positive and negative driving
currents $j_e$ under both positive and negative applied magnetic fields.
In each case, the skyrmion Hall angle saturates for sufficiently large
magnitudes of the driving current.
Reprinted by permission from: Springer Nature,
``Direct observation of the skyrmion Hall effect'',
Nature Phys. {\bf 13}, 162 (2017),
W. Jiang {\it et al.}, \copyright 2017.
}
\label{fig:47}
\end{figure}

Jiang {\it et al.} \cite{Jiang17} performed imaging experiments 
of current driven skyrmions in which they could
observe the drive dependence 
of the skyrmion Hall angle.
The skyrmion motion could not be detected 
by any changes in the topological Hall effect but was
instead deduced from the images.
As illustrated in Fig.~\ref{fig:47}.
four dynamical regimes appear:
a pinned state at low drives,
a state with finite skyrmion velocity but zero skyrmion
Hall angle,
a region in which
the skyrmion Hall angle increases linearly with drive,
and a high drive regime
in which the skyrmion Hall angle 
saturates to the clean limit of $\theta_{\rm SkH}=30^{\circ}$.
The general trend
is  very similar to what is observed
in the particle based simulations \cite{Reichhardt15,Reichhardt15a,Reichhardt16}.
It would be interesting to
consider a system in which the skyrmion Hall angle could be measured directly and
compared to a changing topological Hall effect, since both the
velocity of the skyrmion as well as the direction
of skyrmion motion need to be taken into account when
the magnitude of the topological Hall effect versus drive is measured.

\begin{figure}
\includegraphics[width=\columnwidth]{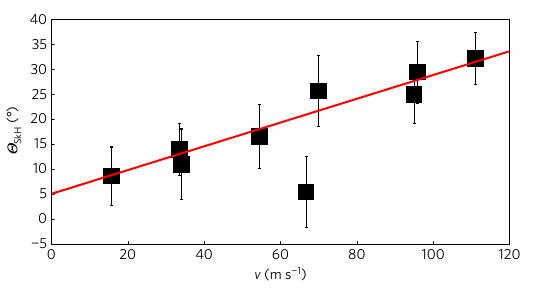}
\caption{
  Image-based experimental measurements of skyrmion Hall angle
  $\theta_{\rm SkH}$ versus
  skyrmion velocity $v$ \cite{Litzius17}, showing a linear dependence.
Reprinted by permission from: Springer Nature,
``Skyrmion Hall effect revealed by direct time-resolved X-ray microscopy'',
Nature Phys. {\bf 13}, 170 (2017),
K. Litzius {\it et al.}, \copyright 2017.
}
\label{fig:48}
\end{figure}

Litzius {\it et al.} \cite{Litzius17} studied skyrmions
under a pulsed drive in the forward and backward directions.
By obtaining
images as a function of
increasing drive amplitude,
they constructed
the 
skyrmion Hall angle
versus current curve shown in Fig.~\ref{fig:48}. 
Here,
$\theta_{\rm SkH}$ is initially low
and increases with increasing drive,
ultimately reaching a value close to $\theta_{\rm SkH} = 40^\circ$. 
Imaging experiments in ferromagnetic systems \cite{Woo18}  also show a similar increase
in the skyrmion Hall angle with drive.
Liztius {\it et al.} \cite{Litzius17} argued that the change of the 
skyrmion Hall angle was due to the ability of the
skyrmion to change
its shape or size with the applied current,
rather than the side jump effect
observed in the particle-based models. 
Using micromagnetic simulations,
Tomasello {\it et al.} found that breathing modes of moving skyrmions
excited by a current could lead to a
change in the skyrmion Hall angle as a function of drive
in the absence of pinning \cite{Tomasello18}. 
More recent studies by Litzius {\it et al.} provide evidence that there
can be a high current pinning dominated regime 
as well as another regime in which excitations
change the skyrmion Hall angle, 
giving rise to different scaling regimes as a function of drive \cite{Litzius20}. 
Current-driven studies of thin-film skyrmions in the 100 nm size range
at speeds of  up to 100 m/s
reveal a strong
dependence of the skyrmion Hall angle on drive,
with an increase 
in the skyrmion Hall angle to
a high velocity saturation value of $\theta_{\rm SkH}=55^\circ$
\cite{Juge19}.
The experimental observations match
the continuum modeling well,
showing a constant $\theta_{\rm SkH}$
in the absence of quenched disorder 
and both a finite depinning threshold and
an increase of $\theta_{\rm SkH}$ up to a saturation value
in the presence of pinning.
Although this work
showed that there were strong shape changes of the skyrmions due to the current
in the absence of disorder,
the authors argued that the changes in the skyrmion Hall angle were
due to the pinning rather than to the shape fluctuations.
Yu {\it et al.} \cite{Yu20} investigated the 
motion and skyrmion Hall effect
of individual and small clusters of 80 nm skyrmions
in FeGe systems with low currents
of $0.96 \times 10^9$ to $1.92 \times 10^9$ A m$^{-2}$.
They found an interesting effect
in which a skyrmion cluster can undergo
rotation as it translates. This
suggests that the Magnus force can induce unusual dynamics in
clusters of moving skyrmions.
Zhang {\it et al.} imaged the motion of half skyrmions,
which have a skyrmion Hall angle that is half as large as that
of a full skyrmion \cite{Zhang20}.
Other experiments found that shape distortions
of half skyrmions could further
reduce the skyrmion Hall angle \cite{Yang20}.

Most of the the experiments performed so far
have been in the single or few skyrmion limit,
so it would be interesting to understand 
what happens in the collective or lattice limit.
Beyond side jump effects, it may be possible
that the pinning is effectively increasing the damping
on the skyrmions through some other mechanism.
Since $\theta_{\rm SkH} \propto \tan^{-1}(\alpha_{m}/\alpha_{d})$,
if $\alpha_{d}$ is itself drive dependent, 
this could produce a drive dependence of $\theta_{\rm SkH}$.

\begin{figure}
\includegraphics[width=\columnwidth]{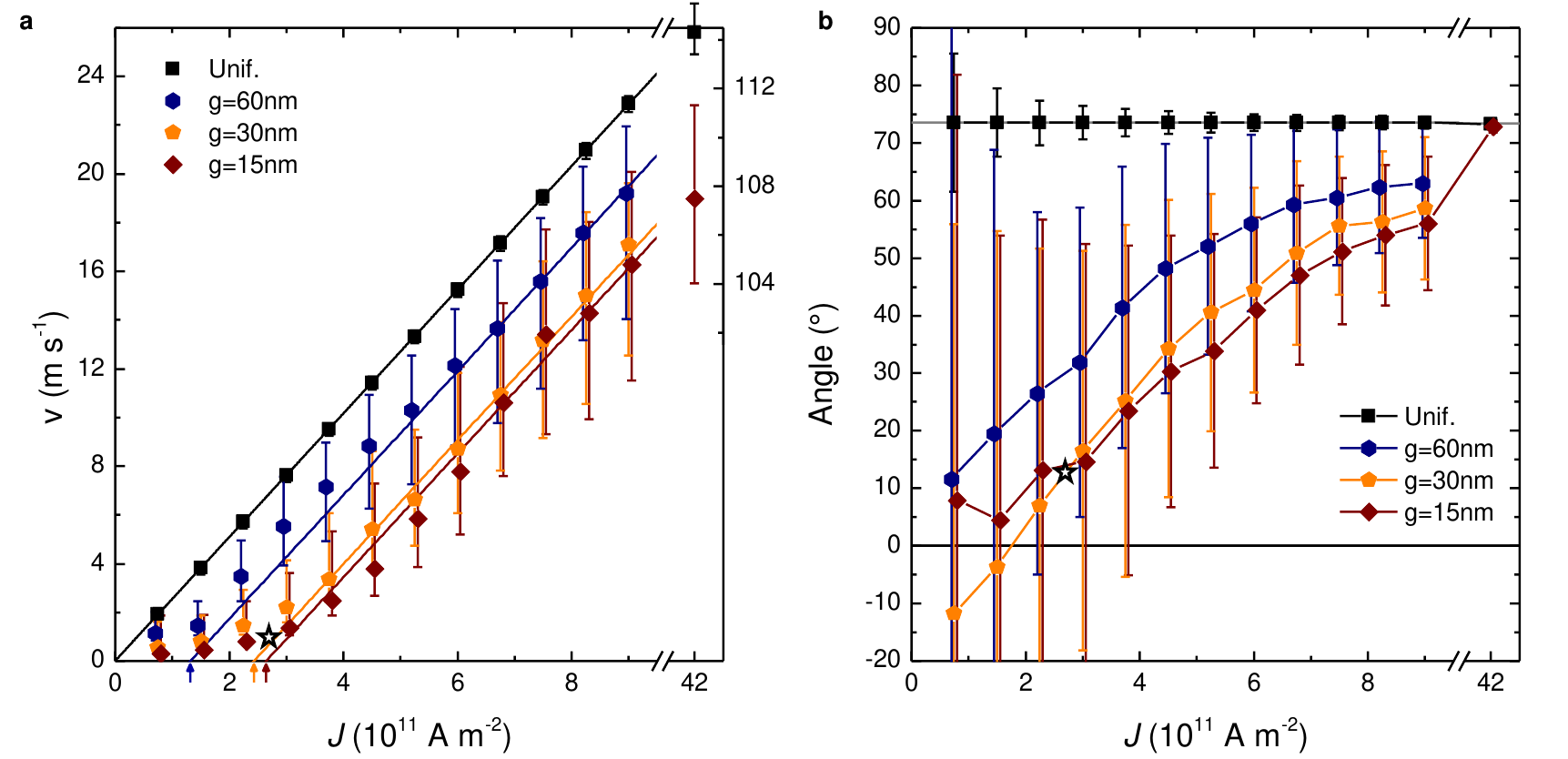}
\caption{
  Continuum simulations of skyrmion motion through a disordered landscape
  composed of grains of different sizes $g$ \cite{Legrand17}.
  (a) Mean skyrmion velocity $v$ vs driving current $J$ showing a finite
  depinning threshold. (b) Skyrmion Hall angle vs $J$ showing that the angle
  increases with increasing $J$ from a value near zero at zero current.
  Reprinted with permission from W. Legrand {\it et al.}, Nano Lett. {\bf 17},
  2703 (2017). Copyright 2017 American Chemical Society.
}
\label{fig:49}
\end{figure}

Several continuum-based simulations have shown a drive  dependence 
of the skyrmion Hall angle as a function of pinning  \cite{Legrand17,Kim17,Juge19}. 
Legrand {\it et al.} \cite{Legrand17} considered pinning produced by
grain boundaries, where small dense grains
correspond to strong pinning.
In this study, a clean system has no depinning threshold
and the skyrmion Hall angle is constant,
while when pinning is present,
there is a finite depinning threshold and the skyrmion
Hall angle is initially small and increases until reaching a saturation
value, as shown in Fig.~\ref{fig:49}.
This work also indicated that there is an optimal grain size for pinning,
meaning that the
relative size of the skyrmions and the pinning sites is important, 
which would be another interesting effect to study more fully.
The optimal grain size could be the result of
a resonance or commensuration
effect, where optimal pinning occurs
when the size of the pinning matches the size of the skyrmion.    
Due to the
limited number of skyrmions
simulated,
the $\theta_{\rm SkH}$ versus drive curves 
contain considerable scatting.
It may be possible that there
are multiple regimes for $\theta_{\rm SkH}$
rather than only a linearly increasing regime and a saturation
regime, which offers another
avenue for future study. 
Numerical work by Juge {\it et al.} \cite{Juge19} produced
results similar to those of Legrand {\it et al.} \cite{Legrand17},
but the scattering
in the data was much smaller.
In these works,
the skyrmion trajectories in regimes with increasing
$\theta_{\rm SkH}$
are similar to
those observed in particle based simulations \cite{Reichhardt16}, 
with a coexistence of pinned 
and moving skyrmions.
Similar dynamics were
observed in the imaging experiments of Montoya {\it et al.} \cite{Montoya18}.
In the continuum simulations, at higher
drives the skyrmions moved in fairly straight trajectories
along a direction close to that of the intrinsic Hall angle \cite{Legrand17}.  
Kim {\it et al.} \cite{Kim17}
performed continuum simulations that
also showed a similar drive dependence of the skyrmion 
Hall angle.

\begin{figure}
\includegraphics[width=\columnwidth]{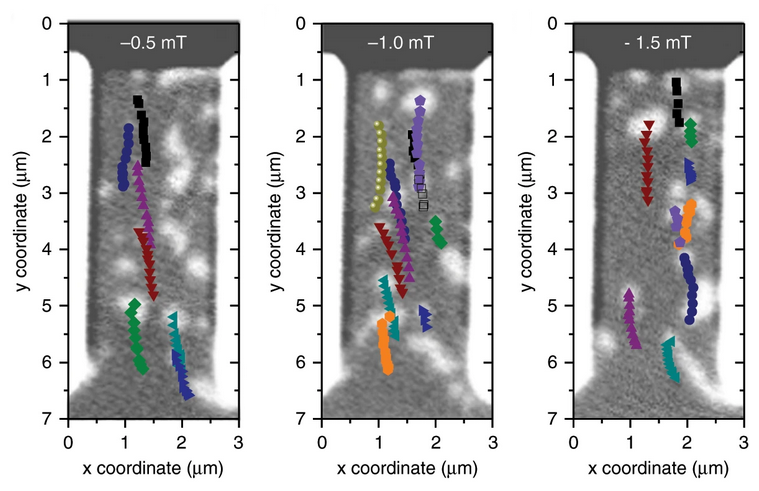}
\caption{
Images of skyrmion motion in a multilayer system at varied magnetic
fields \cite{Zeissler20}.
In each case, the skyrmion Hall angle is close to $\theta_{SkH}=10^\circ$.
As the field is varied, the size of the skyrmion changes, so this result
indicates that the skyrmion Hall angle is independent of the skyrmion
diameter.
Reprinted under CC license from K. Zeissler {\it et al.}, Nature Commun.
{\bf 11}, 428 (2020).
}
\label{fig:fig.new.5}
\end{figure}

Another question is the role of the skyrmion diameter in determining
the skyrmion Hall angle.
Zeissler {\it et al.} \cite{Zeissler20}
examined skyrmions in a magnetic multilayer under a pulsed drive and
found that the skyrmion Hall angle was close to $10^\circ$ and was
independent of the diameter of the skyrmions.
Figure~\ref{fig:fig.new.5}
shows some of the skyrmion trajectories at varied fields,
where the skyrmion diameter increases as the magnitude of the
magnetic field increases but the direction of motion does not change.
This work also revealed that the skyrmion trajectories are deflected by
contact with disorder sites.
In this case it may be possible
that the disorder length scale is much larger
than the skyrmion diameters, placing the system in 
a pinning dominated regime \cite{Reichhardt20}.
It would be interesting to perform a separate
study of the skyrmion Hall angle for varied 
disorder sizes to see if a change occurs
when the effective pinning diameter becomes smaller rather than larger
than the size of the skyrmions.

There have been numerous studies of skyrmions
moving in samples with magnetic grain boundaries 
which
show that in some cases,
the disorder can enhance the skyrmion Hall angle \cite{Salimath19}.
This occurs when the grains are magnetically aligned
in the direction in which
the skyrmions would move in the absence of disorder,
creating a guidance effect in the direction of the
intrinsic skyrmion Hall angle.
This effect depends on the magnitude of
the drive and the orientation of the grains,
but it suggests that it would be possible to control the skyrmion Hall angle
through the proper orientation
of extended defects.

The drive dependence of the skyrmion Hall angle
indicates that a wealth of new dynamical effects can arise that are distinct from those
found in previously studied overdamped systems.
For example, 
when a skyrmion
is driven over a periodic pinning array,
the skyrmion Hall angle increases with drive but becomes quantized 
due to locking with certain symmetry angles of the
periodic substrate \cite{Reichhardt15}.
Until now, the modification of the skyrmion Hall angle by pinning has 
only been studied for ferromagnetic skyrmions, 
so it would be interesting to study antiferromagnetic skyrmions,
polar skyrmions, skyrmionium, antiskyrmions, and
merons to see whether the effect of pinning differs depending on the nature
of the skyrmion.

The antiferromagnetic
skyrmion would be particularly interesting
since it should have $\theta_{\rm SkH}=0$ and,
in principle, its dynamics would be very similar to
those of vortices in superconductors.  If the
antiferromagnetic skyrmion
is interacting with magnetic defects or magnetic pinning,
then a finite skyrmion Hall effect
could arise due to a side jump effect.
It would also be interesting to see whether the lack of a Magnus force
would lead to stronger pinning effects
compared to ferromagnetic skyrmions. 
The skyrmion Hall angle can also be controlled
using various other methods, such as internal modes
\cite{Tomasello18,Chen19}, that can change and even vanish 
at the angular momentum compensation temperature \cite{Hirata19}, or by  
applying particular configurations of gate voltages \cite{Plettenberg20}.
In such cases,
pinning would still play a role in the dynamics, and this area is open for
further investigation.

\subsection{Thermal Effects}

\begin{figure}
\includegraphics[width=\columnwidth]{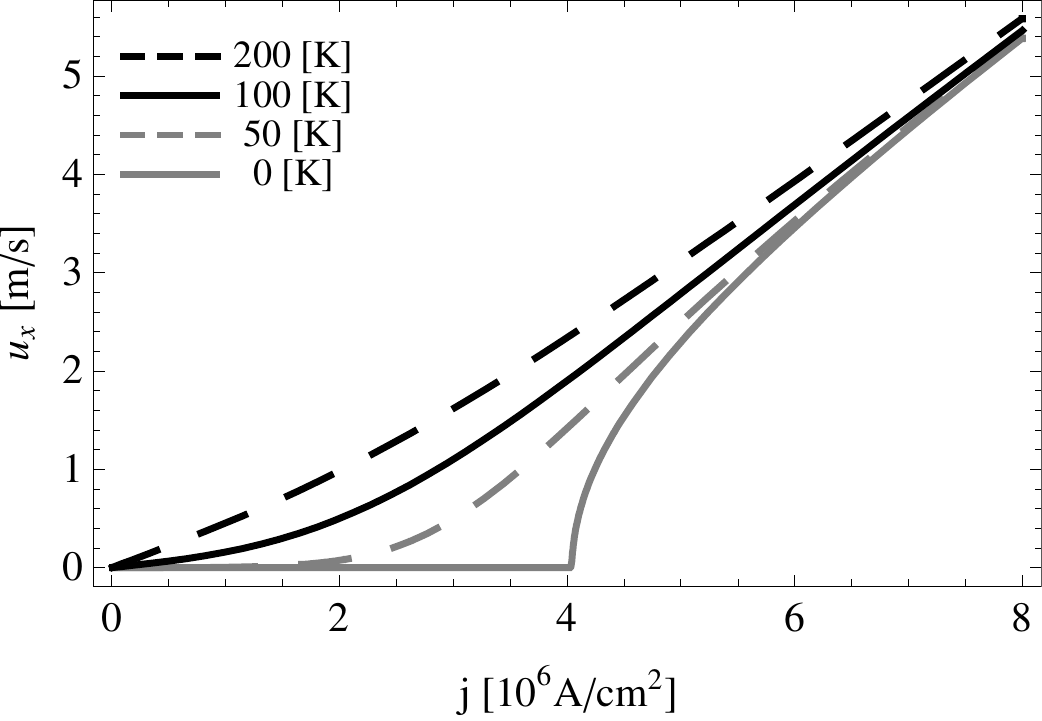}
\includegraphics[width=\columnwidth]{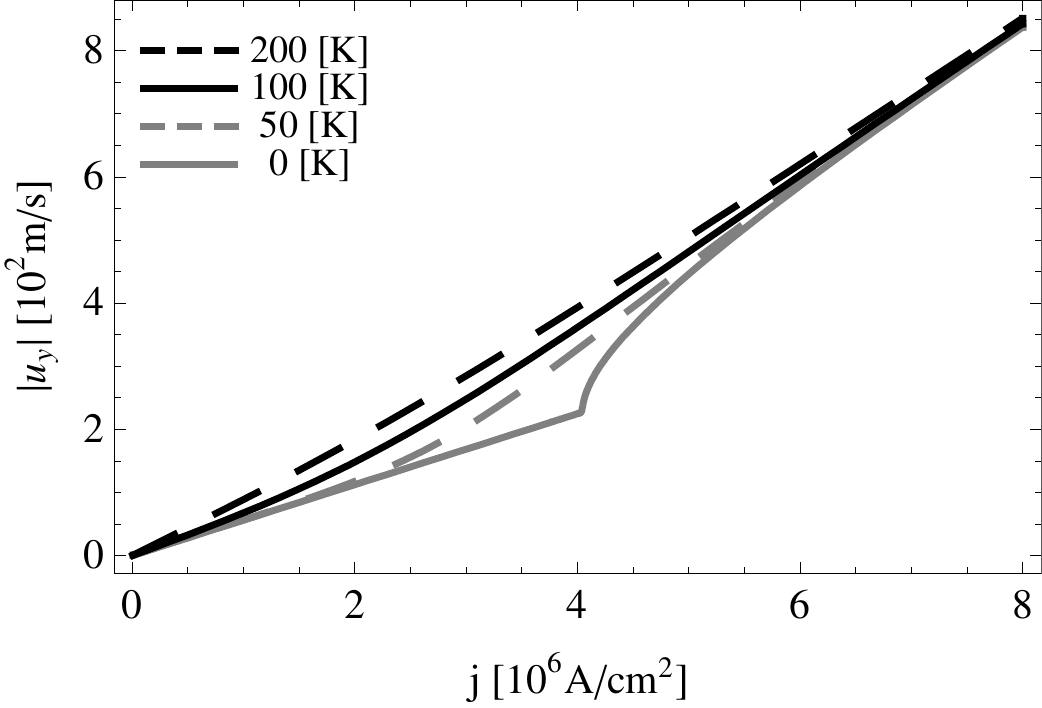}
\caption{
  Theoretical predictions for skyrmion velocity response at different
  temperatures, showing the existence of a creep regime
  below the zero-temperature depinning threshold \cite{Troncoso14}.
  Top panel: longitudinal velocity $u_x$ vs driving current $j$;
  bottom panel: transverse velocity $u_y$ vs driving current $j$.
Reprinted with permission from R. E. Troncoso and A. S. N{\' u}{\~ n}ez,
Phys. Rev. B {\bf 89}, 224403 (2014). Copyright 2014
by the American Physical Society.
}
\label{fig:50}
\end{figure}

\begin{figure}
\includegraphics[width=\columnwidth]{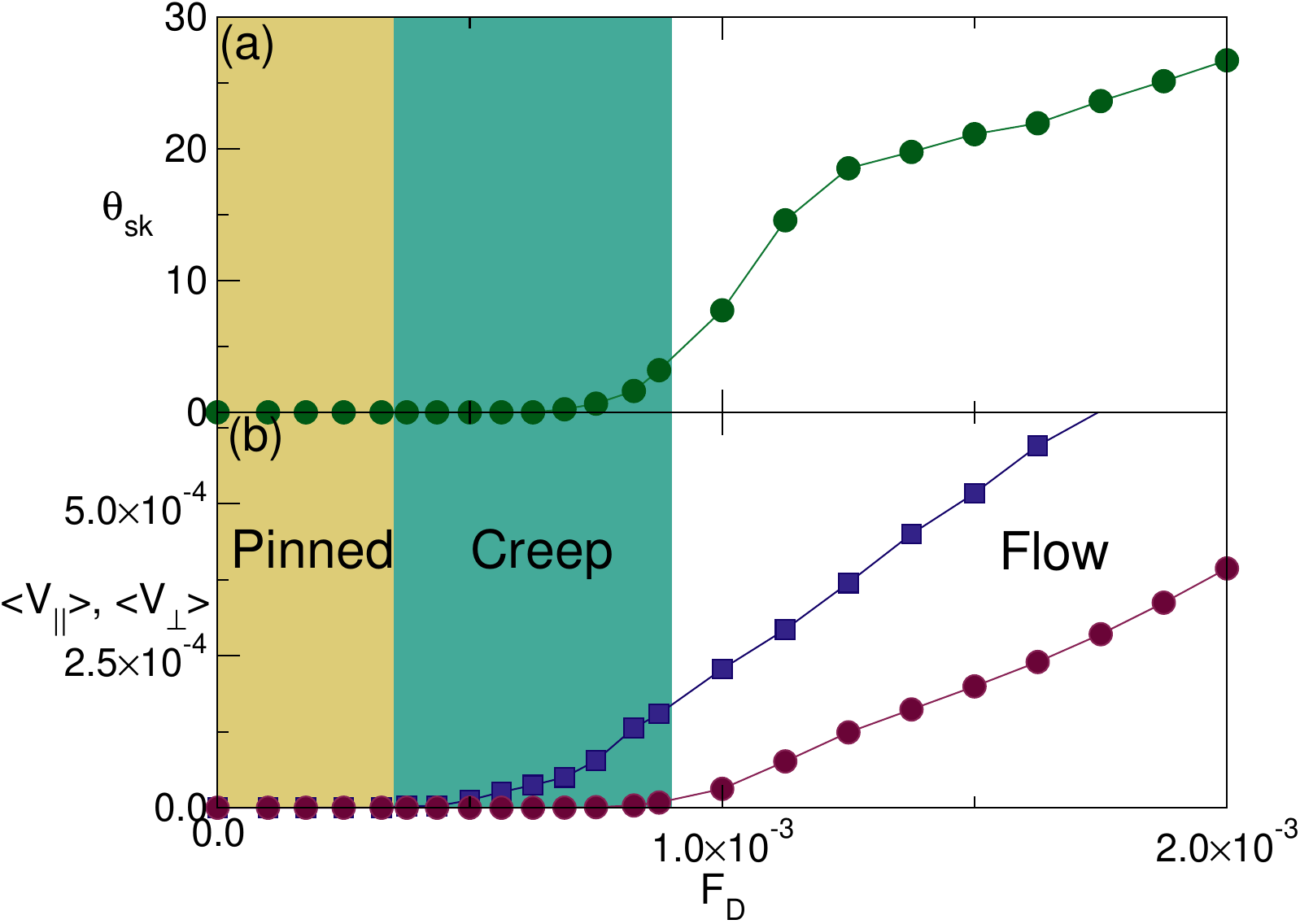}
\caption{
Particle-based simulations of skyrmion motion with finite thermal
fluctuations \cite{Reichhardt19}.  
(a) The skyrmion Hall angle $\theta_{\rm sk}$ vs driving force $F_{D}$.
(b)
The corresponding skyrmion velocity parallel
$\langle V_{||}\rangle$ (blue squares)
and perpendicular $\langle V_{\perp}\rangle$ (red circles)
to the drive vs $F_{D}$.
There is a pinned phase (yellow), a creep phase in which $\theta_{\rm sk}$ is
close to zero (green), and a flowing phase.
Republished with permission of IOP Publishing, Ltd, from
``Thermal creep and the skyrmion Hall angle in driven skyrmion crystals'',
C. Reichhardt and C. J. O. Reichhardt, J. Phys.: Condens. Matter
{\bf 31}, 07LT01
(2019); permission conveyed through Copyright Clearance Center, Inc.
}
\label{fig:52} 
\end{figure}

Experimental observations of
the skyrmion Hall effect have generally been
performed at room temperature, and there are numerous indications 
that skyrmions can exhibit thermal effects such as
Brownian motion 
\cite{Zazvorka18,Nozaki19,Zhao19}, which would
make creep effects and thermally activated hopping
from one pinning well to the next an important process.
It is interesting to ask how 
the depinning threshold and skyrmion Hall angle
behave under the combination of both pinning
and temperature. 
Troncoso and N{\' u}{\~ n}ez \cite{Troncoso14} theoretically studied thermally assisted current driven skyrmion motion 
in the presence of pinning,
and found that the Brownian motion could be described by a stochastic Thiele equation.
They observed a finite
depinning threshold at zero temperature as well as a creep regime
for increasing drive, as shown in Fig.~\ref{fig:50}. 
Reichhardt {\it et al.} \cite{Reichhardt19}
studied the depinning of skyrmions
in random disorder for increasing thermal effects
in the elastic depinning regime, where the skyrmions maintain
the same neighbors. 
At $T = 0$ there is a well defined depinning threshold, while for increasing 
temperature, the depinning threshold decreases and becomes more rounded. The motion
in this case can be divided into a pinned phase, an intermittent or thermally activated
avalanche phase,
and a high drive continuously moving phase.
Figure~\ref{fig:52}(a) illustrates
the measured skyrmion Hall angle $\theta_{\rm SkH}$ versus drive for
a system
at finite temperature
where there is appreciable creep,
and Fig.~\ref{fig:52}(b) shows
$\langle V_{||}\rangle$ and
$\langle V_{\perp}\rangle$ versus $F_{D}$.
There is
a pinned regime with
$\langle V_{||}\rangle = \langle V_{\perp}\rangle = 0$,
a creep regime with finite $\langle V_{||}\rangle$ and
$\langle V_{\perp}\rangle = 0$, giving a skyrmion Hall angle of zero, and 
a flow regime in which the velocity is finite in both directions.
In the latter region,
the skyrmion Hall angle increases with drive
and eventually saturates at the high drive limit.
The appearance of a regime in which
there is a finite longitudinal velocity
but zero perpendicular velocity is also consistent with 
the observations of Jiang {\it et al.} just above depinning \cite{Jiang17}.

\begin{figure}
\includegraphics[width=\columnwidth]{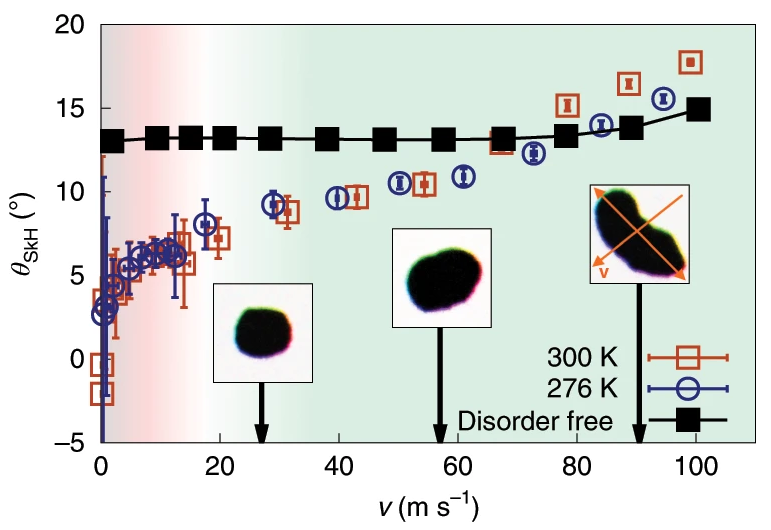}
\caption{
Continuum simulations of the skyrmion Hall angle $\theta_{\rm SkH}$ versus
skyrmion velocity $v$ in a sample with no thermal disorder
(black squares) and at two different finite temperatures (open symbols).
$\theta_{\rm SkH}$ is nearly
independent of velocity in the absence of temperature,
but when thermal fluctuations are present, $\theta_{\rm SkH}$ increases with
increasing velocity. The insets show the change in skyrmion shape from nearly
circular at low velocities to strongly distorted at high velocities.
Reprinted by permission from: Springer Nature,
``The role of temperature and drive current in skyrmion dynamics,''
Nature Electron. {\bf 3}, 30 (2020),
K. Litzius {\it et al.}, \copyright 2020.
}
\label{fig:fig.new.6}
\end{figure}

The behavior of the skyrmion Hall angle with creep suggests that
there could be multiple regimes for 
the evolution of $\theta_{\rm SkH}$ with current and velocity.
Litzius {\it et al.}
\cite{Litzius20}
studied
the impact of thermal 
fluctuations on $\theta_{\rm SkH}$
in both experiment and simulations,
and found
distinct behaviors
in the low and high current regimes.
At lower currents and velocities,
$\theta_{\rm SkH}$ increases rapidly with current,
and at higher drives
there is a crossover to a slower increase with current.
It was argued that at
low drives, the skyrmion behaves more like a particle so that
$\theta_{\rm SkH}$
is controlled by the thermal disorder,
whereas at higher drives, the internal degrees of freedom of the skyrmion
become important and
$\theta_{\rm SkH}$ is
controlled by the distortions or changes of shape of the skyrmions.
As shown in Fig.~\ref{fig:fig.new.6},
where
the skyrmion Hall angle $\theta_{\rm SkH}$ is plotted as a function of
the skyrmion velocity $v$ \cite{Litzius20},
the results of continuum-based simulations are consistent with experiment.
In the absence of thermal disorder,
there is very little change in $\theta_{\rm SkH}$ with velocity except at
the highest values of $v$, where $\theta_{\rm SkH}$ increases slightly.
When thermal disorder is present, there is a sharp increase
in $\theta_{\rm SkH}$ at low velocities rolling over to a more gradual increase
at higher velocities.
The images in the insets of Fig.~\ref{fig:fig.new.6} indicate that
the skyrmion becomes more distorted in shape as the velocity increases.
MacKinnon {\it et al.} \cite{MacKinnon20}
examined the role of additional interfacial spin transfer torques
on driven skyrmion motion
and found that it can strongly reduce the skyrmion Hall angle
for skyrmion diameters that are less than 100 nm.
They also observed that when disorder is present,
$\theta_{\rm SkH}$ 
increases rapidly at low velocities and then
increases more slowly or saturates at high velocities.

There have also been numerical
studies examining the shape distortions of skyrmions at higher
drives \cite{Masell20},
where the skyrmions can start developing a non-circular shape with a 
tail. Here there is a critical current above which the skyrmion
becomes unstable.
It is possible that for
dense skyrmion lattices at higher currents,
lattice transitions
could occur due to the shape changes of the skyrmions, which
could cause the skyrmion-skyrmion interactions to become
more anisotropic. 

\subsection{Future Directions}

It would be interesting to examine the evolution of the skyrmion Hall angle for
different types of pinning,
such as short range versus long range, repulsive versus attractive,
or grain boundary and extended pinning versus point pinning.
Since the skyrmion Hall angle is often considered detrimental to
applications, it may be possible to identify
pinning or defect arrangements that reduce
$\theta_{\rm SkH}$,
or there might even be ways
to exploit the behavior of $\theta_{\rm SkH}$ to
create new devices.
The skyrmion Hall angle 
can also depend strongly 
on the skyrmion type.
For example,
in certain regimes $\theta_{\rm SkH}$ is affected by
the direction of the applied current with respect to
an anisotropy direction contained within the skyrmion itself, meaning that
due to their anisotropy, antiskyrmions could have a rich behavior
under a drive in the presence of pinning \cite{Kovalev18}.
Most
studies have
been performed using dc drives,
but it would also be possible to add
a high frequency ac drive component
which could create breathing modes
that might reduce the pinning, increase the creep, or change the
skyrmion Hall angle. 

Existing studies of
pinning effects and the dynamical motion of skyrmions
have focused on 2D systems; however, there should be a
variety of interesting new effects in 3D systems.
Line-like skyrmions
could undergo 
elastic depinning of the type found for stringlike objects,
but could have distinct modes of motion 
along the length of the line.
There have already been several studies
of
the scaling of certain 
modes in 3D skyrmions \cite{Li19,Seki20}.
It would also be possible to
study the roughening transition of the skyrmion
lines near depinning and
to analyze whether the skyrmions become more
stringlike at higher drives based on changes in the
fractal dimension.
The skyrmions might
maintain
their linelike nature but become entangled,
similar to entangled vortex states,
and the entangled skyrmions could cut themselves free or
be unable to cut and remain tangled.

Skyrmion dynamic phases and the evolution of the skyrmion Hall angle
have been studied for drives arising from an applied current.
It would useful to understand
whether similar or different effects
occur
for skyrmions
subjected to different types of driving, such as thermal gradients or 
magnetic gradients.  
Existing studies have focused on uniform drives;
however, introduction of 
nonuniform drives could produce
interesting effects
due to the velocity dependence of the skyrmion Hall angle.
A system with a
non-uniform current could
exhibit clustering or
other types of new effects not found in overdamped systems. 

\begin{figure}
\includegraphics[width=\columnwidth]{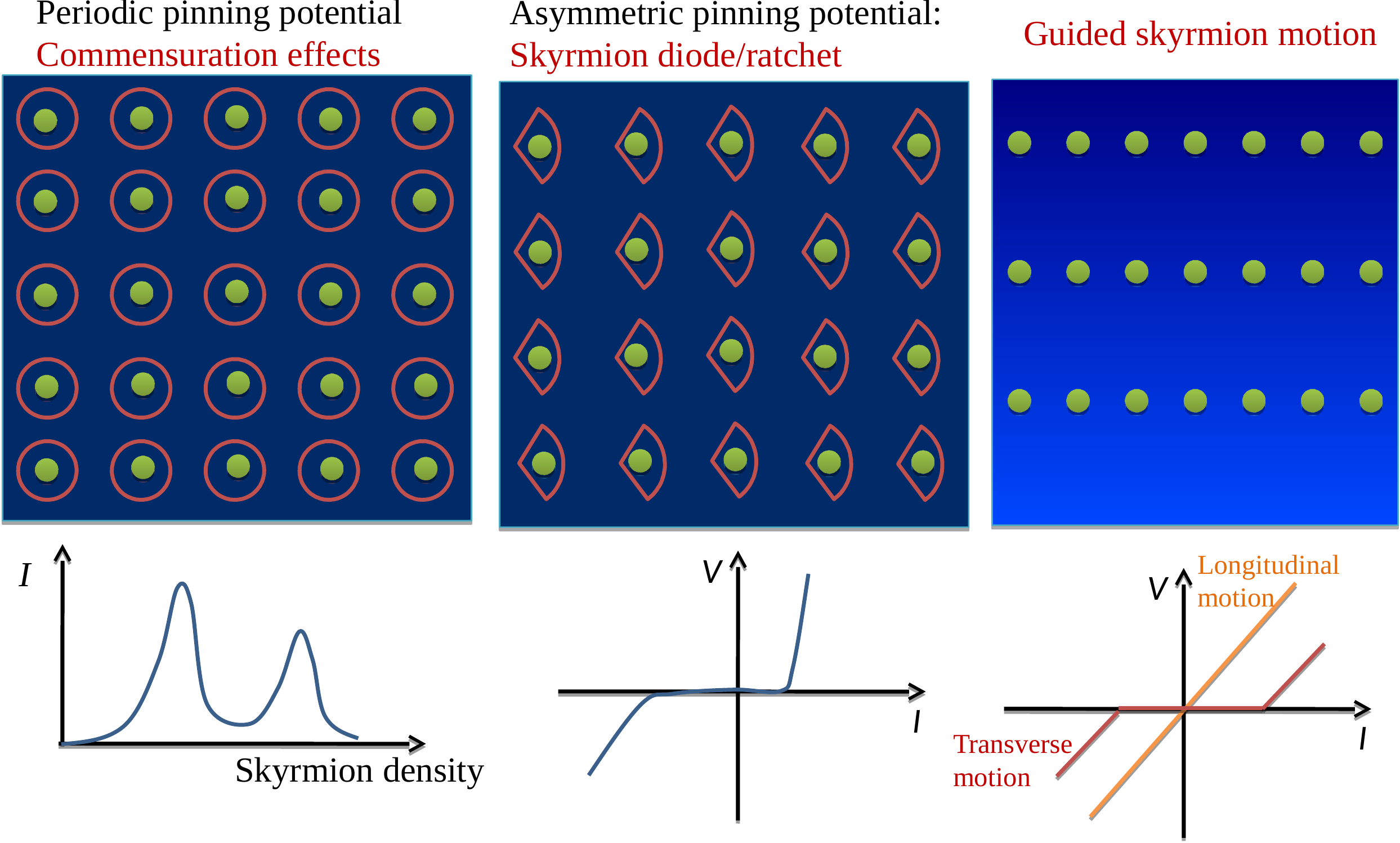}
\caption{
Examples of skyrmions interacting with
different types of nanostructured pinning.  
Left: 2D periodic pinning, where commensuration effects
between the number of skyrmions and the
number of pinning sites can occur.
Center: asymmetric 2D periodic pinning which can
generate
ratchet and diode effects.
Right: 1D periodic pinning.
The lower panels show schematic transport curves that could be observed with
each pinning geometry.
}
\label{fig:53}
\end{figure}
 
\section{Nanostructured and Periodic Landscapes}

There are already a number of proposals for using skyrmions in race
track devices,
which are highly confined geometries for the skyrmion.
Another route 
for generating controlled skyrmion motion is to fabricate
nanostructured pinning arrays, similar to   
those employed for vortices in type-II superconductors
\cite{Reichhardt17,Martin97,Baert95,Harada96,Reichhardt98,Berdiyorov06}, 
vortices in Bose-Einstein condensates with optical traps
\cite{Reijnders04,Tung06}, cold atoms \cite{Buchler03,Benassi11} 
and colloidal particles \cite{Wei98,Brunner02}.
In these systems, the particles can interact with
1D periodic substrates \cite{Reichhardt01,Reijnders04,Wei98,Martinoli75,Dobrovolskiy15,LeThien16}, 
2D square  \cite{Baert95,Harada96,Reichhardt98,Berdiyorov06,Tung06,Bohlein12,Reichhardt02a}, triangular \cite{Reichhardt98,Brunner02,Reichhardt02a},
or quasicrystalline \cite{Kemmler06,Villegas06,Mikhael08}
substrates, or
arrangements with geometric frustration \cite{Libal06,OrtizAmbriz16,Libal09,Latimer13,Wang18a}.
Figure~\ref{fig:53} illustrates three such possible pinning geometries,
including a 2D periodic array of
trapping sites, 
a periodic 1D array, and an asymmetric 2D array
that can generate diode or ratchet effects.     

For assemblies of particles interacting
with either 1D or 2D periodic substrates,
commensuration effects \cite{Bak82} can occur in which 
the periodicity of the lattice matches with
the periodicity of the substrate.
The 
system exhibits strong pinning
under these commensurate conditions
since the particle-particle interaction forces
cancel via symmetry and the entire ensemble of particles behaves similarly
to 
an isolated particle, 
with no additional stress or strain in the lattice caused by defects.  
If, however, there is some lattice mismatch or
an incommensuration is present,
then collective interactions between the particles become important.
For example, at a particle density that is
slightly above commensuration, most of the 
particles are located at the potential energy minima of the substrate just as
in the commensurate situation,
but due to the incommensuration,
a small number of particles
are located
on higher energy portions of the substrate.
Under an applied drive, these
extra particles or kinks depin first at $F_{c1}$,
while the rest of the particles depin at
a higher drive $F_{c2}$,
producing a two step or even multiple step
depinning phenomenon
\cite{Bak82,Benassi11,Bohlein12,Reichhardt97,Gutierrez09,Avci10}. 
A similar effect occurs just below commensuration,
where there are vacancies or anti-kinks that depin first \cite{Bohlein12}.
The commensurate condition is met whenever the number of
particles $p$ is
an integer multiple of the number of
substrate potential
minima $q$,
$p/q = 1,2\ldots N$.
At these integer matching fillings, there is
a local maximum in 
the depinning
threshold $F_{c}$ \cite{Baert95,Reichhardt98,Berdiyorov06,Reichhardt97}. 
There can also be fractional commensuration effects
at fillings such as $p/q = 1/2$ or $1/3$, and signatures of these
fractional fillings depend
on the symmetry of the underlying lattice \cite{Bak82,Grigorenko03}.
In
quasiperiodic or frustrated substrates, other types of commensuration 
effects can arise at
integer and non-integer matchings \cite{Kemmler06,Villegas06,Latimer13,Wang18a}.
Under an applied drive,
these systems can exhibit a rich variety of dynamical behaviors with
well defined transitions between different kinds of plastic flow,
turbulent flow, and ordered flow,
and the extent and number of phases
depends on the commensurability, pinning strength, and direction of 
drive with respect to the periodicity of the substrate 
\cite{Harada96,Benassi11,Bohlein12,Martinoli75,Dobrovolskiy15,LeThien16,Wang18a,Reichhardt97,Gutierrez09,Avci10,Juniper15,Reichhardt11,Bohlein12a}.  

Due to their particle-like nature, skyrmions are ideal candidates
for studying commensurate and incommensurate
effects on various types of substrate geometries.  
The interaction of skyrmions with periodic pinning could be
potentially useful
for creating new types of devices.
Additionally,
the Magnus force and the
internal degrees of freedom could cause skyrmions
to exhibit a variety of new types of
static and dynamic commensurate phases
which are distinct from those found for overdamped systems.   

\subsection{One Dimensional Periodic Substrates and Speed-Up Effects}

\begin{figure}
\includegraphics[width=\columnwidth]{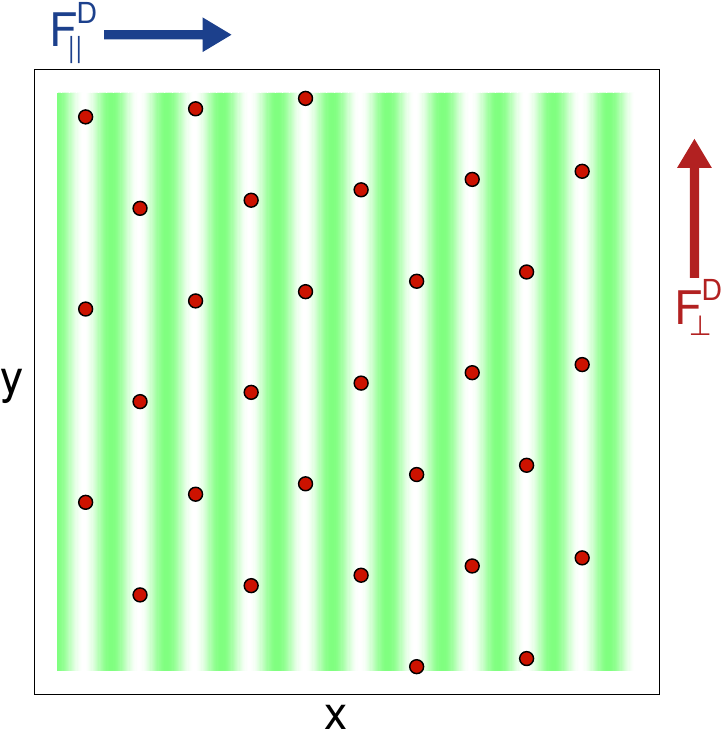}
\caption{
  An example of a periodic
  quasi-one-dimensional substrate for skyrmions
\cite{Reichhardt16a}.
The substrate is sinusoidal along the x direction. 
White areas indicate substrate minima 
while substrate maxima are colored green.
The skyrmions, represented by red dots, can be driven either
parallel to the substrate periodicity by
$F_{||}^{D}$ (blue arrow), or perpendicular to the substrate
periodicity by $F_{\perp}^D$ (red arrow).
Reprinted with permission from C. Reichhardt {\it et al.}, Phys. Rev. B
{\bf 94}, 094413 (2016). Copyright 2016 by the American Physical Society.
}
\label{fig:54}
\end{figure}

\begin{figure}
\includegraphics[width=\columnwidth]{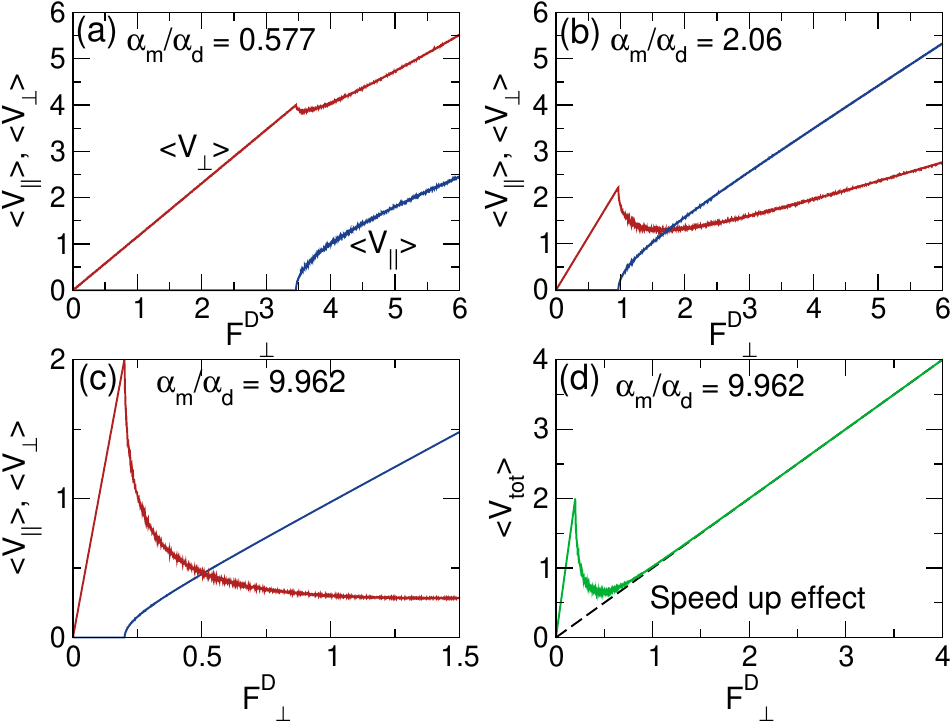}
\caption{
  Illustration of the speed up effect from
particle-based simulations of skyrmion velocities parallel
$\langle V_{||}\rangle$ (blue) and perpendicular
$\langle V_{\perp}\rangle$ (red) to
the substrate periodicity direction for perpendicular driving
$F^D_{\perp}$ in the quasi-1D potential illustrated in Fig.~\ref{fig:54}
\cite{Reichhardt16a}.
(a) At $\theta_{\rm SkH}^{\rm int}=30^\circ$,
the initial skyrmion motion is locked in the perpendicular direction.
There is a drop in $\langle V_\perp\rangle$ at the
critical drive $F_c^{||}$ for
the onset of motion in the parallel direction.
At (b) $\theta_{\rm SkH}^{\rm int}=64^\circ$ and (c)
$\theta_{\rm SkH}^{\rm int}=84.3^\circ$, $F_c^{||}$ shifts to lower drives and
the drop in $\langle V_\perp\rangle$ becomes more pronounced.
(d) The total velocity
$\langle V_{\rm tot}\rangle$ vs $F^D_{\perp}$ at $\theta_{\rm SkH}^{\rm int}=84.3^\circ$.
The dashed line indicates the response
$\langle V_0\rangle$ expected in a system with no substrate.
In the speed up effect,
$\langle V_{\rm tot}\rangle > \langle V_0\rangle$.
Reprinted with permission from C. Reichhardt {\it et al.}, Phys. Rev. B
{\bf 94}, 094413 (2016). Copyright 2016 by the American Physical Society.
}
\label{fig:55}
\end{figure}

We first consider the simplest example
consisting of a skyrmion interacting with
the
1D pinning array illustrated
in Fig.~\ref{fig:54}.
The external
driving can be applied 
either parallel or perpendicular
to the
substrate periodicity,
and the system has very different dynamical responses depending on the
driving direction.
An overdamped system would only have
a finite depinning threshold $F_{c}$ when the driving is parallel
to the substrate periodicity, while driving in the perpendicular direction would
simply cause the particles to
slide along the potential minima. 
This situation changes for
skyrmions
with a finite Magnus force, which
move at an angle with respect to the drive.
The skyrmions exhibit
a finite parallel depinning threshold even when the driving is
perpendicular to the substrate periodicity.
Reichhardt and Olson Reichhardt \cite{Reichhardt16a}
used a 2D particle based simulation to study
skyrmions interacting with 
a periodic 1D substrate,
and found that for driving parallel to the substrate periodicity direction,
the critical depinning force $F_{c}$ is independent of 
the ratio of the Magnus force to the damping strength.
This is in contrast to the case of
random point pinning,
where $F_{c}$ decreases with increasing Magnus force.
The skyrmions
are able to skirt around pointlike pinning sites,
but they cannot avoid passing through a
1D extended pinning site.
When the drive
is
perpendicular
to the substrate periodicity direction,
there is no finite depinning threshold and
the skyrmions
initially move only in the perpendicular direction with
$\theta_{\rm SkH}=0$.
As the drive increases, the
Magnus force on the skyrmions
in the direction
parallel to the substrate periodicity increases
until,
above a critical drive,
the skyrmions begin to jump
over the barriers and move in both the parallel and perpendicular directions.
A drive that is perpendicular
to the substrate symmetry direction produces a situation similar to that
of a skyrmion in a thin race track,
which
moves toward the edge
of the track due to the Magnus
force and leaves the track completely above a critical velocity.
In the case of the 1D periodic substrate in a 2D sample,
the skyrmion hops into the next potential minimum
when the critical velocity is exceeded.

Figure~\ref{fig:55} shows the skyrmion velocity-force curves for
driving in the direction perpendicular to the substrate periodicity
in the system from Fig.~\ref{fig:54} \cite{Reichhardt16a}.
In Fig.~\ref{fig:55}(a), a system with an intrinsic Hall angle of
$\theta_{\rm SkH}^{\rm int}=30^{\circ}$ has
a finite depinning threshold $F^{||}_c$ for
motion in the parallel direction, and
for $0 < F_{D} < F^{||}_c$
the skyrmion motion is locked along the perpendicular
direction, giving $\theta_{\rm SkH}=0$.
For $F_{D} > F^{||}_c$,
the skyrmion begins to move in both directions, and
the onset of finite $\langle V_{||}\rangle$ is accompanied
by a decrease in
$\langle V_{\perp}\rangle$.
In Fig.~\ref{fig:55}(b) and (c),
systems with
$\theta^{\rm int}_{\rm SkH} = 64^\circ$ and $84.3^\circ$, respectively, show
that $F^{||}_c$ drops to lower drives
with increasing $\theta^{\rm int}_{\rm SkH}$, 
while the drop in
$\langle V_{\perp}\rangle$ at $F^{||}_{c}$ becomes more pronounced.
Figure~\ref{fig:55}(d) illustrates the net skyrmion velocity
$\langle V\rangle = (\langle V_{\perp}\rangle^2 + \langle V_{||}\rangle^2)^{1/2}$
versus $F^{D}_{\perp}$
for a sample with $\theta^{\rm int}_{\rm SkH} = 84.3^\circ$, along with a
dashed line indicating the velocity 
$\langle V_0\rangle$ 
expected in the absence of a substrate. 
A pinning-induced speed up effect
appears near $F_c^{||}$ in which
$\langle V\rangle > \langle V_0\rangle$, meaning that
the skyrmion is moving {\it faster} than it would
if the substrate were not present.
This speed up effect,
which does not occur in overdamped systems, is produced
by a combination of
the Magnus force and the pinning potential.
When the skyrmion is constrained
by the pinning potential to move in the direction of the drive,
the Magnus force-induced velocity
component from the pinning $\alpha_m F_p$ is aligned with the drive.
This is added to the velocity component $\alpha_d F_D$ produced by
the drive, giving a total velocity of
$\langle V\rangle = \alpha_{d}F_{D} + \alpha_{m}F_{p}$.
Here the nonconservative Magnus force has turned the pinning force into
an effective additional driving force.
Speed up effects are the most prominent on 1D 
substrates, and
have been studied numerically
for a single skyrmion moving along domain walls \cite{Xing20}. 
They can also occur for random and 2D periodic pinning arrays. 
Gong {\it et al.} \cite{Gong20} numerically
studied skyrmion motion in random disorder and found that
the skyrmion velocity can be boosted
in regimes where motion in the transverse or
skyrmion Hall angle direction is suppressed.
This indicates that whenever the skyrmion motion along $\theta^{\rm int}_{\rm SkH}$
is impeded, the
Magnus force can transfer
part or all of the component of motion in that direction
to the direction along which the skyrmion is constrained to move.

\begin{figure}
\includegraphics[width=\columnwidth]{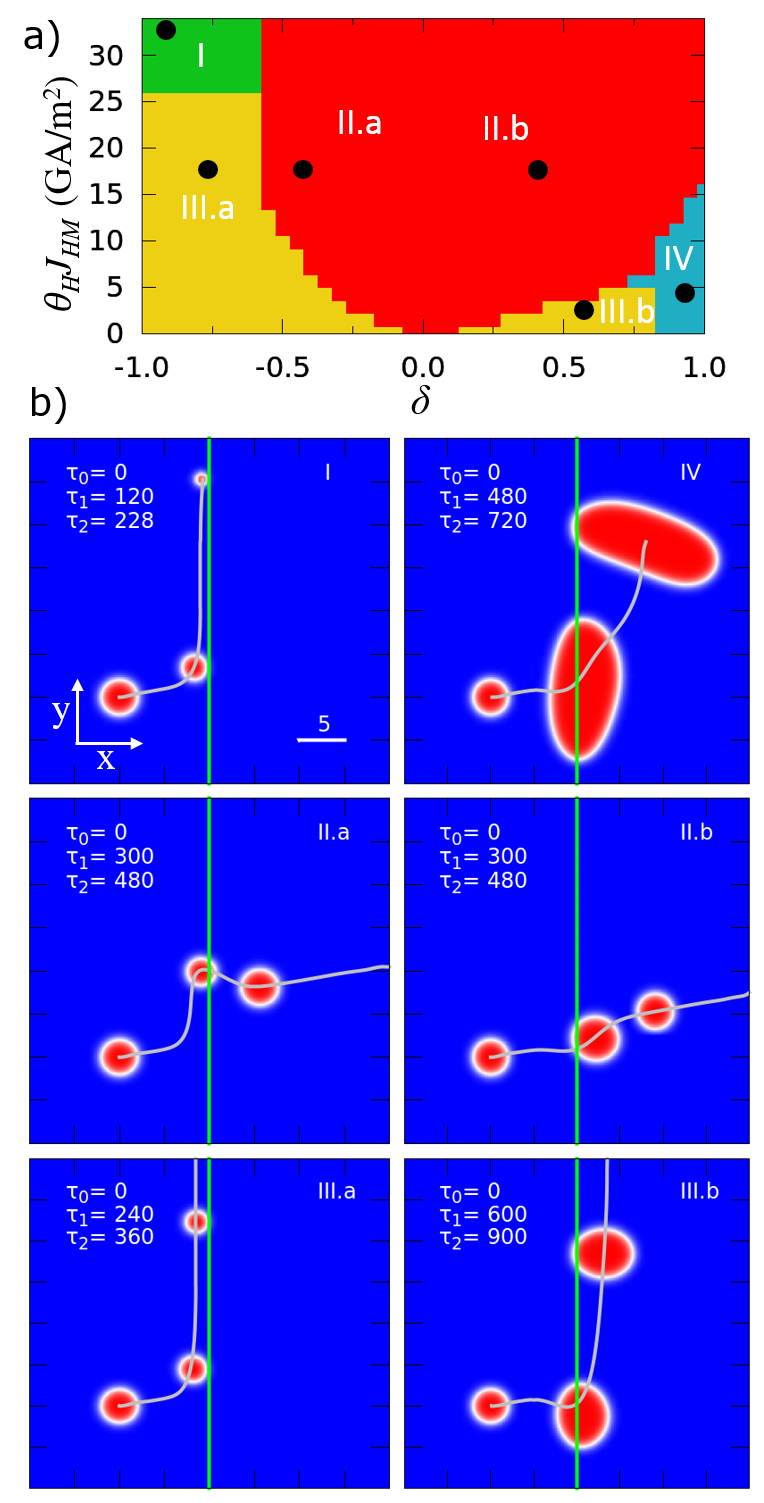}
\caption{
Results from continuum simulations of a skyrmion interacting with a
line along which the DMI has been changed by an amount $\delta$ compared
to the rest of the sample \cite{CastellQueralt19}.
(a) A phase diagram as a function of the product of the skyrmion Hall
angle $\theta_H$ and current $J_{HM}$ vs $\delta$.
(b) Illustration of motion in the six different regimes.
The defect is indicated by a green vertical line and the skyrmion
trajectory is given by the gray line.
Republished with permission of the Royal Society of Chemistry, from
``Accelerating, guiding, and compressing skyrmions by defect rails'',
J. Castell-Queralt {\it et al.}, Nanoscale {\bf 11}, 12589
(2019); permission conveyed through Copyright Clearance Center, Inc.
}
\label{fig:56}
\end{figure}

Skyrmion speed up effects have been observed in micromagnetic simulations 
of race tracks \cite{Sampaio13} 
and for scattering off a single 
pinning site in both continuum and Thiele based approaches \cite{Muller15}. 
Iwasaki {\it et al.} used both a Thiele approach and micromagnetic simulations
to examine the large velocity enhancement near a boundary
and showed that it is
related to a colossal spin transfer torque effect \cite{Iwasaki14}. 
Here the velocity is enhanced by a factor of $1/\alpha$, where $\alpha$ is the
Gilbert damping, and the maximum velocity is determined by the
magnitude of the confining 
force produced by the sample edge.
Several other works also describe the acceleration of skyrmions
along sample edges \cite{Martinez18,CastellQueralt19}.
Castell-Queralt {\it et al.} \cite{CastellQueralt19}
examined the dynamics of a skyrmion moving across 
a rail where, in addition to skyrmion acceleration along the edge,
they observed guiding and compressing effects.
They found that speed ups of as much as an order
of magnitude are possible compared to motion in a system without defects.
Figure~\ref{fig:56} shows the results from micromagnetic simulations
\cite{CastellQueralt19} of  a
skyrmion approaching a single defect line along which the DMI interaction strength
$\delta$ has a modified value.
Here $\delta = -1$ indicates
a complete suppression of the DMI and $\delta = 1$
is the unaltered DMI interaction strength,
so that $\delta < 0$ produces a skyrmion repulsion
and $\delta > 0$ causes the line to attract the skyrmion.
The dynamic phase diagram
in Fig.~\ref{fig:56}(a) shows the
behavior as a function of the product of the
skyrmion Hall angle and current
versus $\delta$,
while Fig.~\ref{fig:56}(b) illustrates the six different phases of motion.
In phases I and III.a, the skyrmion
is guided along the line and shrinks.
In phase IV the skyrmion moves across the line but experiences
strong distortion. 
In phases II.a and II.b the skyrmion crosses the line and is weakly
deflected,
while in phase III.b
the skyrmion crosses the line and is strongly deflected.
The same work also
demonstrated skyrmion guidance
using a combination of two line defects, one of which is repulsive and the
other attractive, and showed that a strong acceleration effect occurs in this case.

The work of Reichhardt and Olson Reichhardt \cite{Reichhardt16a} also
focused on collective effects
for skyrmions moving over 1D periodic arrays.
For driving perpendicular to the substrate periodicity direction,
a number of dynamical phases arise,
including a pinned smectic state similar to
that observed for colloidal
particles and superconducting vortices in periodic 1D substrates,
a disordered plastic flow state just above depinning, a moving hexatic state,
and a moving crystal state.
All these phases produce signatures in
the different velocity components as
well as changes in $\theta_{\rm SkH}$,
and they could be detected experimentally
via neutron scatting, changes in the topological Hall effect, or noise measurements.       

For a skyrmion moving over either a 1D or 2D substrate,
various interference effects can arise. 
In general, a particle
moving over a periodic 
substrate
has a time dependent velocity modulation 
at a 
washboard frequency $\omega_{d}$ which increases
with increasing dc drive $F_{D}$ or current $J$.
When an ac drive $F_{ac}=A\sin(\omega_{ac}t)$ is added to the dc drive,
at certain values of $F_{ac}$ a matching between the ac drive
frequency and $\omega_d$ occurs,
creating a resonance.
Such effects have been
observed experimentally for
superconducting vortex lattices moving over random disorder
\cite{Harris95,Okuma07,Fiory71,Okuma11}.
Since the dc velocity must remain locked at a fixed value for
the resonance condition to persist,
a region of  constant or locked velocity 
appears over an interval of $F_{D}$ values.
When the difference between the two frequencies
becomes too large,
the system
jumps out of the velocity locked step;
however,
additional velocity locking steps
appear whenever $\omega_{d}/\omega_{ac}$ is an integer.
The steps
in the velocity at the resonant condition and its higher harmonics
are known as Shapiro steps
\cite{Shapiro63,Benz90}. 
If the ac amplitude $A$ is large, additional
nonlinear effects can occur, producing fractional steps and strongly 
fluctuating regions.
Shapiro steps have been observed in a wide variety of systems 
that exhibit depinning on periodic substrates, such
as sliding charge density waves \cite{Coppersmith86} and 
vortices in type-II superconductors with 1D and 2D periodic substrates
\cite{Martinoli75,VanLook99,Reichhardt00},
All of these systems are either overdamped or have inertial effects,
but none of them include Magnus
forces.

\begin{figure}
\includegraphics[width=\columnwidth]{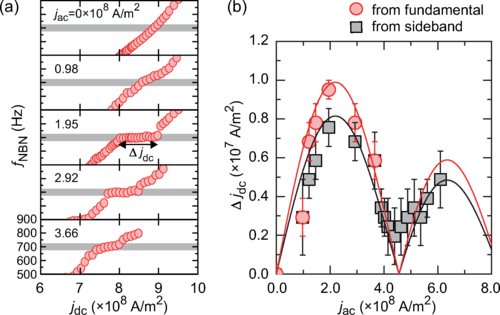}
\caption{
Phase locking and Shapiro steps for current driven skyrmions in MnSi
under combined dc and ac driving \cite{Sato20}.
(a) Magnitude of the narrow band noise $f_{\rm NBN}$ as a function of dc driving
current $j_{\rm dc}$ for different values of the ac current
$j_{\rm ac}$, showing the
emergence of a locking step when $j_{\rm ac}=1.95\times 10^{8}$ A/m$^2$.
(b) Dependence of the locking step width $\Delta j_{\rm dc}$ on ac current
amplitude $j_{\rm ac}$ showing oscillations in the form of a Bessel function,
as expected for a Shapiro step.
Reprinted with permission from T. Sato {\it et al.}, Phys. Rev. B
{\bf 102}, 180411(R) (2020). Copyright 2020 by the American Physical Society.
}
\label{fig:fig.New.4}
\end{figure}

In skyrmion systems,
the Magnus force should produce a variety of new types of
phase locking  phenomena.
For example, the mixing of the velocity components by the Magnus force
makes it possible for
locking steps to occur for any driving direction,
as demonstrated in a particle
based model for skyrmions moving over a periodic 1D potential with a dc
drive applied parallel to the substrate periodicity and an ac drive applied either
parallel or perpendicular to the dc drive
\cite{Reichhardt15b}.
Here Magnus indicted Shapiro steps appear in the velocity-force
curves, and the step widths $\Delta F_{ac}$ 
oscillate according to the Bessel function
$\Delta F_{ac} = |J_{n}(F^{ac}_{x})|$,
as expected for Shapiro steps.
The skyrmion orbits on a locking step
are considerably more complex than those found
in overdamped systems.
Sato {\it et al.} \cite{Sato20} measured voltage fluctuations for current
induced skyrmion lattice motion in MnSi.
They
found a narrow band noise signal that shifted
to higher frequency with increasing current, 
indicating that the speed of the skyrmion motion was increasing.
When they added an ac driving current,
a clear mode locking signal emerged with strongly enhanced
narrow band noise.
The plots of
the dependence of the narrow band noise magnitude on
the dc current density in
Fig.~\ref{fig:fig.New.4}(a)
include a step-like regime where the narrow
band signal is locked to the washboard frequency.
For zero applied ac current, no step is present, but as the amplitude of the
ac current increases, the width of the narrow band step $\Delta j_{dc}$
in Fig.~\ref{fig:fig.New.4}(b)
increases and decreases according to the Bessel function form
of Shapiro steps.

Other combinations of drives for skyrmions on 1D periodic arrays 
produce unusual collective effects.
For example, the dc drive can be applied perpendicular to the substrate
periodicity while the ac drive is either parallel or perpendicular to the
dc drive.
In an overdamped system, this orientation of the dc drive
does not produce any interference effects; however, for the
skyrmion system,
phase locking effects appear, 
including a new phenomenon in which spikes
rather than steps appear in the velocity force curves.
This Shapiro spike
structure
occurs
when the ac and dc drives are both perpendicular to the
substrate periodicity
\cite{Reichhardt17a}.
In some cases the phase locking can cause the skyrmion
to move at $90^\circ$ with respect to the dc drive,
an effect known
as absolute transverse mobility \cite{Reichhardt03a}.
There can also be regions
in which $V_{\perp}$ is negative, indicative of absolute
negative mobility \cite{Eichhorn02,Ros05}
where the skyrmion is actually moving
against the direction of the external drive.

\begin{figure}
\includegraphics[width=\columnwidth]{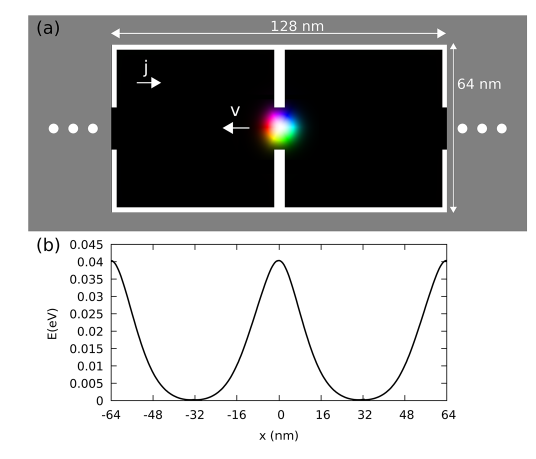}
\caption{
(a) Micromagnetic simulation of a skyrmion moving in
an infinitely long
nanotrack containing periodic notches (white regions) \cite{Leliaert19}.
The DMI is reduced in the notch regions compared to the rest of the
nanotrack. (b) The energy landscape experienced by the skyrmion as a
function of position.
Republished with permission of IOP Publishing, Ltd, from
``Coupling of the skyrmion velocity to its breathing mode in periodically
notched nanotracks'', J. Leliaert {\it et al.}, J. Phys. D: Appl. Phys.
{\bf 52}, 024003
(2019); permission conveyed through Copyright Clearance Center, Inc.
}
\label{fig:60}
\end{figure}

\begin{figure}
\includegraphics[width=\columnwidth]{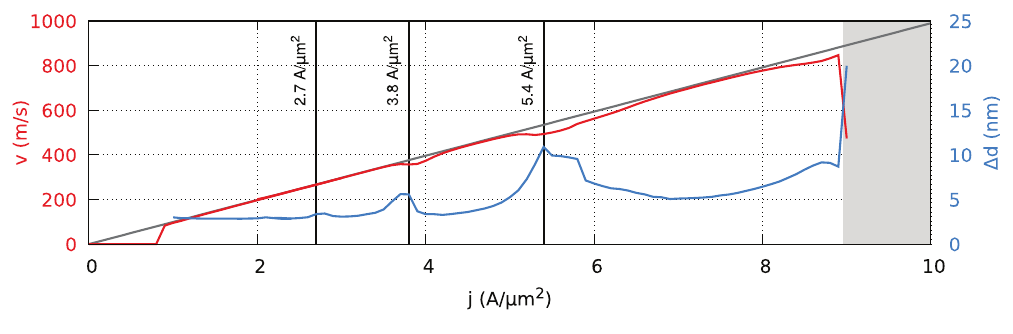}
\caption{
From micromagnetic simulations \cite{Leliaert19}
of the periodically notched system in
Fig.~\ref{fig:60}, the skyrmion velocity $v$ (red) and
skyrmion size fluctuations $\Delta d$ (blue) vs driving current $j$.
Resonances appear when the washboard frequency of the skyrmion motion
couples with the skyrmion breathing mode.
Republished with permission of IOP Publishing, Ltd, from
``Coupling of the skyrmion velocity to its breathing mode in periodically
notched nanotracks'', J. Leliaert {\it et al.}, J. Phys. D: Appl. Phys.
{\bf 52}, 024003
(2019); permission conveyed through Copyright Clearance Center, Inc.
}
\label{fig:61}
\end{figure} 

Since skyrmions have internal modes
with their own intrinsic frequencies,
there should be a wealth of
possible resonances involving the coupling of these modes to
an external
ac frequency,
a substrate frequency produced by dc motion over periodic pinning,
or
the intrinsic washboard frequency of the skyrmion lattice.
These
dynamics would be quite different from those typically
found for overdamped or rigid particles.
There is already some work along these lines
by Leliaert {\it et al.} \cite{Leliaert19},
who performed micromagnetic simulations of skyrmions 
moving through a wire with a periodic modulation of notches,
as shown 
in Fig.~\ref{fig:60}.
The notches, produced by varying the DM interaction,
induce a periodic modulation of the
skyrmion motion which couples to the skyrmion breathing mode,
producing a series of resonances
in the velocity-force curves, as illustrated
in Fig.~\ref{fig:61}. 

\subsection{Further Directions for 1D Periodic Substrates}

\begin{figure}
\includegraphics[width=\columnwidth]{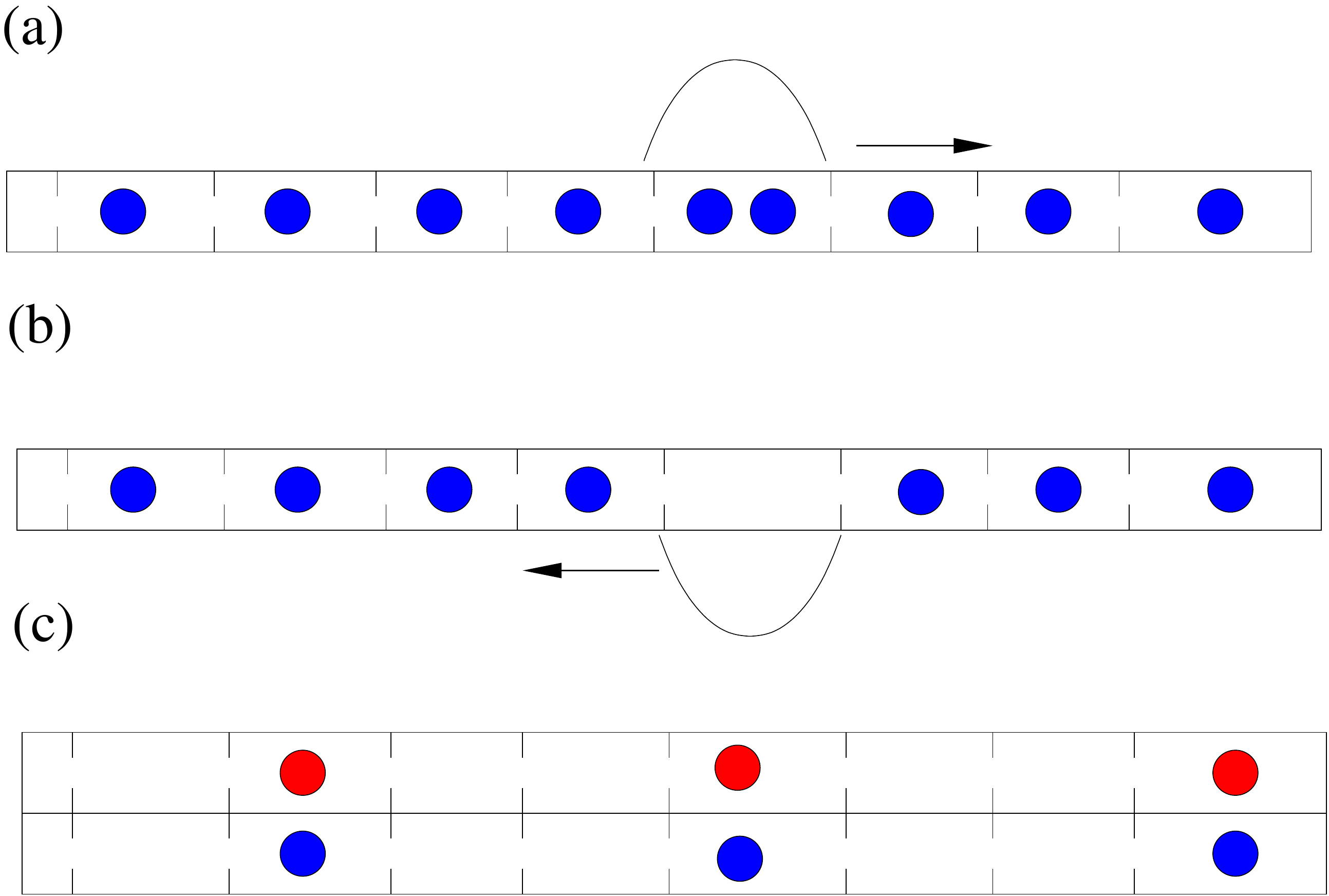}
\caption{ (a)  Schematic of skyrmions
in a nanowire interacting with a 1D periodic substrate
at a filling just above the 1:1 matching.
The additional skyrmion creates a mobile
kink that moves in the direction of drive.
Every time the kink moves through the system,
the skyrmion translates by one lattice constant.
(b) The same, but for an anti-kink just below the 1:1 matching, which 
moves in the opposite direction.
(c) Two coupled wires with different species
of skyrmions could form skyrmions and antiskyrmions
that could bind together and
create skyrmion excitons.  
}
\label{fig:62}
\end{figure} 

There are a variety of potential
race track memory applications of 1D periodic substrates
for both bulk and thin films,
including situations in which
multiple interacting skyrmions could be coupled inside
a nanowire with
a periodic modulation.
In this case,
mobile kinks in the 1D skyrmion chain
could reduce the effects of the skyrmion Hall angle. 
An example is shown schematically in
Fig.~\ref{fig:62}(a), where a constriction with a periodic modulation is
filled with skyrmions at a ratio just above the 1:1 matching condition.
The extra skyrmion forms a kink that travels in the direction of 
drive.
Conversely, if the system is just below the 1:1 matching, a vacancy appears
which
moves
in the opposite direction, as shown in Fig.~\ref{fig:62}(b).
The periodic modulation could be created using a
periodic array of notches \cite{Marchiori17},
variations of the DMI, spatially varying damping \cite{Zhang17a,Zhou19},
or periodic thickness modulations \cite{Loreto19}. 
For
coupled colloidal particles on 1D periodic substrates,
it was shown
that the kinks can act like emergent particles with
their own internal frequency, making it possible to observe
phase locking of kinks under a combined dc and ac drive \cite{Juniper15}. 
The kinks themselves could serve as information carriers instead of the
actual skyrmions.
The 1D substrate need not be static;
a dynamic substrate
can be created using arrays of different gate voltages \cite{Liu19} 
that can be turned on and off to create a flashing potential for the skyrmions.  
In this case, an additional frequency
arises from the periodic flashing that could couple with the
internal frequencies of the skyrmions.
In most overdamped systems, 
Shapiro steps appear when a dc drive is combined
with a single ac driving frequency;
however,
for skyrmions it was shown
that biharmonic ac forces \cite{Chen19} can lead to directed skyrmion motion even 
in the absence of a dc drive.
Thus, having a
skyrmion move over a 1D substrate
under both dc and biharmonic ac drives could
produce a variety of new phenomena.
It would also be possible to couple together nanowires of different materials
such that the skyrmions interact between the nanowires, leading to
skyrmion drag effects
as shown schematically in Fig.~\ref{fig:62}(c).
For example,
a nanowire containing antiskyrmions that 
couples to another nanowire
containing regular skyrmions
could produce an effective 
skyrmion exciton. 

\subsection{Skyrmions with Two Dimensional Periodic Pinning}

\begin{figure}
\includegraphics[width=\columnwidth]{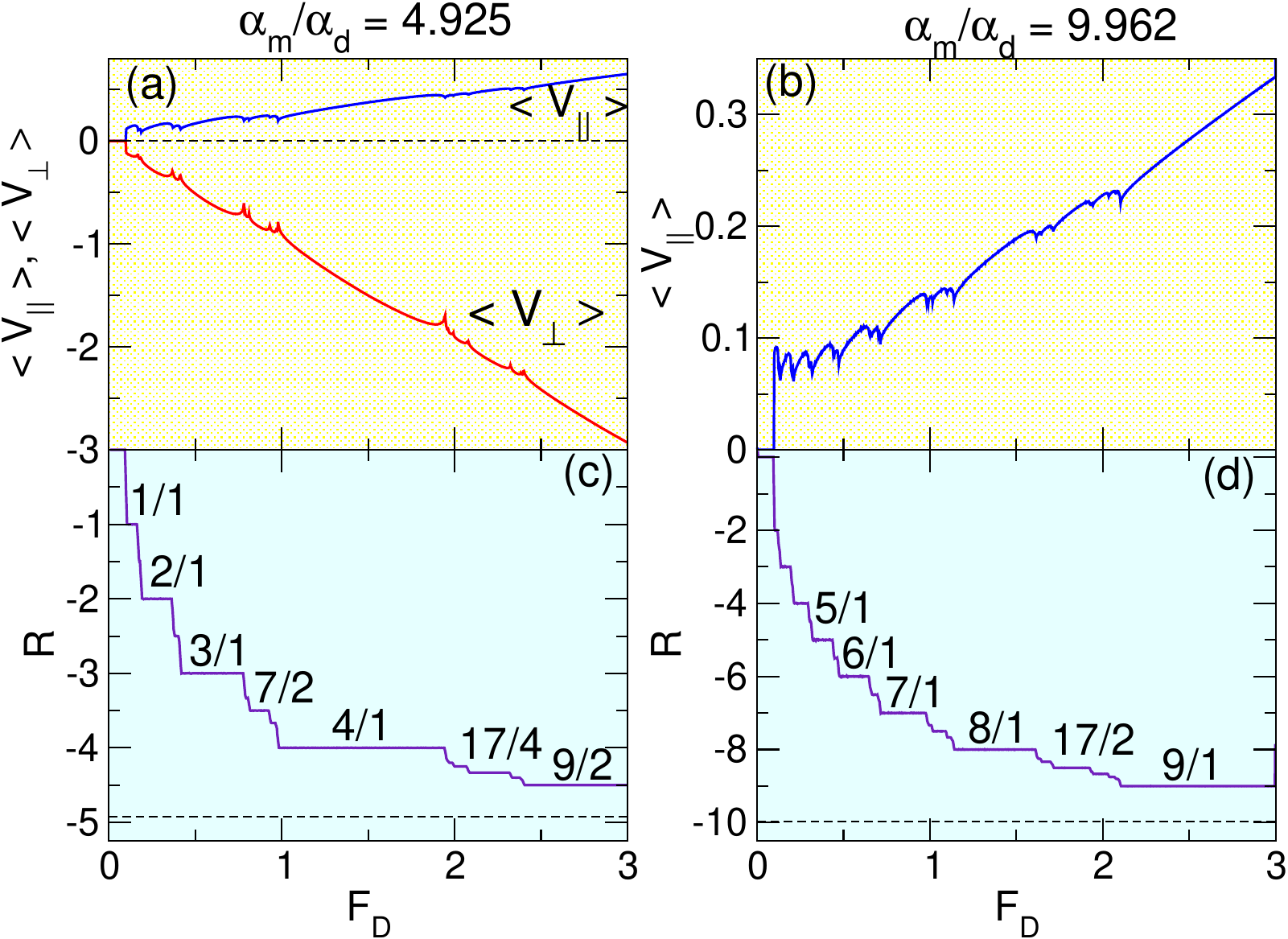}
\caption{
Particle-based simulations of skyrmions moving over a square array
of pinning sites showing quantization of the skyrmion
Hall angle \cite{Reichhardt15}.
(a) The velocity parallel, $\langle V_{||}\rangle$ (blue),
and perpendicular, $\langle V_{\perp}\rangle$ (red), to the driving direction
vs the dc drive amplitude $F_D$ at a Magnus ratio to damping
ratio of $\alpha_m/\alpha_d=4.925$.
(b) $\langle V_{||}\rangle$ vs $F_D$ for
a larger ratio $\alpha_m/\alpha_d=9.962$.
(c) The ratio $R=\langle V_\perp\rangle/\langle V_{||}\rangle=\tan(\theta_{\rm SkH})$
vs $F_D$ for the sample in panel (a), where steps appear at rational
fractions.
(d) $R$ vs $F_D$ for the sample in panel (b) also exhibits a
series of steps.
Reprinted with permission from C. Reichhardt {\it et al.}, Phys. Rev. B
{\bf 91}, 104426 (2015). Copyright 2015 by the American Physical Society.
}
\label{fig:63}
\end{figure}

\begin{figure}
\includegraphics[width=\columnwidth]{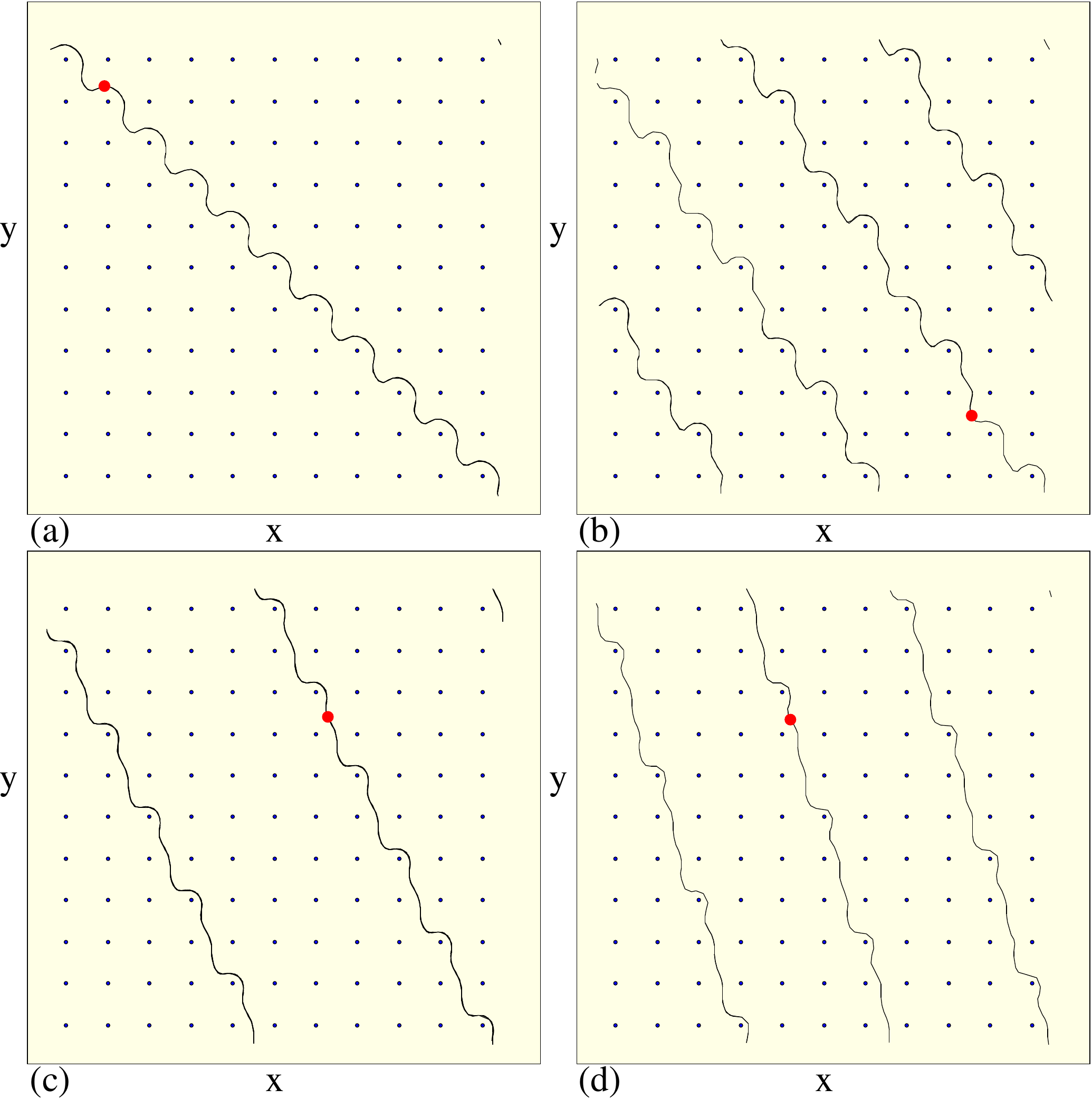}
\caption{
Skyrmion trajectories (lines) for particle based simulations of
the system in Fig.~\ref{fig:60} with a skyrmion
(red circle) moving through a periodic array of pinning sites (black dots)
at (a) $|R|=1$, (b) $|R|=5/3$, (c) $|R|=2$, and (d) $|R|=3$
\cite{Reichhardt15}.
Reprinted with permission from C. Reichhardt {\it et al.}, Phys. Rev. B
{\bf 91}, 104426 (2015). Copyright 2015 by the American Physical Society.
}
\label{fig:64}
\end{figure}

Reichhardt {\it et al.} used 
a particle-based model to study a single
skyrmion moving over  a 2D square periodic potential \cite{Reichhardt15}. 
This system has a finite depinning threshold and
a drive-dependent skyrmion Hall angle $\theta_{\rm SkH}$,
similar
to what is observed for random pinning as discussed above
\cite{Jiang17,Litzius17,Legrand17,Reichhardt15a,Reichhardt16,Kim17};
however, due to the square substrate symmetry,
the skyrmion motion preferentially locks to certain directions
$\theta_{\rm SkH}=\tan^{-1}(n/m)$ with integer $m$ and $n$.
For example, locking at
$\theta_{\rm SkH}=45^\circ$ occurs when $n = 1$ and $m = 1$ while 
locking at
$\theta_{\rm SkH}=23^\circ$ corresponds to $n = 1$ and $m = 2$.
For increasing drive,
the skyrmion can only remain locked in its
direction of motion if its net velocity $\langle V\rangle$ decreases,
so each locking step is associated with
a window of negative differential mobility
in which $d\langle V\rangle/dF_{D} < 0$.
Cusps in both
the parallel and perpendicular velocities,
$\langle V_{||}\rangle$ and $\langle V_{\perp}\rangle$,
appear at the transition from one
directional locking step to the next,
as shown
in Fig.~\ref{fig:63}(a,b).
Figure~\ref{fig:63}(c,d) illustrates the ratio
$R = \langle V_{\perp}\rangle/\langle V_{||}\rangle=\tan(\theta_{\rm SkH})$,
indicating that
the skyrmion Hall angle is
quantized as a function of increasing drive.   
On the
$|R| = 1$
step,
the particle
is constrained to move
along $\theta_{\rm SkH} = 45^{\circ}$,
as illustrated in Fig.~\ref{fig:64}(a).
The skyrmion trajectories for
motion on the $|R|=5/3$, 2, and 3 steps
appear in Figs.~\ref{fig:64}(b), (c), and (d), respectively.
In general, the integer steps are more pronounced  than the 
fractional steps. 
Such directional locking should be a generic feature of ferromagnetic skyrmions 
moving over periodic pinning arrays.  
A similar directional locking effect with steps in the velocity force curves
was studied 
for
superconducting vortices \cite{Reichhardt99} and colloidal particles
\cite{Korda02,MacDonald03,Risbud14} moving over 2D periodic substrates,
but in these overdamped systems,
the external drive must change direction in order to generate the locking steps,
whereas in the skyrmion system, the direction of the
dc drive remains fixed.

\begin{figure}
\includegraphics[width=\columnwidth]{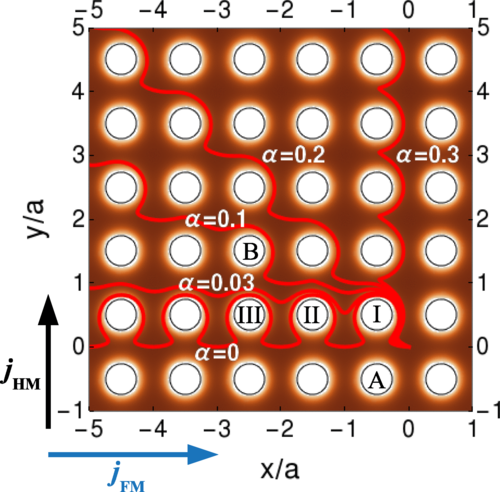}
\caption{
Combined micromagnetic and analytic calculations of skyrmion trajectories
(red lines) in a square array of magnetic dots for different values
of the damping coefficient $\alpha$ \cite{Feilhauer20}.
By varying the direction of the applied current pulse, the skyrmion can be
steered to any selected position in the array.
Reprinted with permission from J. Feilhauer {\it et al.}, Phys. Rev. B
{\bf 102}, 184425 (2020). Copyright 2020 by the American Physical Society.
}
\label{fig:fig.new.9}
\end{figure}

Feilhauer {\it et al.} \cite{Feilhauer20}
employed a combined micromagnetic and Thiele equation approach
to study skyrmion motion in a magnetic antidot array.
They found that the skyrmion motion locks to the symmetry angles of the
array and that the skyrmion Hall angle
can be controlled by varying the damping,
as shown in
Fig.~\ref{fig:fig.new.9}.
By careful choice of the direction of the current pulse,
a skyrmion can be steered to
move into almost any other plaquette position,
suggesting that this drive protocol could be a useful method for applications. 
There have already been some experimental efforts to create
a similar type of substrate using antidot lattices \cite{Saha19}. 

Locking of the skyrmion motion to
particular symmetry directions
of 2D periodic arrays
could be harnessed to create a topological sorting device
for different species of skyrmions with slightly different
values of $\theta_{\rm SkH}^{\rm int}$.
Here, one species would lock to a symmetry direction of the substrate while
the other species would not, producing a lateral separation of the species
over time.
A demonstration of this separation effect
was achieved in simulations by
Vizarim {\it et al.} for a bidisperse
assembly of skyrmions driven through a square
obstacle array
\cite{Vizarim20}.
This procedure is
similar to that used in microfluidic systems,
and suggests that skyrmion bubbles with a carefully selected
size could be separated from skyrmion bubbles of other sizes.
Using particle-based simulations,
Vizarim {\it et al.} also showed that a skyrmion interacting with
a 2D periodic array under a dc drive and one or more ac drives can
undergo a variety of controlled motions \cite{Vizarim20c}
and can exhibit non-monotonic behavior
of the skyrmion Hall angle \cite{Vizarim20d}. 
Another interesting effect observed in particle based models
of skyrmions driven over periodic arrays
is skyrmion clustering or segregation.
This is similar to the segregated states found for strong
random pinning in
both continuum \cite{Koshibae18}
and particle based simulations \cite{Reichhardt19a}.

Reichhardt {\it et al.} also studied
collective static arrangements of
skyrmions interacting with square pinning arrays
as a function of skyrmion density
using a particle based model \cite{Reichhardt18}.
When the number of skyrmions $N_{sk}$ is an integer
multiple of the number of pinning sites $N_p$,
a series of commensurate states
appear
in which different 
types of skyrmion crystals can be stabilized,
including square or triangular lattices.
Ordered skyrmion lattices
can also form at rational filling fractions
$f \equiv N_{sk}/N_p$ such as 
$f =1/2$, where the skyrmions adopt a checkerboard pattern.
The $f=1.65$ and $f=2.0$ configurations
were also
observed
in continuum-based simulations
for a square array of pinning sites
produced by 
local changes in the anisotropy \cite{Koshibae18}.  

\begin{figure}
\includegraphics[width=\columnwidth]{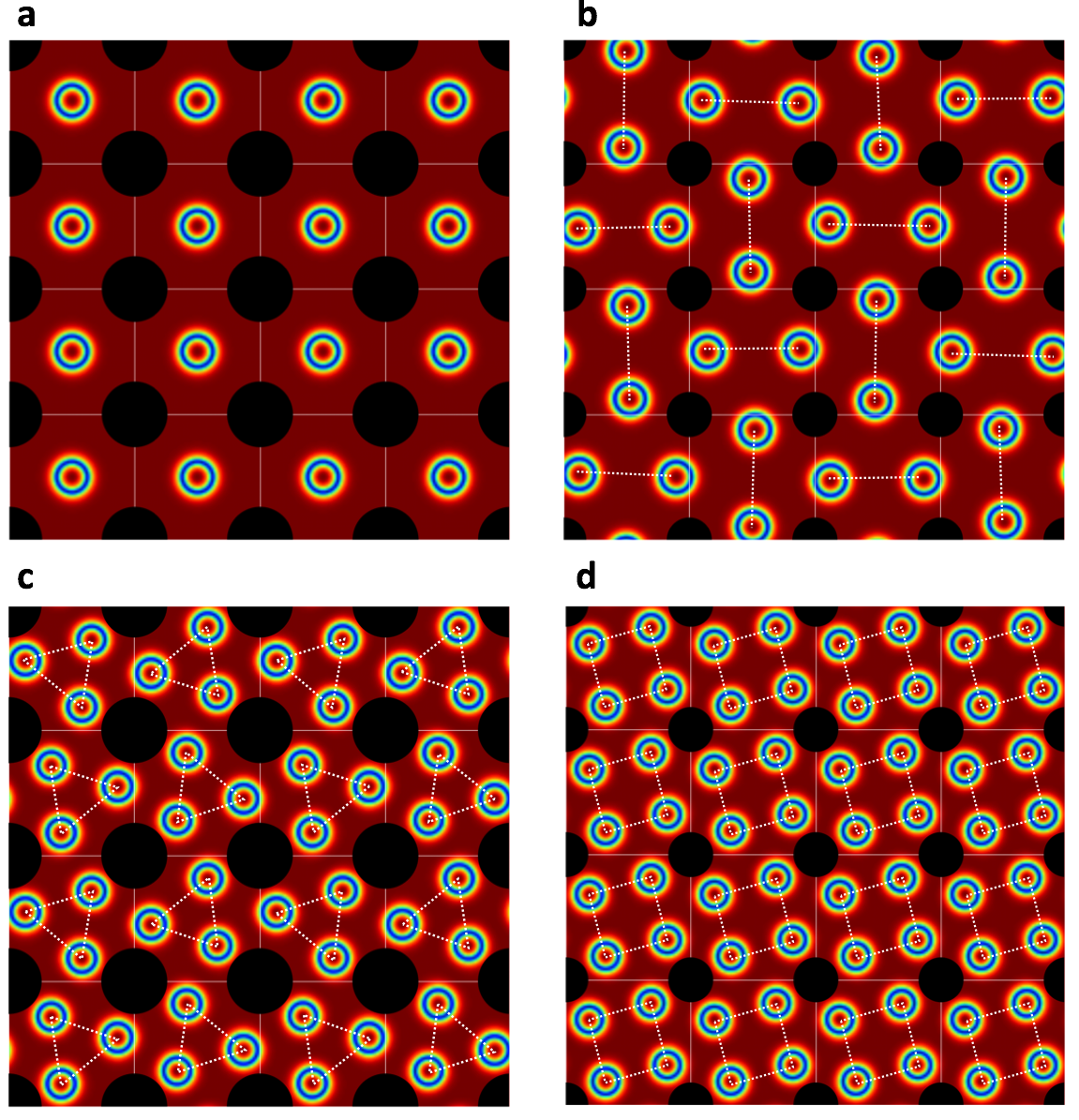}
\caption{
Continuum simulations of chiral liquid crystal skyrmions (blue rings)
interacting with a periodic array of obstacles (black circles) \cite{Duzgun20}.
(a) A filling ratio of $f=1$ where the skyrmions form a square lattice.
(b) Alternating dimer ordering for $f=2$.
(c) A trimer arrangement at $f=3$.
(d) An ordered quadrimer state at $f=4$.
Republished with permission of the Royal Society of Chemistry, from
``Commensurate states and pattern switching via liquid crystal skyrmions trapped
in a square lattice'', A. Duzgun {\it et al.}, Soft Matter {\bf 16}, 3338
(2020); permission conveyed through Copyright Clearance Center, Inc.
}
\label{fig:fig.new.8}
\end{figure}

Duzgun {\it et al.} explored the ordering of liquid crystal
skyrmions interacting with a square array of defects 
using continuum based simulations \cite{Duzgun20}.
At a one-to-one matching of $f=1$,
the skyrmions form a square lattice as illustrated
in Fig.~\ref{fig:fig.new.8}(a).
Fillings of $f=2$, 3, and 4 produce dimer, trimer, and quadrimer states as
shown in Fig.~\ref{fig:fig.new.8}(b-d).
At some filling fractions such as $f=2$,
the skyrmions deform into elongated states
in order to match the substrate symmetry better.

\begin{figure}
\includegraphics[width=\columnwidth]{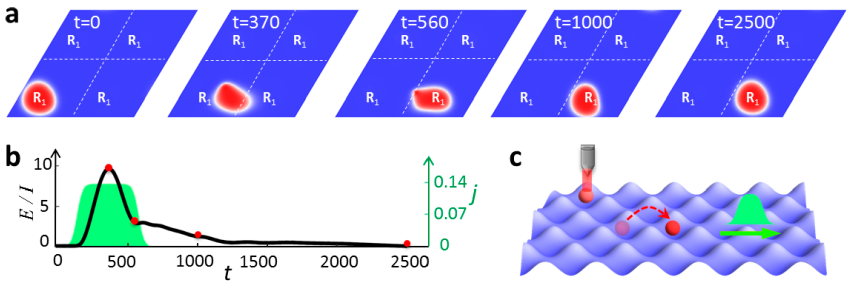}
\caption{
Numerical model for the motion of a skyrmion over a moir{\` e} pattern
formed by a  
  van der Waals 2D magnet \cite{Tong18}.
  (a) The red region indicates the location of the skyrmion as a function
  of time.
  (b) The time profile of the applied current $j$ (green) and the energy
  of the skyrmion $E/I$ during the motion illustrated in panel (a).
  (c) A schematic of the use of a spin-polarized scanning tunneling
  microscopy tip (gray) to write a skyrmion, which is
  then moved from one substrate minimum to another with a current pulse (green).
Reprinted with permission from Q. Tong {\it et al.}, Nano Lett. {\bf 18},
7194 (2018). Copyright 2018 American Chemical Society.
}
\label{fig:67}
\end{figure}

Observation of skyrmion motion in systems with two periodic surfaces
can be achieved using moir{\`e} patterns in
van der Waals 2D magnets \cite{Tong18}.
The moir{\`e} patterns are generated by introducing
a lateral modulation of the interlayer magnetic coupling for different
atomic angles.
In the case of weak
interlayer coupling, a skyrmion can be viewed as
moving over a periodic
substrate composed of trapping sites formed by the moir{\`e} pattern.
Figure~\ref{fig:67}(a)
shows the periodic motion
that can be induced by the pattern.
In Fig.~\ref{fig:67}(b),
application of a current pulse
can cause the skyrmion to jump from
one side of a trapping barrier to the other.
Tong {\it et al.}
proposed that the 2D
moir{\` e} trapping array could be used to 
create a stable background substrate for controlled skyrmion motion 
for various applications \cite{Tong18}.

\begin{figure}
\includegraphics[width=\columnwidth]{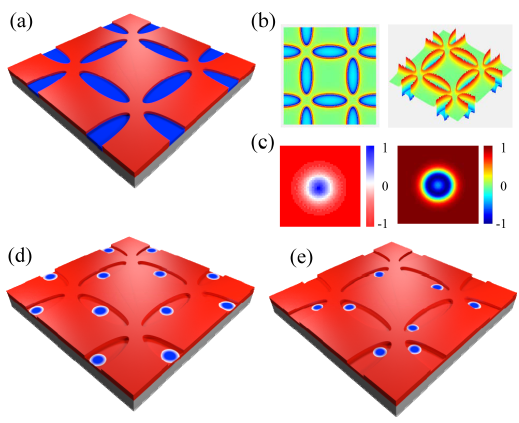}
\caption{
An artificial ice geometry for skyrmions \cite{Ma16}.  
(a) The geometry is constructed using elliptical blind holes with
opposite magnetization directions inside and outside the holes.
(b) The perpendicular or $z$ component of the resulting stray field.
(c) Images of the spin configuration (left) and the topological
density distribution (right) of an isolated individual skyrmion.
(d) Large skyrmions
sit at the
center of each blind hole to form a non-frustrated configuration.
(e) Small skyrmions
sit at
one end of each blind hole and form a frustrated state.
Reprinted with permission from F. Ma {\it et al.}, Phys. Rev. B
{\bf 94}, 144405 (2016). Copyright 2016 by the American Physical Society.
}
\label{fig:68}
\end{figure}

Skyrmions have also been studied
in 2D
arrays of artificial spin ice geometries.
Here, a skyrmion is placed
in a double well potential, and the position of the skyrmion on either
end of the well
can be mapped onto an effective spin direction.
Figure~\ref{fig:68} shows  schematically
how such structures could be made 
using a thickness modulation \cite{Ma16}. 
The skyrmions form a spin ice ordering very similar to that observed
for
superconducting vortices \cite{Libal09} and
colloidal particles \cite{Libal06} on square and hexagonal
double well artificial ice arrays.
Unlike the vortices and colloidal particles,
the skyrmions can change size or deform.
As a result, a transition can occur
from a frustrated state in which each skyrmion occupies
only one side of the double well
to an unfrustrated state in which
a single skyrmion
stretches out and occupies the center of the well, as shown
in Fig.~\ref{fig:68}(d-e).
There have also been studies of so-called artificial skyrmion lattices in 
a 2D array of magnetic dots, where the individual dots
contain skyrmion states
\cite{Gilbert15,Zhang16b}.
The next step in such work would
be to see whether skyrmions in adjacent dots could be coupled,
or if the entire system could be
placed on a ferromagnetic substrate that would permit the skyrmions to
hop directly from one dot to the next.

\subsubsection{Further Studies with Periodic Substrates}

\begin{figure}
\includegraphics[width=\columnwidth]{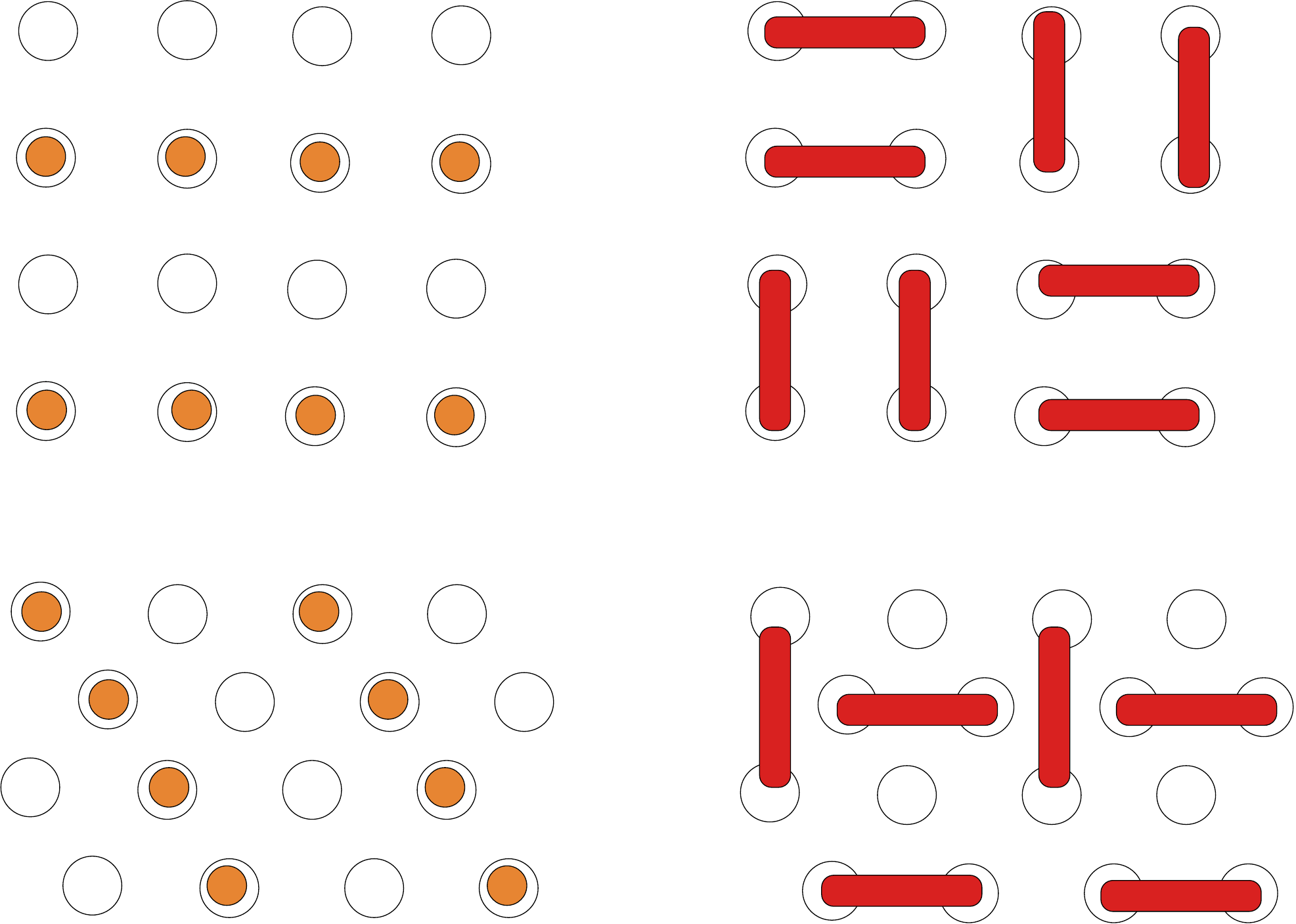}
\caption{
Schematic of possible orderings on square
and triangular pinning arrays (open circles) at half filling.
Left: Skyrmion (orange circles) orderings. 
Right: The skyrmions elongate into meron pairs (red rectangles)
to create a 1:1 filling for
the square pinning array, but still leave unoccupied pins in the
triangular pinning array.
}
\label{fig:69}
\end{figure}

A wide variety of
avenues of study are available for
skyrmions on 2D periodic substrates created by a range of methods.
New types of skyrmion-based memory devices could be produced by
storing information in certain skyrmion configurations
that could
be changed by the application of a current.
At fillings slightly away from commensuration,
a well defined number of kinks or antikinks are present that
act like
emergent particles with their own dynamics.
It would be interesting to explore
whether the Magnus force or the internal degrees of freedom of the skyrmions
would change the dynamics of kinks and antikinks
compared to what is observed
in overdamped or rigid particle systems.
When thermal fluctuations become relevant,
the kinks or antikinks could form their own lattice and exhibit melting
phenomena.
Up to now,
numerical work
on incommensurate states has focused on particle-based models,
so new studies based on micromagnetic calculations could reveal
many additional effects
related to the
ability of the skyrmion to change its shape.
These could include
a variety of new types of commensuration and dynamical effects that
do not appear in overdamped or inertial systems.
For example,
a system containing twice as many pinning sites
as particles normally forms a square
or striped sublattice as illustrated
in Fig.~\ref{fig:69}.
If the
pinning is strong enough,
however, the skyrmions can elongate to form pairs of merons
that cover
each lattice site,
representing an effective dimer covering model that has numerous possible
ordered states.
Triangular substrates at half filling
would form strongly frustrated states if
the skyrmions elongate into meron pairs.

The strong gyroscopic motion of skyrmions makes it possible
to explore coupled skyrmions oscillating in 
dense 2D arrays of dots where each dot
can have different
materials properties.
The coupled oscillations could pass through a series of
locking transitions
as a function of some form of ac driving.
The sliding dynamics of skyrmions
over a periodic array would also be an interesting avenue of study.
For example, Koshibae and Nagaosa 
\cite{Koshibae18} showed
that skyrmion creation and annihilation occurs at certain drives and pinning
strengths when skyrmions are moving through random arrays.
On periodic arrays, such events may be much better controlled.
For instance,
a skyrmion could move a specific number of lattice sites before it is
annihilated, or a new skyrmion is created, or some combination of these
effects occurs.
Under superimposed ac and dc driving,
a resonance could arise between the ac drive and the motion of the
skyrmions over the substrate or the skyrmion breathing modes.
Similar effects could be studied for other systems
such as merons, combined meron-skyrmion lattices, antiskyrmions,
and antiferromagnetic skyrmions.
Periodic pinning arrays could be created in bulk systems as well,
where the pinning could either be
only present on the surface
or could pass through the bulk in the form of
columnar defects of the type that can be generated
using patterned irradiation.

\subsection{Asymmetric Arrays, Diodes, and Ratchets}

In a ratchet device, an applied ac drive leads to the net dc motion of
a particle.
Ratcheting motion in overdamped systems is typically achieved
using an asymmetric pinning potential \cite{Reimann02,Hanggi09}.
The flashing of an asymmetric substrate in a thermal system can
generate stochastic ratchet transport,
while higher dimensional ratchet effects
can occur on symmetric substrates if time symmetry is broken by
a chiral ac drive.
Ratchet effects have been studied extensively in particle-based systems such 
as colloidal particles \cite{Rousselet94,Xiao11}, 
vortices in type II superconductors
\cite{Lee99,Olson01,Villegas03,Lin11,Shklovskij14}, and cold atoms \cite{Salger09}. 
There are also examples of ratchet effects in 
magnetic systems, where domain walls interacting with asymmetric 
dot arrays
undergo ratcheting motion
under various types of external ac driving \cite{Marconi11,Franken12,HerreroAlbillos18}.    
Ratchet effects have generally been studied in overdamped systems;
however, additional effects appear
when terms such as inertia are included in the equation of motion \cite{Reimann02,Hanggi09}.
Skyrmions, as particle-like objects,
represent
a natural system in which to study ratchet effects,
and
their strong non-dissipative
Magnus force
can
produce new effects distinct from what has been observed previously
in other ratchet systems.

\begin{figure}
\includegraphics[width=\columnwidth]{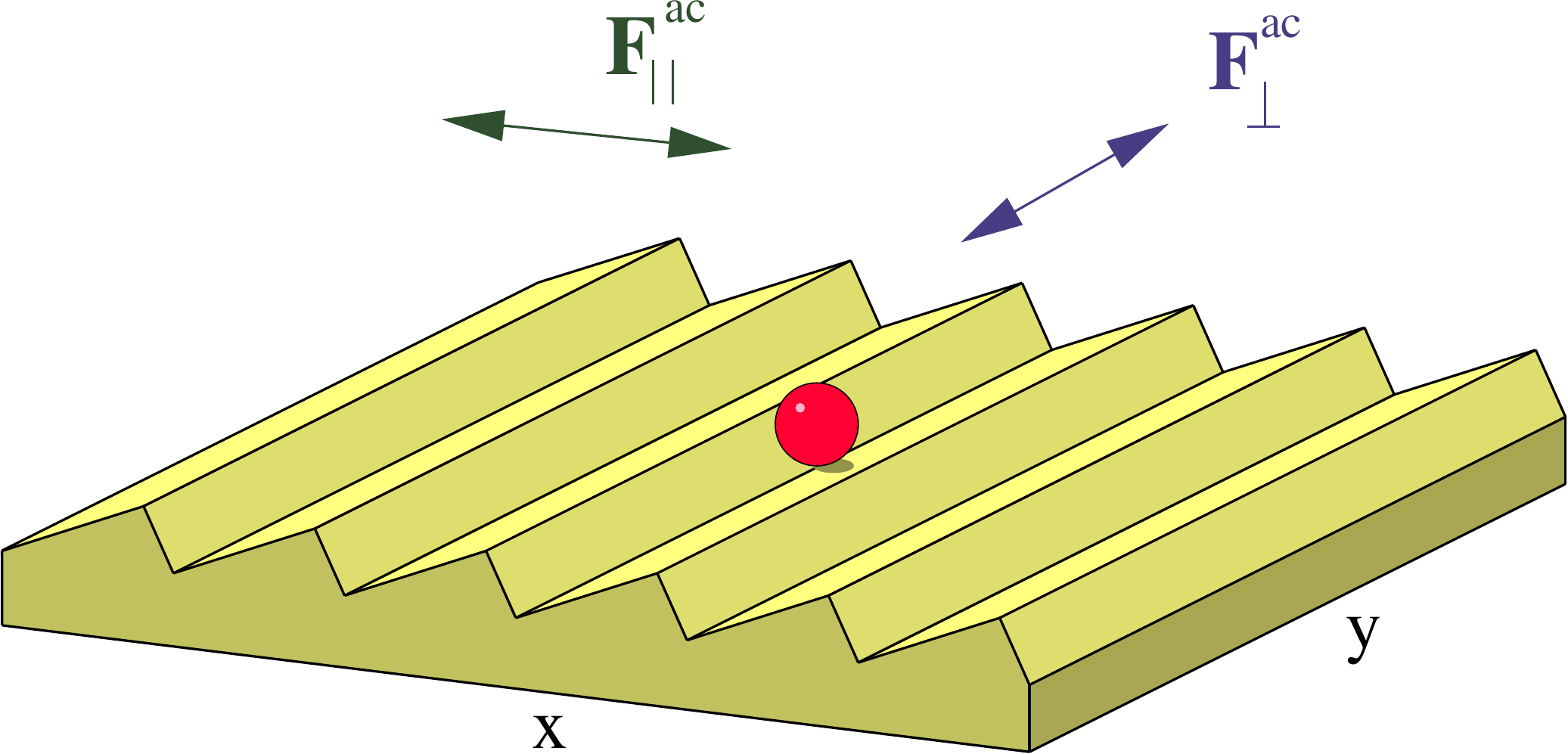}
\caption{
Schematic of a quasi-one-dimensional asymmetric ratchet potential
\cite{Reichhardt15c}.
A skyrmion (red circle) can be driven by an ac current applied
parallel ($F^{ac}_{||}$, green arrow) or perpendicular ($F^{ac}_{\perp}$,
blue arrow) to the substrate periodicity direction. An overdamped
particle would exhibit no ratcheting effect under $F^{ac}_\perp$, but due
to the Magnus effect, a skyrmion can
undergo ratcheting motion under either ac driving
direction.
Reprinted under CC license from C. Reichhardt {\it et al.},
New J. Phys. {\bf 17}, 073034 (2015).
}
\label{fig:70}
\end{figure}

\begin{figure}
\includegraphics[width=\columnwidth]{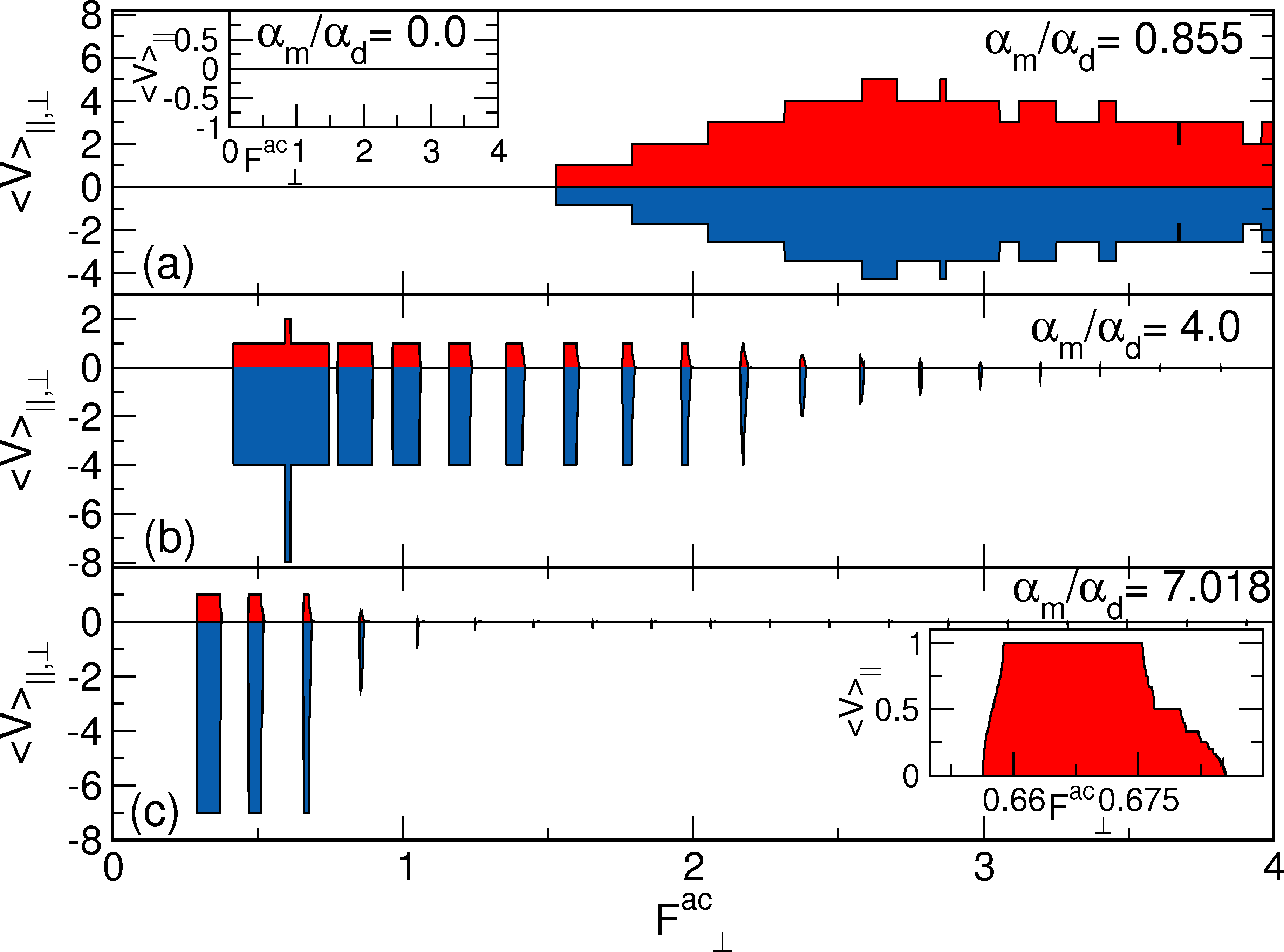}
\caption{
Particle based simulations of skyrmion ratchet motion under
perpendicular driving $F^{ac}_{\perp}$ 
on the asymmetric substrate illustrated in Fig.~\ref{fig:70}
\cite{Reichhardt15c}.
Panels (a,b,c) show
velocities parallel, $\langle V_{||}\rangle$ (red),
and perpendicular, $\langle V_{\perp}\rangle$ (blue), to the
substrate asymmetry as a function of ac driving force magnitude
$F^{ac}_{\perp}$
for different values of the
Magnus force to damping force ratio $\alpha_m/\alpha_d$.
Ratcheting with quantized velocity values occurs in both the parallel
and perpendicular directions above a threshold value of
$F^{ac}_{\perp}$, and there can be drive windows in which no ratcheting
motion occurs.
Inset of (a):
For an overdamped system, no ratcheting occurs
in either direction at any value of $F^{ac}_{\perp}$.
Inset of (c): a blow up of
panel (c) highlighting the presence of fractional velocity steps.
Reprinted under CC license from C. Reichhardt {\it et al.},
New J. Phys. {\bf 17}, 073034 (2015).
}
\label{fig:71}
\end{figure}

The first proposal for a skyrmion ratchet
involved a 1D asymmetric substrate, studied by
Reichhardt {\it et al.} \cite{Reichhardt15c} using a particle based approach.
The skyrmions move in 2D on 
the substrate potential illustrated in Fig.~\ref{fig:70}, which
has the form
\begin{equation}
U(x) = U_{0}[\sin(2\pi x/a) + 0.25\sin(4\pi x/a)]
\end{equation}
where $a$ is the substrate periodicity.
In the overdamped limit, if an ac drive  
is applied in the direction of the substrate periodicity
(the $x$-direction),
a standard ratchet effect arises
in which the particle can translate by one or more substrate periods
in the easy ($+x$) direction under each cycle of an ac drive.
The depinning threshold
is finite for both the easy  ($+x$) and hard ($-x$)
directions; however, the
depinning threshold is larger in the hard direction,
so that in the dc limit, 
the system acts as a diode.
If the ac drive is applied in the perpendicular
or $y$-direction in an overdamped system,
there is no ratchet effect since no symmetry is broken.
In the case of
skyrmions with a finite Magnus force,
which move at an angle $\theta_{\rm SkH}$ with respect to the driving direction,
a ratchet effect can occur even when the ac drive is
purely perpendicular to the
substrate periodicity direction.
This is termed a Magnus ratchet effect.
Figure~\ref{fig:71} shows the velocity component in both
the parallel and perpendicular directions for the system in
Fig.~\ref{fig:70} under perpendicular ac driving $F^{ac}_{\perp}$.
The inset of Fig.~\ref{fig:71}(a) indicates
that an overdamped system produces
no ratchet effect,
while Fig.~\ref{fig:71}(a,b,c) illustrates ratcheting motion in
samples with various intrinsic skyrmion Hall angles.
The ratchet velocities
have well defined quantized values,
and there are regions of ac amplitude over which no ratchet effect occurs.
The inset in 
Fig.~\ref{fig:71}(c) shows a blowup of a single step
where there are also fractional ratchet steps. In this 
system, the skyrmions execute complex 2D orbits while ratcheting.

Ma {\it et al.} \cite{Ma17}
used particle based simulations to consider
skyrmions interacting with 2D asymmetric arrays
in which the pinning sites have a density gradient.
They found that, depending on whether the ac drive is applied parallel or
perpendicular to the substrate periodicity direction,
an entirely new type of ratchet effect called a vector ratchet can appear, in which
the direction of
motion of the skyrmions can be tuned by up to 360$^\circ$ by varying the
amplitude of the ac driving. 

\begin{figure}
\includegraphics[width=\columnwidth]{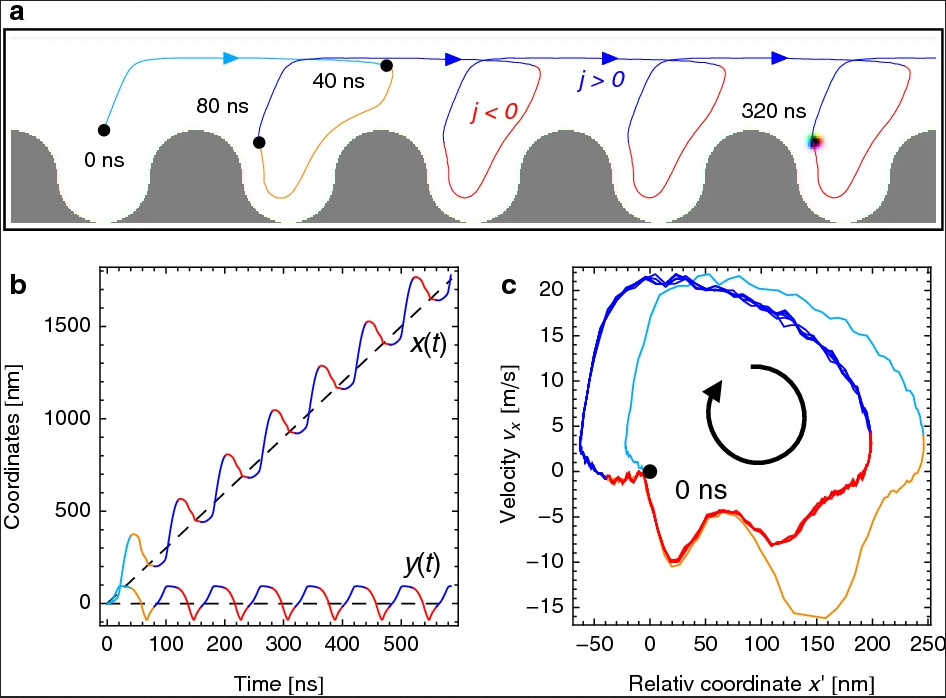}
\caption{
Thiele-based simulations showing the operation of a ratchet
mechanism in a skyrmion racetrack \cite{Gobel21}.
(a) An asymmetry in the racetrack edge combines with the Magnus force to
produce a 2D orbit that translates over time.
(b) A plot of the position of the skyrmion as a function of time showing a
deterministic ratcheting motion in the positive $x$ direction.
(c) The shape of the skyrmion orbit as a function of
$x$ direction velocity $v_x$ vs the relative displacement in $x$ from the
average position.
Reprinted under CC license from B. G{\" o}bel and I. Mertig, Sci. Rep.
{\bf 11}, 3020 (2021).
}
\label{fig:fig.new.10}
\end{figure}

G{\" o}bel and Mertig \cite{Gobel21}
performed numerical continuum modeling
of skyrmions
interacting with a patterned race track to show that the
skyrmion Hall angle can be used to create a skyrmion ratchet.
Figure~\ref{fig:fig.new.10}(a)
illustrates the race track geometry with 
a ratcheting skyrmion orbit appearing as
a function of time under an oscillating drive.
The Magnus force is responsible for creating
the 2D orbit that is necessary to induce the ratchet effect.
Figure~\ref{fig:fig.new.10}(b) shows that the skyrmion propagates
deterministically as a function of time,
while Fig.~\ref{fig:fig.new.10}(c) illustrates the skyrmion velocity
versus relative position.
G{\" o}bel and Mertig
explain that the skyrmion ratchet differs from a standard overdamped
ratchet
due to the fact that the Magnus force allows
velocity components to be created perpendicular
to the confining force produced by the sample edges.  

Migita {\it et al.} \cite{Migita21} considered continuum
simulations of skyrmions in asymmetric constricted geometries
under an oscillating magnetic field. Here the diameter of the skyrmion
oscillates as a function of time, producing a unidirectional
translation of the skyrmion.

Even in the absence of a substrate, a
skyrmion ratchet effect can emerge.
Chen {\it et al.} used continuum based modeling
to show that it is possible to produce a skyrmion 
ratchet effect using biharmonic ac driving 
\cite{Chen19}.
Under certain conditions,
directed motion appears when
the internal modes of the skyrmions induce
an asymmetric shape oscillation.
The motion can be controlled by varying the ac drive parameters.
Further studies by Chen {\it et al.}
extended this same mechanism by
coupling the skyrmion
to a linear defect
in order to take advantage of the
speed up effect and create an ultrafast ratchet effect \cite{Chen20}.  

Wang {\it et al.} \cite{Wang15}
found that under an oscillating field,
the changing shape of the skyrmion
can produce directional motion
in the absence of a
substrate.
A similar wiggling skyrmion propagation mechanism
based on parametric pumping in an oscillating electric field
was studied by Yuan {\it et al.}
\cite{Yuan19}.
There have also been proposals to
drive gyroscopic skyrmion motion
by means of
steps in
the magnetic anisotropy 
\cite{Zhou19a}. 
These results indicate
that in skyrmion systems, there are many possible ways in which
to achieve the temporal or spatial symmetry breaking that is required
to produce a ratchet effect.
It would
also be interesting to consider
the breathing modes of skyrmions produced by biharmonic drives.
These could be coupled to 1D, 2D periodic, or asymmetric periodic 
substrates in order to determine
whether the breathing could strongly enhance the
directed motion or make it easier to control.

Due to the rich dynamics of skyrmions
in terms of both Magnus force and
the internal modes, there are many more types
of ratchet effects that could be studied.
One effect that has 
only been considered briefly is collective ratchets.
In overdamped systems, collective interactions between particles
can produce incommensurate states in which the solitons themselves
can undergo ratcheting motion.
In these collective states, it is possible for
ratchet reversals to 
occur \cite{Hanggi09}.
If there
are variations in the skyrmion
sizes or different coexisting species of skyrmions,
it could be possible to
realize a ratchet system
in which one skyrmion size or species
is ratcheted more effectively or in a different
direction than the other sizes or species.
It would be very interesting to consider ratchet effects in
collectively interacting skyrmion assemblies.
Since the skyrmions have internal modes, it may 
even be possible to realize propagating skyrmion breathing modes. 
Such modes could have low dissipation
and could be used as another method for transmitting information. 
Experimentally, asymmetric substrates could be created
using periodic gradients in the sample thickness, 
DMI, doping, irradiation, or the magnetic field. 

\subsection{Coupling Skyrmions to Other Quasiperiodic Lattice Structures}

\begin{figure}
\includegraphics[width=\columnwidth]{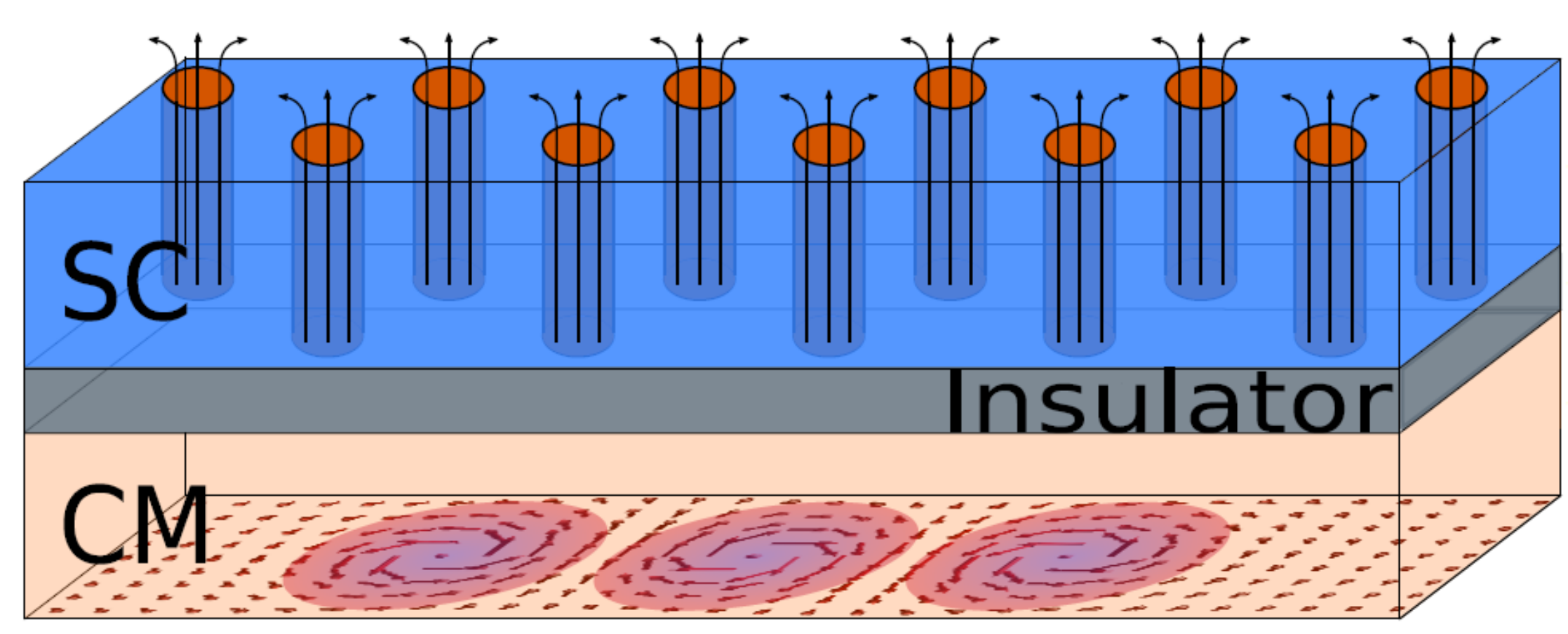}
\caption{
Schematic of the coupling between a chiral ferromagnet (CM, tan)
containing a skyrmion crystal (purple) and a superconducting film (SC, blue)
\cite{Dahir19}.
The materials are separated by a thin insulating barrier (gray) to
ensure that only the magnetic fields from the skyrmion lattice pass into
the superconductor. The attractive interaction between vortices and
skyrmions generates vortices (orange) in the superconductor.
Reprinted with permission from S. M. Dahir {\it et al.}, Phys. Rev. Lett.
{\bf 122}, 097001 (2019). Copyright 2019 by the American Physical Society.
}
\label{fig:73}
\end{figure}

\begin{figure}
\includegraphics[width=\columnwidth]{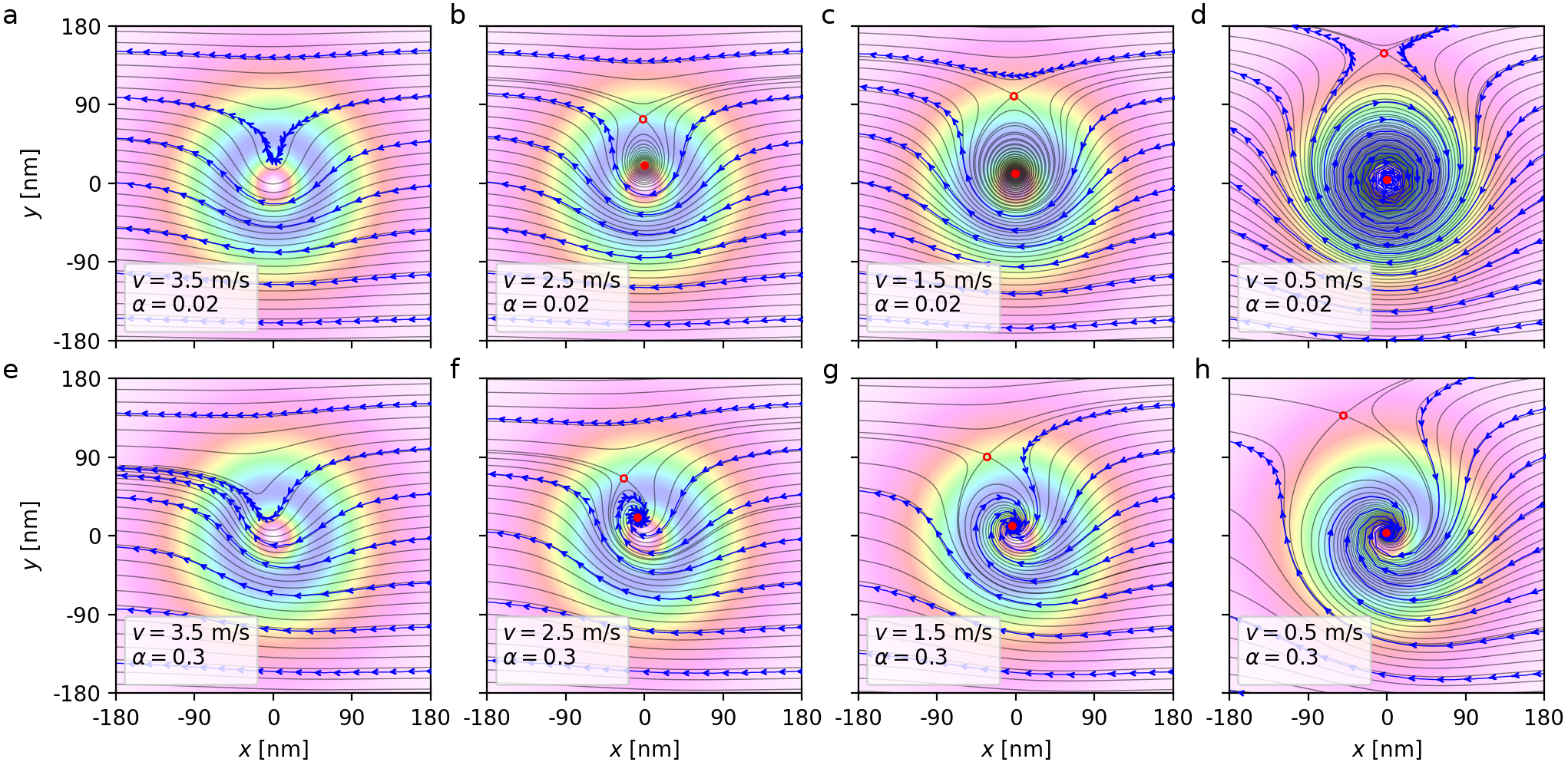}
\caption{
Micromagnetic calculations (arrows) and Thiele equation calculations
(thin lines)  
of skyrmion trajectories in the moving frame
produced by interactions with a moving superconducting vortex
\cite{Menezes19a}.
The background coloring represents the $z$ component of the magnetization
from the vortex that would appear in the absence of the skyrmion.
Open dots represent fixed saddle points and filled dots indicate
stable spiral points.
Reprinted with permission from R. M. Menezes {\it et al.}, Phys. Rev. B
{\bf 100}, 014431 (2019). Copyright 2019 by the American Physical Society.
}
\label{fig:74}
\end{figure}

Another method for creating
periodic structures that can act like pinning sites
is to cause the skyrmions
to interact with other topological objects, such as
vortices in a type-II superconductor.
More generally, there is interest in coupling skyrmions to
superconductors in order to control certain topological aspects of
the superconductor \cite{Mascot21}.
Several studies have already examined
interactions between superconducting vortices and skyrmions
\cite{Hals16,Baumard19,Dahir19}. 
Figure~\ref{fig:73} shows a schematic
from Dahir {\it et al.}
of a chiral ferromagnet coupled to
a superconducting thin film through an insulating layer
\cite{Dahir19} 
In this case, the skyrmions produce a
vortex-antivortex lattice in the superconductor.
Baumard {\it et al.} \cite{Baumard19} considered 
a thin film superconductor
in which Pearl vortices are induced by the skyrmions.
In these systems, the ratio of
the number of skyrmions to the number of vortices
can be tuned with a magnetic field,
and the skyrmions experience an effective
periodic substrate produced by the periodicity of the 
vortex lattice.
Further approaches include studying the effect of
a driving current
on such systems,
where the voltage response in the superconductor could be used as a
means to detect the motion of the skyrmions.
The effects of either naturally occurring or
artificially nanostructured
pinning could also be explored.
Menezes {\it et al.} \cite{Menezes19a} calculated the dynamics of skyrmions
interacting with a moving superconducting vortex
using both micromagnetic simulations and the Thiele equation.
In Fig.~\ref{fig:74}, the skyrmion 
trajectories in the moving frame
exhibit gyroscopic spiraling motion,
and in some cases skyrmions are captured by the vortex core.
Recently Palmero {\it et al.}
demonstrated experimentally that skyrmions could be used to
tailor a pinning potential
for vortices in a type-II superconductor \cite{Palermo20}.  
Future directions include analyzing different types of skyrmions
interacting with vortices, or considering
vortices in bulk rather than thin film systems. 

\subsection{Single Skyrmion Manipulation}

By
dragging
a single particle through a random disordered 
background or through a bath of other particles,
it is possible to produce
a local probe of
colloidal assemblies \cite{Reichhardt08a,Puertas14}
or superconducting vortices
\cite{Kafri07,Straver08,Reichhardt09,Auslaender09}.
The features in the velocity-force curves obtained for the probe particle
can be used to gain insight about the behavior of the bulk system,
such as changes in the viscosity and pinning force
as well as the existence of cutting or entanglement.
A similar local probe technique could be applied to a skyrmion system
by dragging individual skyrmions with some form of tip or
by coupling an individual skyrmion to a driven object.
Wang {\it et al.}
\cite{Wang20b} have proposed to use an optical tweezer to
manipulate skyrmions by optically trapping and dragging the skyrmion.
In this case, if the tip speed is too fast, the skyrmion could break away
from the tip. Other possible methods for creating a local probe
include dragging a skyrmion with a magnetic tip or
even dragging a group of skyrmions
with an array
of optical traps or magnetic tips.  

Along these lines, there is also the possibility
that skyrmion systems could host Majorana fermion states
\cite{Yang16,Rex19}.
Thus, one application
of skyrmion
manipulation could be to drag Majorana-containing
skyrmions around one another on a patterned
substrate in order to create braided Majorana states
for qubit operations.
Such operations have been proposed for
superconducting vortex systems with Majorana states
in the vortex core \cite{Ma20}.
The vortices are coupled to a periodic pinning array and a
magnetic tip is used to perform
a representative set of braiding moves that contain
all of the necessary operations for quantum logic gates.
A similar approach could be used for skyrmions.  

\section{Summary} 
Skyrmions are attracting increasing interest
as new materials continue to be identified
which support 
different skyrmion species as well as related topological objects.
Since skyrmions can be manipulated
or driven by a variety of techniques,
the role of pinning or quenched disorder will become a more important
aspect of future skyrmion studies.
There is already considerable evidence that
skyrmions can experience both weak and strong
pinning effects depending on the sample thickness or material type,
and it has been demonstrated that 
skyrmions exhibit a rich phenomenology
of dynamics, including gyroscopic motion and the skyrmion Hall
angle, all of which appear to depend on
the nature of the disorder as well as on the drive.
Due to the presence of the Magnus force, both
individual and collective skyrmion states
undergo new types of pinning and depinning phenomena
that are distinct from those previously studied in overdamped systems.
Pinning and
dynamic effects of skyrmions
interacting with disordered or ordered substrates
are of technological importance for
skyrmion applications, and the Magnus effects in the skyrmion system
open a new
field in equilibrium and nonequilibrium statistical mechanics. 

\begin{acknowledgments}
We gratefully acknowledge the support of the U.S. Department of
Energy through the LANL/LDRD program for this work.
This work was supported by the US Department of Energy through
the Los Alamos National Laboratory.  Los Alamos National Laboratory is
operated by Triad National Security, LLC, for the National Nuclear Security
Administration of the U. S. Department of Energy (Contract No. 892333218NCA000001).
\end{acknowledgments}

\bibliography{mybib}

\end{document}